\documentclass[review,jog]{igs}

\usepackage[utf8]{inputenc}

\usepackage{mathabx}
\usepackage{graphicx}
\usepackage{siunitx}
 \usepackage{hyperref}
\usepackage{float}
\usepackage{booktabs}
\usepackage{geometry}
\usepackage{amsmath}
 \usepackage{amssymb}
\usepackage{lineno}

\jourvolume{X}
\jourissue{X}
\jourpubyear{2025}

\begin{document}

\title{Mapping Supraglacial Water as a Window into Surge Hydrology: Linking Surface Water, Drainage Efficiency, and Surge Dynamics on Negribreen, Svalbard}

\author[Middleton and others]{Rachel Middleton,$^{1,2}$ Ute Herzfeld,$^{1}$ Thomas Trantow $^{1,2}$}

  \affiliation{%
  $^1$Geomathematics, Remote Sensing and Cryospheric Sciences Laboratory, Department of
Electrical, Computer and Energy Engineering, University of Colorado Boulder, Boulder, CO, USA\\
$2$Department of Civil,Environmental and Architectural Engineering, University of Colorado, Boulder, CO, USA\\
Correspondence: Rachel Middleton
\email{rachel.middleton@colorado.edu}}

\begin{frontmatter}

\maketitle

\begin{abstract}
We analyze the dynamics of Negribreen Glacier System, a polythermal glacier in Svalbard, during its ongoing surge and investigate the role of supraglacial (surface) water as both an indicator of ice-dynamic processes and a driver of surge evolution. We identify three distinct surge phases: the initial acceleration phase, mature phase, and return to quiescence. Comparing the quiescent supraglacial hydrological state to each of the surge phases, we observe a sudden increase in hydrological connectivity between the glacier surface and base during initial acceleration, followed by a gradual return to quiescent water extent. In the mature surge phase, emergent water-filled crevasses coincide with regions of compressive forcing and extensive deformation, follow local accelerations, and preceded smaller, secondary accelerations. Additionaly, rapid drainage of surface ponds is observed in the mature surge.  A data-fusion approach, using Maxar WorldView(c) imagery, ICESat-2 altimetry, and Sentinel-1 Synthetic Aperture Radar, is taken to create a time series of supraglacial water maps, water volumes, surface velocity changes, and spatial ice surface roughness. 
 These observations provide a qualitative (process understanding) and quantitative (water time series) basis for supraglacial water sources as a driver and indicator of surge activity for Arctic glaciers.
\end{abstract}

\end{frontmatter}

\section{Introduction}

\subsection{Glacial Hydrology}

Water on the surface of the glacier, called supraglacial water, can take the form of melt ponds, flow in melt streams and fill crevasses \cite{Decaux-2019}. Water trapped within the glacier itself, called englacial water, can fill voids in the ice  and flow down small channels that connect the surface to the base of the glacier \cite{barrett-2008}. Water beneath the glacier, called subglacial water or basal water, can flow through subglacial channels, draining from the glacier, or become trapped beneath the glacier.

This paper is motivated by the current surge of NGS, an Arctic, polythermal glacier. At least two types of surge mechanisms exist \cite{murray-2003}, and the current NGS surge provides an opportunity to examine the role of water in the surge of a polythermal Arctic glacier, in contrast to a temperate surge glacier. We focus first on a solution for detecting and mapping water, of which we find supraglacial water is mappable water. We then consider mapped water as an indicator of processes internal to the glacier during the surge. This necessitates a comprehensive introduction to several components of this work: mapping supraglacial water, considering sources of supraglacial water, and identifying geophysical processes of the surge by introduction of other remote sensing techniques to build on existing surge theory.

Liquid water critically alters surge dynamics, as the role of subglacial/basal water has been clearly demonstrated with numerical modeling of glacier dynamics \cite{flowers-2002, sommers-2018, SHMIP, anderson-2004,trantow-2024}. Trapped basal water lubricates the bed below the glacier, acting as a friction-reducing agent and modifying the basal boundary condition in a numerical model of the glacier body \cite{flowers-2002, sommers-2018,SHMIP}. Flowing basal water removes both mass and heat from the glacier, and erodes subglacial drainage channels \cite{SHMIP, Kamb-1987}. However, the thermodynamic and mechanical role that supraglacial water plays in surge glacier dynamics, both as a source for subglacial water and in the surface enthalpy and stress regime have received less attention, with studies mainly focus on supraglacial water for non-surge-type glaciers \cite{gilbert-2020, chudley-2021}.

Sources of supraglacial water include rainfall, runoff water from nearby areas, and melting of snow or firn, partially compacted, porous snow \cite{firn}, on the glacier's surface.

At the bed, basal water originates from the following sources: (i) surface water inputs are routed through crevasse networks \cite{Decaux-2019}, (ii) pressure-melting caused by the reduction of ice’s melting point under high overburden pressure \cite{irvine-2011}, (iii) frictional heat, generated by ice motion over the bed, melting ice at the ice-bed interface \cite{SHMIP}, and (iv) geothermal flux from the bed \cite{SHMIP}; flow separation around bedrock bumps additionally creates low-pressure cavities through cavitation \cite{greve-2009}, enlarging subglacial storage.

Sudden deformation of the glacier can dam the glacier's subglacial hydrological drainage system, or high amounts of overburden pressure can slow the development of channel opening, also daming the drainage system, resulting in large amounts of water stored subglacially or englacially \cite{irvine-2011} and preventing drainage of supraglacial water. While it is known that subglacial channels are largely dependent on changes in the supraglacial drainage system \cite{Decaux-2019}, the extent to which supraglacial water changes based on the dynamics itself has received less attention. 
In this work, we focus on understanding how supraglacial water evolves as the glacier of interest, Negribreen, experiences extreme deformation during its current surge (2016-current(2025)) \cite{trantow-2025, herzfeld_campaign_2019, herzfeld_icesat2.crev}.

   \subsection{Surging is a Unique Glacial Acceleration Mechanism}
    
    Surging is a glacial acceleration mechanism observed in at least 1148 glaciers and expected to occur in at least 2317 glaciers globally \cite{sevestre-2015}. However, it is the least well understood acceleration mechanism and was attributed to be the largest source of uncertainty in estimating future sea-level rise \cite{IPCC5, herzfeld_campaign_2019}. 
Surge-type glaciers exhibit a rare type of acceleration in which the glacier accelerates in accordance to phases: the quiescent phase, in which little movement is observed and mass is accumulated, and the surge phase, in which the glacier accelerates to 10-1000 times more than its quiescent velocity \cite{murray-2003, Clarke-1984,sevestre-2015, trantow-2024}. 

Typical periods of quiescence are 20-100 years and surges typically last 5-10 years\cite{sevestre-2015}. Mass loss experienced during a single surge period exceeds the mass loss experienced during the much longer quiescent phase. 
While surge glaciers are rare, they are limited to specific geographic clusters, including portions of Svalbard, Alaska, Canada, Greenland, and High Mountain Asia \cite{sevestre-2015}.  
Several shared surge mechanisms are responsible for the surges of these glaciers and surges occur on a quasi-cycle, but the periodicity of surges varies between regions (3-5 years for temperate and 5-10 years for arctic, polythermal glaciers \cite{murray-2003}. Because polythermal, arctic glaciers surge for longer periods than their temperate counterparts, it is common to segment the surge phase into its own ``sub-phases'': the initial acceleration phase (peak velocity, first appearance of crevasses, and acceleration along the entire glacial body), the mature surge phase (velocities begin to slow and complex deformation features occur including water-filled crevasses \cite{murray-2003, herzfeld-2004, mcdonald_2013_IGS_lidarberingbagley}), and a return to quiescence (subsidence of kinematic and deformation activity) \cite{herzfeld_campaign_2019, trantow-2025, haga-2020, benn-2019}.

    \cite{murray-2003} compared surges of temperate glaciers to surges of polythermal glaciers and initially proposed that at least two different surge mechanisms existed. 
    Temperate glacier surges could be enabled by the ``linked cavity mechanism", proposed by \cite{Kamb-1987}. 
    This mechanism assumes high water pressures are introduced at the glacier bed when an inefficient ``linked cavity" drainage system develops as a result of the freeze-closure of existing drainage channels. The theory of this mechanism was developed based on dye-tracing experiments at Variegated Glacier, Alaska \cite{Kamb-1987}. The surge of Variegated Glacier ended abruptly, with the sudden collapse of linked cavity drainage system and an outburst of subglacial water at the terminus, further leading credibility to the surge mechanism proposed by \cite{Kamb-1987}. 
    
    Polythermal glaciers are observed to terminate their surges more gradually, as is the case for Negribreen and other surges in Svalbard \cite{murray-2003,benn-2019, barrett-2008}, implying that a sudden collapse of a subglacial drainage system would not explain the termination of the surge. Additionally, supraglacial water on temperate surge-type glaciers is more extensive, filling marginal lakes and many water-filled crevasses, lending more evidence for the presence of a large inefficient drainage system. Previous surge studies, have focused on the role of basal water in surge initiation \cite{murray-2003, Kamb-1987, Kavanaugh-2000,Clarke-1984}, with relatively little focus on the role of the supraglacial drainage system in the continuation of the surge \cite{copland-2003}. 
    Several mechanisms for surging of polythermal glaciers have been proposed. 
         \cite{sevestre-2015} argued that all surging could be explained by their oscillation of (1) mass, (2) enthalpy, and (3) velocity, and a glacier which has a limiting factor of either mass balance or enthalpy balance can surge\cite{benn-2019}. 

            Considerably, water plays the most critical role in the enthalpy balance scheme of a glacier moving at slow, quiescent speeds: allowing for warming of the ice and the conversion of cold ice to temperate ice in polythermal glaciers \cite{gilbert-2020}. 
            
    \cite{fowler-2001} proposed that surging could also be enabled by thermal deformation of a soft bed, in which enough pressure at the bed of the glacier, which is a result of ice overburden by continual building of mass during the quiescent phase, brings ice at the glacier bed interface to the pressure melting point, and heat is lost to melted water at the bed. As a result, basal water deforms the soft bed (often till) in regions of inefficient drainage, and further allow for progression of the surge over the resultant wet, deforming bed sediments, as was seen in the case of a nearby surging glacier, Bakaninbreen \cite{barrett-2008}. Where then are these regions of inefficient drainage and are they indirectly observable from surface water ponding in regions of large deformation? This is the focus of this study, as we will focus on trends in surface ponding during the development of the Negribreen surge.

\subsection{The Hydrology of Surge-Type Glaciers}
    
    Observational studies supported the theory that the englacial hydrologic system for temperate surge-type glaciers experienced critical changes prior to a surge \cite{Kamb-1987, Clarke-1984, copland-2003, trantow-2024}. During the quiescent phase, surge glaciers efficiently drain water through a basal network of large, discrete channels. During the surge, the englacial hydrological network is inefficient and consists of many distributed smaller channels or cavities. Prior to a surge, a transition between this efficient system and inefficient system occurs, and is strongly related to increased ice overburden disrupting efficient drainage channels.
    
    Hydrological drainage configurations of polythermal glaciers has received less attention relative to their temperate counterparts. Polythermal glaciers contain both a temperate ice layer, often a bottom or internal layer, and cold ice layer, which exists in thinner sections of the glacier or on top of the temperate ice layer \cite{irvine-2011}. 
    Temperate ice and cold ice in a polythermal glacier are seperated by the cold transition zone (CTZ), which was previously viewed as a ``thermal dam" preventing regelation (pressure melting) and meltwater flow from occurring beneath the CTZ \cite{Blatter-1991, Clarke-1984}.
    However, observations of basal water pressure \cite{Murray-2001, Wohlleben-2009}, englacial water \cite{barrett-2008, kendrick-2018}, and water outflow \cite{hagen-2000} suggest polythermal glaciers can develop basal hydrologic drainage systems equally complex to temperate glacier hydrological drainage systems. Subglacial drainage structures of polythermal glaciers follow basal hydraulic potential \cite{hagen-2000}, similar to temperate glaciers. 
    
    The routing and residence time of supraglacial glacial liquid water could be governed by the glacier’s stress regime \cite{chudley-2021}. Compressive stress arises where ice flow decelerates downstream (negative strain-rate) \cite{herzfeld_mayer1997bagley, mayer_herzfeld_clarke2002_crevegs}, forcing ice to thicken and encouraging water to pond or become trapped. Examples of regions with compressive stress include the up-glacier side of over-deepenings and shear margins of slow-moving lobes. Extensional stress develops where flow accelerates (positive strain-rate), thinning the ice and opening pathways that promote rapid drainage, this is typical of fast-flowing ice streams, crevasse fields, and many marine-terminating glaciers near their grounding lines. 
Whether surface and basal waters are stored or drained at a given location depends on this dynamic balance between compressive and extensional zones, linking glacier mechanics directly to hydrological partitioning and surge potential.

    When basal water storage occurs (inefficient subglacial drainage),  the basal effective pressure, $N$, is reduced by the increasing basal water pressure \cite{SHMIP}.  Effective pressure is defined as the difference between the ice overburden pressure, $p_i$, and basal water pressure, $p_w$: $N = p_i - p_w$. Effective pressures near zero allow for low friction conditions in which the ice can simply slide over the base\cite{anderson-2004}.
Theoretical configurations of inefficient hydrological drainage systems \cite{SHMIP} represent a linked cavity drainage system \cite{Kamb-1987} as a porous sheet \cite{Glads,hewitt-2011}, distributed cavities \cite{anderson-2004,kessler-2004}, or variations of either \cite{sommers-2018}. 
Surge glaciers experience abrupt and rapid deformation, which may be aided by the reduction of effective pressure, and hence, increased basal sliding \cite{irvine-2011, copland-2003}.

    A multitude of observational studies support the persistence of efficient, channelized drainage for Svalbard-type glaciers moving at slow velocities: 
    several studies produced maps of subglacial and englacial channels from direct cave entry and exploration in drainage systems of cold-based Svalbard glaciers \\ \cite{benn-2009,naegeli-2014,mankoff-2017} and in other studies terrestrial laser scanning  (TLS) and ground penetrating radar (GPR) surveys have revealed dendritic channels or large conduits \cite{kamintzis-2023, hansen-2020, karuss-2022}.

     Fewer studies have documented inefficient drainage systems, particularly leading up to or during a surge. \cite{porter-2001}  measured water pressure and strain rate in subglacial till during the late surge phase of Bakaninbreen, Svalbard to better capture the role of the basal till layer. Water-pressure and pore-pressure records with till deformation show hydraulically inefficient conditions (high water-pressures).  \cite{smith-2002} interpreted these measurements in conjunction with complementary ground-penetrating radar data to conclude that high water pressures maintained the propagation of the surge in Bakaninbreen, and that the surge was terminated by reduction of water pressure by the escape of water through a portion of the bed which was thawed. Similar observations have been made for Kronebreen, Svalbard \cite{how-2017} and Kongsvegen, Svalbard \cite{bouchayer-2024}, but reflect that changes to subglacial drainage can occur within a season, even during quiescence, enhancing sliding from elevated water pressures when influx to the bed is sufficiently high.

\subsection{Remote Sensing of Supraglacial Water}

Remote sensing of supraglacial water features and their subsequent detection and monitoring from satellite data is common in the cryospheric science community for features such as such as melt ponds in arctic sea ice\cite{dda-bif-seaice, webster-2022, spergel-2021} and glacial lakes \cite{stokes-2019,huang-2018,legg-2019}. However, smaller-scale water features, such as water-filled crevasses, require higher-resolution data and highly tuned or automated algorithms, to map. This work establishes a framework for observing small-scale supraglacial  water features by combining spectral classification of Maxar WorldView (c) multispectral images with ice and water surface heights determination from ICESat-2 altimetry. 
Furthermore, this framework is utilized to create a time series of supraglacial water maps and water volume estimates on Negribreen Glacier System (NGS), providing insight into the state of the hydrological system during the evolution of a surge event.

\section{Approach}

\subsection{Study Area: Negribreen Glacier System (NGS)}

Negribreen is a surge-type glacier located in west-central Svalbard, an archipelago which lies approximately midway between Norway and the North Pole (Fig. 1), and more specifically on the Filchnerfonna, an ice area named after explorer, Wilhelm Filchner. Negribreen and its tributary glaciers, Ordonnansbreen, Akademikarbreen, and Transparentbreen to the north, and Filchnerfallet and Rembebreen to the south \cite{trantow-2025}, are collectively referred to as the Negribreen Glacier System (NGS). 

Negribreen is currently experiencing a surge \cite{herzfeld_campaign_2019}, which began in 2016, and reached maximum velocities (22 m/day) during the height of the acceleration phase in July 2017 \\ \cite{trantow-2025,herzfeld_campaign_2019,twickleret2025_geoclass2_frontiers,haga-2020}. In June 2024, Negribreen was still moving at surge speeds, approximately 5 m/d (Fig S2).

The study area is analyzed in three separate regions, which were assigned based on the mature phase deformation state, an upper-crevasse region, a middle-crevasse region, and a lower crevasse region (Fig. 1). In the mature phase of the NGS surge, the upper crevasse region has water-filled crevasses and some complex crevasse structures, the middle region has extensional, simple crevasses, and the lower crevasse region has chaotic crevasses. This segmentation approach aids in creation of accurate and temporally resolved water maps, as imagery often needs to be stitched together to cover the entire glacier, and images are often not coincident across all three regions. Additionally, the relative occurrence of water is consistent within each of these regions, but varies more between them, making it intuitive to examine water occurrence in each of the regions separately.

Large and extensive crevasse fields with unique surface water features on NGS have resulted from the many years of quick movement, which are observable from satellite imagery and airborne campaigns \cite{herzfeld_campaign_2019}. The last surge of Negribreen occurred in 1935-1936 \cite{lafauconnier-1991}. Debate over surge trigger mechanisms warrants a detailed description of the glacier's geometry, thermal regime, bed, and climatology.

\subsubsection{Geometry}  Negribreen has a length of $\sim$ 27 km along its flow direction and  width of $\sim$ 6 km perpendicular to its flow direction. Its area is $\sim$ 160 sqkm, not including its source areas or tributary glaciers. NGS has an approximate area of 330 sqkm, including the tributary glaciers and source areas. 
Negribreen's observed surface heights range from 60 m at the calving front to 800 m at its source area \cite{trantow-2025}, and hence Negribreen has a $\sim$ 2.7\% rise from its terminus to source area.
Prior to its surge, Negribreen had a median thickness of 171 m and standard deviation in thickness of 74 m,  reaching a thickness of 400 m in the middle crevasse region (see Fig. 1), according to bed and surface heights reported in \cite{furst-2018}.
Between 2019-2020, the upper $\sim$ 2/3 of Negribreen experienced surface lowering, up to -30m in some areas, while the portion of the glacier we call the lower crevasse region experienced general thickening, up to +30m in some areas \cite{trantow-2025}.

\subsubsection{Thermal regime} 
Radio echo sounding of Negribreen in 1984 by \cite{Dowdeswell-1984} detected higher internal reflectors 80-190 m below the glacier surface, implying that a temperate layer of ice existed below the cold overlying ice.  Analysis of nearby marine-terminating surge glaciers \cite{Murray-2001, sevestre-2015} with similar geometry to NGS have a two-layer thermal regime consisting of a temperate ice layer in contact with the bed superimposed by a cold ice layer, with cold ice margins and a small region of cold ice at the glacial front. Extensive crevassing during Negribreen's surge has undoubtedly disturbed this simple thermal composition \cite{gilbert-2020}, allowing at least for transient storage of water in the top layer of melting and deforming cold ice and also potentially allowing for hydro-fracture to penetrate the CTZ increasing the hydraulic connectivity of Negribreen.

\subsubsection{Climatology} 
Svalbard's precipitation is low by global standards, but varies across the archipelago, having a typically dry west coast ($<$400mm/yr) and relatively wet eastern coast (reaching 1000 mm/yr) (see Fig. 2). A positive eastward gradient in precipitation reflects exposure to moisture from the Barents Sea and reduced topographic prevention of rainfall (less rain-shadow effect from Spitsbergen's mountains). Annual precipitation observations across all of Svalbard suggest an increasing trend in annual precipitation \cite{das-2025}.

The Filchnerfonna region, containing NGS, and located in eastern Svalbard is expected to receive higher amounts of precipitation than those observed at the Svalbard Airport, which are available for download at \url{https://seklima.met.no}. Reanalysis data \cite{NCEP-II} that covers the NGS location also reflects higher precipitation overall, 75.1 mm more annually on average, or a 39.3\% increase.
Negribreen has a southwest-facing slope resulting in higher amounts of solar radiation than its northern-facing counterparts, implying further amounts of surface warming and melt. 
Seasonal trends in Svalbard precipitation typically reflect that most precipitation falls from late July to December, some precipitation from January to March, and very little precipitation from April to June.
Recent extreme events include record wet years, 2016-2017, which brought mudslides \cite{lapointe-2024, das-2025}, coinciding with the start of the Negribreen surge.

\subsection{Data}
Observed water features on the Negribreen Glacier System (NGS) include melt ponds, melt streams, and, most importantly, water-filled crevasses. 
The observational capability of these water features is dependent on the resolution of multi-spectral imagery and the accuracy of the classification tool.

\subsubsection{Worldview Imagery}

High resolution multi-spectral Maxar WorldView (c) imagery was acquired for both visual inspection and mapping of observable water features. 
The imagery acquired consisted of imagery from Worldview 1, 2, and 3 satellites which share the same spectral bands and have an optimal resolution of 1.85 m \cite{wv-specs}. 
Water is potentially detectable during the warm season (May-September) of each year, and imagery is acquired for these months between the years 2015 and 2023 to capture the glacial surface prior to the surge, during the initial rapid acceleration phase, and the mature surge phase. 
Only images with less than 10\% cloud cover over Negribreen are retained for the spectral classification of the ice surface for purposes of mapping water features.

\subsubsection{Sentinel-1 Synthetic Aperture Radar Data for Deriving Velocity Maps}

Synthetic aperture radar (SAR) data, collected by the European Space Agency's (ESA) Copernicus Mission Sentinel-1 satellite, is utilized to generate velocity maps of NGS using the SNAP toolbox \\ \cite{SNAP}. Here, we base the velocity maps on ground range detected (GRD) SAR images, with digital number (DN) values representing the amplitude of the measured radar backscatter signal. The GRD data were radiometrically calibrated in the ESA SNAP toolbox \\ \cite{SNAP}. We plot differences of these velocity maps to depict areas of acceleration or deceleration. Velocity maps are reported in supplementary information.

\subsubsection{ICESat-2 Photon Point Clouds for Determining Crevasse Depth and Thickness Evolution}

To determine ice surface heights, we analyze data collected by the National Aeronautic and Space Administration (NASA) Ice, Cloud, and Land Elevation Satellite-2 (ICESat-2) Mission.

With the Advanced Topographic Laser Altimeter System (ATLAS), NASA's ICESat-2 Mission has been collecting multi-beam, micro-pulse, photon-counting laser altimeter data since September 15, 2018 \cite{ATL03-ATBD}. Hence, ICESat-2 data is concurrent with most of the surge of the NGS, which enables the surface altimetry component of this study. 
With an along-track nominal resolution of 0.7 m along-track (under clear-sky atmospheric conditions), ICESat-2 records ice surface returns at a resolution that is sufficient to capture complex surface heights of crevassed glaciers. This information (the geolocated photon point cloud) is reported on the ICESat-2 ATLAS Data Product 3 (ATL03) \cite{ATL03-ATBD}.

Retrieval of surface heights, including crevasses and other morphologically complex features, is facilitated by the Density-Dimension Algorithm for ice surfaces (DDA-ice)\cite{herzfeld_surface-height_2017,herzfeld_icesat2.crev}, see section 2.3.4.

ATL03 granules used for derivation of crevasse depths are reported in supplementary information.

\subsection{Methods}

The Negribreen surge expresses as coupled changes in ice deformation, ice motion, and the glacier’s hydrologic routing. These processes are monitored by mapping supraglacial water, spatial ice surface roughness (SISR), and changes in summer velocity. We interpret these spatial signatures jointly to make theoretical inferences about surge glacier dynamic processes and the hydrologic state applicable to Svalbard-type surge glaciers, similar to Negribreen. 
While the types geophysical processes that can be revealed by SISR and velocity changes are well established, the potential geophysical processes that can detected with supraglacial water maps has received relatively little attention. By explicitly considering the timing of supraglacial water occurrence within the context of the geophysical processes depicted by roughness and velocity changes in addition to the spatial similarities between the three spatial signatures, we establish the geophysical processes that are detected or related to supraglacial water occurrence and, in doing so, develop a comprehensive theory for surge mechanisms applicable to Svalbard-type surge glaciers.

\subsubsection{Spatial Ice Surface Roughness (SISR) from Sentinel-1 SAR}

To characterize deformation associated with the surge, we quantify spatial ice-surface roughness (SISR) directly from Sentinel-1 SAR backscatter using geostatistical vario functions, which are spatial functions defined in a discrete mathematics framework that capture spatial variability similar to probabilistic variograms, but generalize to residual functions and functions of higher order and are calculated numerically in a straightforward way from experimental data \cite{herzfeld2002_varvar}. Differences of intensity values are formed and averaged over all pairs of pixels separated by a given lag $h$ to compute the first-order vario function,
\begin{linenomath*}
\begin{equation}
\label{eq:v1}
v_1(h)=\frac{1}{2n}\sum_{i=1}^{n}\big[z(x_i)-z(x_i+h)\big]^2,
\end{equation}
\end{linenomath*}
and, to reduce sensitivity to large-scale radiometric trends, the residual vario function,
\begin{linenomath*}
\begin{equation}
\label{eq:res}
m(h)=\frac{1}{n}\sum_{i=1}^{n}\big[z(x_i)-z(x_i+h)\big], \qquad
res_1(h)=v_1(h)-\tfrac{1}{2}m(h)^2,
\end{equation}
\end{linenomath*}
where $z(\cdot)$ denotes calibrated SAR intensity and $n$ is the number of pixel pairs with separation $h$ \\ \cite{herzfeld2002_varvar,herzfeld2008_master,herzfeldet_2014_iceroughjakjglac}.

We summarize roughness with the geostatistical parameter $\mathit{pond}_{res}$, defined as the maximum of the experimental residual vario function as in \cite{herzfeldet_2015_seaicemodel.frampaper} within a moving analysis window,
\begin{linenomath*}
\begin{equation}
\label{eq:pondres}
\mathit{pond}_{res}=\max\{\,res_1(h_i)\mid i=1,\ldots,n\,\},
\end{equation}
\end{linenomath*}
with $h_i=i\,h_{unit}$. In this study we use a $40\times 40$\,pixel moving window, a unit lag of $h_{unit}=20$\,pixels, and compute directional vario functions along eight azimuths; we map SISR as the directional average of $\mathit{pond}_{res}$ across these directions. This residual formulation suppresses broad radiometric or incidence-angle trends and emphasizes roughness produced by crevasse fields and other high-frequency surface texture associated with surge deformation \cite{herzfeld2008_master,herzfeldet_2014_iceroughjakjglac}.

Sentinel-1 Level-1 GRD data (detected amplitude) were radiometrically calibrated in the ESA SNAP toolbox \cite{SNAP} to backscatter, and $\mathit{pond}_{res}$ is computed per window and direction on the calibrated backscatter values, then averaged to produce a scalar roughness map. Generally, high $\mathit{pond}_{res}$ marks strongly crevassed provinces; low $\mathit{pond}_{res}$ corresponds to smooth, undeformed or snow-covered ice. In Negribreen, the emergence and spatial expansion of elevated $\mathit{pond}_{res}$ tracks the advance and maturation of the surge and aligns with the development of crevasse provinces discussed below \cite{herzfeldet_2014_iceroughjakjglac}.

\subsubsection{Ice Surface Velocity Changes}

Velocity maps were derived from pairs of Sentinel-1 synthetic aperture radar (SAR) level-1 Ground Range Detected (GRD) images with a 12 day difference in acquisition time. The ESA SNAP program was used to conduct offset tracking \cite{SNAP} between the pairs of SAR images, which provided a series of seasonal velocity maps \cite{trantow-2025}.

To identify regions of acceleration or deceleration during the surge, we calculate the difference between July velocities for years 2016-2020. To identify regions of seasonal acceleration prior to the surge (during the quiescent phase), we also calculate the difference in velocities between January 2015 and August 2015.

\subsubsection{Binary Occurrence of Supraglacial Water/ Spectral Classification}

Multi-spectral imagery with sufficiently high resolution, as exemplified by the imagery used in this study (1.85 m optimal resolution) is capable of capturing intensity values within small-scale water features, such as water-filled crevasses.
Furthermore, pixel values across multiple spectral bands allow for a spectral classification approach to identify pixels containing water.

One commonly used classification approach is to calculate a spectral index, such as the Normalized Difference Water Index (NDWI) (see Eq. 4), for the entire image and classify pixels as containing water if they exceed a specified threshold. 
Spectral indices are combinations of reflectance values in specific bands designed to highlight particular surface properties. The concept dates back to the early days of remote sensing \cite{Kriegler-1969} and was applied to vegetation monitoring in the 1970s with the introduction of the Normalized Difference Vegetation Index (NDVI) \cite{rouse-1973}. NDVI exploits the fact that healthy green vegetation strongly reflects near-infrared (NIR) light but absorbs red light, yielding high NDVI values for healthy vegetation and low NDVI values for bare ground or water. The success of NDVI spurred the development of many other normalized differences tailored to detect water \cite{mcfeeters-1996}, snow \cite{dozier-1989}, burn scars \cite{burn-scars-spectral-index}, and more. 

Among these indices, NDWI was introduced in the mid-1990s to detect open water features in multispectral imagery \cite{mcfeeters-1996} which originally used green and near-infrared bands. Other modified versions of NDWI have been proposed, \cite{xu-2006} uses the green and SWIR bands, for example. 
Applying NDWI to glacier surfaces requires special consideration as snow and glacier ice have very different spectral characteristics than vegetated or urban landscapes, which affects how we design a useful water index. Clean ice and snow are highly reflective across the visible (red, green, blue) spectrum and only begin to absorb significantly in NIR \cite{kutuzov-2021}. The spectral reflectance curve for pure snow/ice typically peaks in the blue-green (400-600 nm) and then decreases toward the red and NIR \cite{qunzhu-1983}. Water shows an opposing spectral behavior, with low reflectance in blue and green and almost no reflectance in the red or NIR portions of the spectrum \cite{qunzhu-1983}. 

Thus, a standard green-NIR NDWI would likely fail on a glacier, because both water and snow/ice would yield low NIR reflectance, but a more suitable NDWI would used reflectances in the blue and red edge bands (see Eq. 4). A ponded water pixel on a glacier will appear very dark in red wavelengths, whereas adjacent ice or snow is relatively bright in red, and in the blue band, both water and ice reflect more similarly. The net effect is that water pixels have a much higher NDWI value, while ice/snow will have low NDWI values. \cite{yang-2013} demonstrated this on the Greenland Ice sheet by comparing the conventional green-NIR NDWI to a blue-red NDWI, and the blue-red formulation produced sharper lake boundaries and higher contrast between meltwater and the ice background. 

The choice of WorldView's blue (450-510 nm) and red edge bands (704-744 nm) is thus grounded in this prior knowledge.

 \begin{equation}
NDWI = \frac{\rho_{blue} - \rho_{red}}{\rho_{blue} + \rho_{red}}
\end{equation}

This work capitalizes first on the low spectral reflectance of small water bodies, as compared to surrounding glacial ice, in the red edge (704-744 nm) band alone. 
As a second measure, the blue/red NDWI is used to further isolate water features from spectrally similar glacial ice, which were not initially removed by the red threshold. Using NDWI alone was not found to be effective as shaded regions of the glacier were often misclassified as water. 

To reduce misclassification error, which is typically a result of moraine material and shaded glacial ice having spectral similarities to water, areas in topographic shadow are masked and the glacial surface is classified within four subregions, which mostly avoid medial moraines carrying source material.
The topographic shadow analysis employs a line-of-sight ray tracing algorithm to identify areas obscured from direct solar illumination. We compute the illumination intensity over NGS in the given image based on surface orientation relative to the sun position. 
For each pixel with elevation $h_0$ and position ($x_0,y_0)$, as given in the 30-m Arctic DEM \cite{Arctic-DEM}, a ray is traced along the sun's azimuth angle ($\theta_a$) and elevation angle ($\theta_e$). At distance $d$ along this ray, coordinates of illumination ($x_d, x_y$) for a pixel resolution of $res_x$ m in the x direction and $res_y$ m in the y direction are calculated as:

\begin{equation}
x_d = x_0 + d \cdot \cos(\theta_a) / res_x
\end{equation}
 
\begin{equation} 
y_d = y_0 - d \cdot \sin(\theta_a) / res_y
\end{equation}

The ray's height at distance $d$ is given by:

\begin{equation}
h_d = h_0 + d \cdot \tan(\theta_e)
\end{equation}

A pixel is classified as shadowed if any terrain along the ray path exceeds the ray height:

\begin{equation}
\text{shadow} =
\begin{cases}
	0, & \text{if $\exists d : h_{\text{terrain}}(x_d, y_d) > h_d$ }\\
        1, & \text{otherwise}
 \end{cases}
\end{equation}
A dilation buffer of 13 pixels is applied to the shadow mask to account for partial shadowing effects at shadow boundaries, producing a conservative mask that preserves data quality in subsequent water detection procedures.

Thus, the spectral classification proceeds as follows: for a given image with less than 10\% cloud cover over NGS, pixels with reflectances in the red band less than a user-defined threshold and an NDWI greater than another user-defined threshold were labeled as water. Areas in topographic shadow were masked out from the resulting labeled image, which were calculated by conducting a line of sight analysis based on the solar azimuth and zenith and NSIDC's 10m resolution arctic-wide digital elevation model.

Despite a careful selection red and NDWI thresholds for creation of water maps, we found this spectral classification approach to not demonstrate robustness. In section S3, we test the sensitivity of the thresholding approach by minimally perturbing the selected thresholds and measuring the percent change in returned surface area of pixels classified as surface water, with the highest percent changes greater than 150\%.


\subsubsection{Determining Crevasse Depth using the DDA-Ice Algorithm}

To determine ice surface heights for a given year, all warm season (MJJAS) ATL03 granules from ICESat-2 are processed with DDA-Ice using the parameters identified to be optimal for land ice \cite{herzfeld_icesat2.crev}, see table S1.

The DDA-ice algorithm was designed for surface height determination of crevassed and otherwise morphologically complex ice surfaces in addition to smooth ice surfaces \cite{herzfeld_surface-height_2017,herzfeld_icesat2.crev}, making it an ideal candidate for determining crevasse depths on NGS. 

All DDA methods are built around the calculation of the density field, using the Gaussian radial basis function,  and a density-based separation between signal and background (``noise") that, motivated by the notion that a geophysically valid reflector (such as the ice surface) has a higher density of photons in the received point cloud than background. 
More specifically, we utilize two variations of the DDA-ice algorithm: DDA-ice-1 and DDA-ice-2. The DDA-ice-1 performs the following steps, (0) cloud-ground separation,  (1) signal-noise separation, 
(2) calculation of the density field, (3) auto-adaptive threshold function, (4) surface-height determination (using a roughness-adapting ground follower). \cite{herzfeld_surface-height_2017, herzfeld_icesat2.crev}. 
The DDA-ice-1 for ATLAS data utilizes the geolocated photon cloud as reported in the ATLAS product ATL03 (but not the photon classification given in ATL03). Data postings of the density field  have the same spatial resolution as the original data, and surface heights are reported at near-sensor resolution, for example, at 2.5~m along-track for crevassed surfaces and 5~m along-track for smooth surfaces.
To detect secondary surfaces, such as the bottom of water in crevasses or melt ponds, the DDA-ice-2 can be applied. In essence, this algorithm performs two passes of the core steps of the DDA \cite{icesat2.water}.
The DDA-ice has been validated for surface-height determination of an Arctic  surge glacier during surge \cite{herzfeld_airborne_2022} and applied to two years of ICESat-2 data to derive glaciological processes during the surge \cite{trantow-2024}.

Although a total of 9 ICESat-2 reference ground tracks (RGTs) and 27 strong beams, three for each RGT, cross the NGS \cite{trantow-2025}, only one coincidence with a water-filled crevasse in RGT 594 GT1L (see Fig. 6) occurs, a result of both the spatial sparsity of water for this Arctic glacier and because there are spatial gaps (90m) between each ICESat-2 beam. Therefore, direct derivation of water depth is available for this coincidence only, which is derived by taking the elevation difference between the water surface and crevasse bottom, as determined by the DDA-ice-2 algorithm \cite{DDA-ice-2}.  Water depth is estimated to be a conservative 50\% of the total crevasse depth, which is reflected by the direct water coincidence (see Fig. 6).

We extract a single depth estimate for each crevasse which has been intersected by ICESat-2's ATLAS measurements and for which ice surface heights have been determined by the DDA-ice algorithm using a simple detection method for minima and maxima in the ice surface height profile and taking the mean depth for each crevasse feature (see Eq. 9).

\begin{equation}
{depth} = \frac{1}{2}[(max_{-1} - min) + (max_{+1} - min)]
\end{equation}
Where $max_{-1}$ is the local maximum at the closest along track distance before the local min and $max_{+1}$ is the closest local maximum after the local min.

Indirect derivation of crevasse depth is conducted for all mapped supraglacial water features in imagery by interpolating crevasse depth from the DDA-ice profile to the observed feature location with an inverse-distance weighted average (see interpolate\_depths\_to\_polygons function in ultility/dda\_depth.py in the github repository).

In summary, for a given year, all ICESat-2 ATL03 granules over NGS and between May-September  are converted to ice surface heights (see Fig. S4) . These ice surface heights are then used to calculate depth for each crevasse feature (see Fig. 7), resulting in a series of coordinates with associated depths. The crevasse depth of each water feature is then the inverse distance-weighted average of the $k=5$ nearest neighboring points with known depths.

\subsubsection{Supraglacial Water Volume}

To approximate the volume of supraglacial water detected in a given binary image water map produced from the spectral classification, a raster to vector conversion is performed, resulting in discrete water polygon features for each WV image date. We interpolate the crevasse depth for each polygon from the crevasse depth data as described in the previous section for the same year.

The rasterio library in python was used to perform the raster to vector conversion, without making any generalizations to the water feature shapes, meaning that the shapes generated with exact outlines of the water pixels. Touching pixels were outlined with one polygon. 
To significantly reduce the cost of further computations, and to eliminate any potential misclassifications, shapes which did not outline more than two pixels were eliminated.
The surface area of each feature is the product of the number of pixels it contains and the pixel area, which is approximately 3.42 km$^2$ for the multispectral WorldView imagery \cite{wv-specs} used. Pixel areas are taken from each image's metadata.

For any given polygon feature for which the surface area and depth has been determined, calculation of water volume is estimated with a simple geometric approach.

Volume, $v [m^3]$, can be expressed as a function of crevasse depth, $d [m]$, crevasse surface area, $a [m^2]$, and a unit-less conversion factor, $k$, which is based on the assumed crevasse shape.

\begin{equation}
v = k*d*a
\end{equation}

Thus, the variance of water volume is constrained by the variance of $k$, depth, and surface area. We will consider the variance of each of these three components separately to create a feature space which captures the potential range of volume estimates.

Consider $\hat{k}$ to be a reflection of crevasse shape which is assumed to most appropriately represent the true crevasse shape, $k$. And $e_k$ represents the difference between $k$ and $\hat{k}$, or the error.

This work relies on the assumption that each water feature, which in the case study here is typically a water-filled crevasse, has a morphologically simple geometry similar to a triangular prism, as exemplified by newly opened surge crevasses (see Fig. 3).
Crevasses with a true shape more convex than a triangular prism would imply $k > \hat{k}$, and hence have $e_k > 0$, while a more concave shape would have $e_k <0$. 

For a water-filled crevasse with the shape of an inverted triangular prism whose base has an area equal to the water feature area and height axis equal to the depth of the water feature (which is assumed to be $\frac{1}{2}$ the total crevasse depth), $k = \frac{1}{4}$. Then $v_{crev}$ is the volume of the water-filled crevasse:

\begin{equation}
v_{crev} = \frac{1}{2}v_{prism}=\frac{1}{4}a*d
\end{equation}

Where $v_{prism}$ is the volume of a triangular prism ($\frac{1}{2}b*l*d$), $a$ is the surface area of the feature detected from imagery, and $d$ is the crevasse depth determined by the DDA-ice algorithm's ice surface heights. 

Hence, we generalize this formula to Equation 12, with $\hat{k} = \frac{1}{4}$ and allow for variation in shape an depth by assigning $e_k = 0.15$.

This results in a finalized formula for volume of each mapped water feature:

\begin{equation}
v = (\frac{1}{4} \pm 0.15) a*d
\end{equation}

Determining a more accurate estimate of volume for a feature is difficult without extensive data validation through field campaigns. A more accurate estimate of water volume, however, would not necessarily result in a significant improvement to how realistic the total volume estimate could be or have a significant impact on subsequent modeling of supraglacial water as an input to a glacial model. This approach provides a general estimate for the volume of each water feature. Additionally, the general shape of the crevasses which contain water, as revealed by deriving ice surface heights (see Figs. 4,7) over NGS with the DDA-Ice 2 algorithm \cite{DDA-ice-2}, reveals that this generalization is warranted, given the simplistic crevasse morphology of surge crevasses.

\subsubsection{Combined Analysis of Spatial Signatures (SISR, Velocity Changes, Supraglacial Water)}

Crevassing, conveyed in this work as spatial ice surface roughness, is a form of surface signature that can be utilized to identify regions on a glacier that have experienced the same type of glaciological processes, including deformation and  water occurrence. 
 Crevassing occurs when a threshold in the kinematic forces resultant from glacial acceleration is exceeded and the material (ice) responds with brittle deformation 
\cite{mayer-2000,mayer-2008,herzfeldet_2014_iceroughjakjglac,trantow-2018} 
 Observations show that crevasses form in fields of similar spatial characteristics with clear boundaries to neighboring crevasse fields \cite{herzfeld_icesat2.crev, herzfeldet_2014_iceroughjakjglac}.
This observation motivates the introduction of the concept of a crevasse province, or more generally, an ice surface province, defined as an area  with the same spatial characteristics (homogeneity) and the maximal area for which this homogeneity  of spatial characteristics holds (maximality). More generally, deformation patterns are defined by the strain states experienced by the ice, which are stable in space, despite downstream moving ice, as the boundary conditions which determine the stress field are also relatively stable \cite{mayer-2000, herzfeldet_2014_iceroughjakjglac}.
We rely on the concept of crevasse provinces to validate conclusions about spatial similarities between surface signatures, as the chaotic nature of glacial mechanics introduces a random component to the manifestation measured acceleration and water ponding. More specifically, we consider that surface ponded water for water-limited Svalbard-type glaciers, represents geophysical processes generalizable to that crevasse province, rather than the individual crevasse in which it occurs.

While it is understood that, for surges, accelerations in the summer velocity indicate regions of low basal effective pressure and spatial ice surface roughness depicts regions of deformation, little is known about the geophysical processes that can be inferred from supraglacial water ponding.

Surface velocity fields are a direct observation of strain rates, which must be linked to stress through a flow law. Under the standard Stokes formulation for slow, viscous glacier flow \\ \cite{greve-2009}, the Cauchy stress tensor $\boldsymbol{\sigma}$ satisfies
$\nabla \cdot \boldsymbol{\sigma} + \rho \mathbf{g} = \mathbf{0}$,
with $\boldsymbol{\sigma} = -p \mathbf{I} + \boldsymbol{\tau}$, where $p$ is pressure, $\mathbf{I}$ is the identity tensor, and $\boldsymbol{\tau}$ is the deviatoric stress tensor. The deviatoric stress is related to the strain‐rate tensor $\dot{\boldsymbol{\varepsilon}}$ via Glen’s flow law,
$\dot{\varepsilon}_{ij} = A(T),\tau{\mathrm{e}}^{,n-1},\tau_{ij}$,
where $\tau_{\mathrm{e}} = \left(\tfrac{1}{2}\tau_{ij}\tau_{ij}\right)^{1/2}$ is the second invariant of $\boldsymbol{\tau}$, $n \approx 3$, and $A(T)$ is a temperature‐dependent rate factor \cite{greve-2009}. Horizontal surface velocities $u(x,y)$ (and $v(x,y)$) derived from Sentinel‐1 offset tracking provide estimates of the dominant components of $\dot{\boldsymbol{\varepsilon}}$ (longitudinal, transverse, and shear) through spatial velocity gradients. Under the assumption of slowly varying rheology and thickness over the spatial scales considered, spatial changes in strain rate are therefore proportional to changes in deviatoric stress. In the following, we interpret velocity differences and their gradients as proxies for evolving extensional and compressive stresses, and use them in combination with crevasse and roughness maps, which more directly reflect the stress regime, to infer changes in the mechanical forcing during the surge.

The geophysical process that is most unconfined, and which we seek to describe by mapping supraglacial water, is the hydrological connectivity between the glacier surface and the base, often represented by an assumed melt source parameter in subglacial hydrological models \cite{SHMIP}. To inform this process and potentially other geophysical processes, we investigate both spatial and temporal trends of supraglacial water during the evolution of the surge, in the context of the timing and spatial patterns of the geophysical processes indicated by roughness and acceleration of the ice surface. 

\section{Results}

\subsection{Temporal Evolution of Supraglacial Water}


We quantify occurrence, area, and volume of mapped supraglacial water from Maxar WorldView (c) scenes spanning quiescence (2015), the initial/rapid acceleration (2016–2018), the mature surge (2019–2023), and the onset of return to quiescence (2023+) in the upper crevasse region, where the majority of supraglacial water appears. Despite gaps in imagery (late June to early August 2018 and limited clear scenes in 2023), a clear pattern emerges: an abrupt loss in supraglacial water occurs in the upper crevasse region in the initial acceleration phase, but water appears earlier in each year and persists longer as the surge matures. The time at which each location of the glacier, including the upper crevasse region, is affected by the surge varies locally, as the upper crevasse region does not accelerate until 2017 (see Fig. 15c).

\subsubsection{Phase 0: Quiescence (1936-2015).} 

%
The spatial maps presented in this section provide clear evidence of the transition between surge phases. We define first the timing of each surge phase and provide evidence using velocity differences (Fig. 15), the appearance of supraglacial water (Fig. 11), and the deformation state of the glacier (Fig. 16).

In 2015, NGS was still in its quiescent phase since its last surge in 1936-37 \cite{lafauconnier-1991}. Accelerations shown by differencing January 2015 and August 2015 velocities are seasonal, on the scale of $\Delta$ velocity $\approx 0.6 $ m/d. 
The state of the supraglacial system echos NGS's quiescence. 
WV imagery from July of 2015 (Fig. 10) depicts a supraglacial drainage system consisting of melt streams that traverse over the glacial surface approximately parallel to each other. Melt ponds in the upper crevasse region form near the marginal moraine of Akademikarbreen and Negribeen in regions of topographic lows. Their collection in topographic lows is clear as water accumulates below ice falls. There is no obvious remotely observable evidence (moulins or deep crevasses) of hydrological connectivity to a basal drainage system, with crevasses only present on NGS's floating tongue, at the terminus of the glacier. However, observation of NGS's quiescent state gives us a baseline to compare other water volumes to and represents the supraglacial water that is readily available for storage on the glacier surface. We estimate the total surface area of supraglacial water in the upper crevasse region was equal to approximately 0.5 km$^2$ (0.48 km$^2$ for July 1-10 and 0.56 km$^2$ for July 11-20) prior to surge initiation.

\subsubsection{Phase 1: Rapid Acceleration (2016-2017).}


NGS then experiences its first surge phase, which we term the "rapid acceleration phase". Differences in velocity from August 2015 and July 2016 depict accelerations at the calving front on the scale of $\Delta$ velocity $\approx 3 $ m/d. This initial acceleration and the first appearance of crevasses in the lower crevasse region, indicated by two deformation centers in the SISR map in 2016 (see fig. 16b). Propagation of this acceleration up-glacier occurs in 2017, with peak changes in velocity at the calving front reaching $\Delta$ velocity $\approx 16 $ m/d and changes in velocity in the upper crevasse region reaching $\Delta$ velocity $\approx 7 $ m/d.

Water is sparse and short-lived (late-June to late-July), with surface areas notably below quiescent values, in the rapid acceleration phase (2016-2017). 
Water was first observed to pond in late-June and last observed to pond in late-July of 2016, but the timing is consistent with melt signatures of 2015 (prior to the surge). Similarly, water is observed to pond only in late-June of 2017. We estimate the total surface area of supraglacial water in the upper crevasse region  during the initial acceleration phase was equal to approximately 0.24 km$^2$, 0.28 km$^2$, and 0.19 km$^2$ for mid-June 2016, mid-July 2016, and mid-June 2017 respectively. 

The lack of occurrence of supraglacial water, especially water-filled crevasses, is consistent with other studies, as water-filled crevasses mark the start of the mature surge phase \cite{lingleet1993} \\ \cite{trantow-2024}. 


\subsubsection{Phase 2: Mature Surge (2018-2021).}

The conclusion of the rapid acceleration phase is evident as the up-glacier wave of momentum that manifested as extreme velocities is no longer evident in velocity differences between July 2017 and July 2018. Rather, marginal and localized accelerations that are smaller in magnitude dominate acceleration signatures for 2018 and later (see Fig. 15d).   The mature phase is best described as the expansion of the surge into different parts of the glacier system caused by the initial rapid acceleration and mass transfer of Negribreen Glacier, which itself begins to slow down though elevated speeds large-scale deformations persist. This is reflected by the appearance of water-filled crevasses (see Fig. 11) first in the upper crevasse region (see Fig. 1) and later in the marginal regions of the middle crevasse region. While supraglacial water ponding appears in the upper crevasse region in 2018, the lateral margins of Negribreen, which begin to accelerate in 2018, do not experience ponding until 2019, i.e. locally, they do not enter the mature surge phase until 2019.


Compared with the rapid acceleration surge phase, the early mature surge phase (2018–2019) shows that supraglacial water forms earlier, by mid-June, and persists longer, until mid-August (Fig. 9). 
Earlier occurrences of water as the surge progresses may reflect the glacier's ability to store available water inputs with a more heavily crevassed surface and a resistance to supraglacial drainage. Later occurrences of supraglacial water with the progression of the surge suggest lowered connectivity between the supraglacial and subglacial hydrological systems. 

We estimate the total surface area and derive volume for the  early mature surge phase. In 2018 and 2019, we approximate that supraglacial water covered an approximate area of 0.2 km$^2$ at most (mid-June, 2019 and mid-August 2018). However, the remainder of water maps made during this period reflect much lower surface areas (0.02-0.15 km$^2$). 

A distinct separation is apparent between this early mature surge phase and later mature surge phase by the sudden increase in surface water area and volumes, beginning in 2020. From 2020 to 2021, the glacier exhibited an extended period of supraglacial water presence, ranging from early June to late September 2021. During this phase, water areas and volumes match or exceed areas from the quiescent phase, reaching 0.55 km$^2$ in late-June 2020 and 0.64 km$^2$ in early-September 2021. 


\subsubsection{ Phase 3: Return to Quiescence (2022-current(2025)).}

The return to quiescence is marked by slowed ice deformation rates (see section 3.4 on spatial ice surface roughness) and the subsidence of local accelerations coupled with overall slowing of the glacier's velocity. This phase begins in 2022. Supraglacial water storage in 2022 has the earliest appearance observed so far in the surge (early May 2022). In 2022, area and volume estimates of supraglacial water remain high ($Area_{water} \approx 0.46 km^2$) and ($Volume_{water} \approx 9.6 e 5 m^3$). In the 2023 and later, little imagery was available or analyzed, and hence further analysis is needed to draw conclusions about the hydrological drainage state during this stage of the surge. SISR maps (Fig. 16) show surface roughness is still severe in 2022 and 2023, but do not show large expansion of roughness centers, only higher magnitudes of roughness where deformation has occurred.



\subsection{Spatial Observations of Supraglacial Water}

\subsubsection{Evolution of Supraglacial Water in the Upper Crevasse Region}

At the beginning of the mature phase (2018-2019) water begins to pond in the upper crevasse region, at the very upper extent of surge activity, as indicated by roughness maps (see section 3.2). As the mature phase continues, (2020-2024), water-filled crevasses appear more extensively in the upper crevasse region and also appear in the glacier's lateral margins, following local accelerations. In figure 11, the extent of supraglacial water is depicted as a heat map for each year in which imagery existed and water is observable. The heat maps in Fig. 11 represent the probability of water occurrence, given $n$ binary water raster maps. For example, $n = 15$ binary water rasters were created from WV imagery for the year 2019. These rasters are stacked and the average value in each pixel is calculated. For visual clarity, a gaussian kernel is applied to the probability map, creating a more general and smooth heat map overall.

Ponding of supraglacial water occurs primarily in the upper crevasse region and is observed only marginally in the middle crevasse province and in a single crevasse in the lower crevasse province.
Supraglacial features observed in the upper crevasse region include melt ponds and several clusters of water-filled crevasses. 
Prior to the surge and also during the surge, a large, melt pond forms annually along the medial moraine, in approximately the same location. It has a maximum surface area of $5.25e4$ $m^2$. This melt pond is documented each year when imagery is available - reaching its peak extent in July of 2015, and late June of 2019-2022. In 2019-2022 the pond slowly begins to diminish through July and then rapidly drains, which is surprisingly indicative of a locally efficient drainage mechanism. 
The prevalence of this large melt pond in a region of a topographic low implies that the glacial drainage system may still develop efficient englacial mechanisms which are capable of draining a large melt pond within a few days as observed between July 10, 2019 (when the pond is present) and July 14, 2019 (the pond is drained). 


Ponding within crevasses occurs in shared spatial clusters which are also observable from 2019-2022. The most ponding occurs in the upper-most crevasse province, to the southwest of  the medial moraine. This region demonstrates behavior that would be expected of an inefficient drainage system, as the ponding is prolific across the entire crevasse province and increases in spatial extent from its first occurrence each year.


Surface water observations support the theory that a subglacial drainage system of heterogeneous efficiency is present in the mature phase of the surge. In regions where crevassing occurs and water is ponded, the drainage system would be considered to be inefficient, while the development of a partially efficient drainage system may still be possible, as evidenced by the rapid drainage of a large melt pond feature annually in mid-July.

%
%

\subsubsection{Evolution of Supraglacial Water in the Middle Crevasse Region}

At surge initiation, in 2016, the middle crevasse region remains relatively untouched, and melt streams still persist. By 2017, newly formed crevasses occur, but supraglacial water does not appear until 2019 at the beginning of the mature phase of the surge.
In 2020, very small water-filled crevasses appear near Negribreen's southern lateral margin and lower northern lateral margin. In the subsequent year, 2021, the water-filled crevasses on the southern lateral margin are still present and the crevasses on the northern lateral margin expand up glacier. In later years, 2021, 2022, more water-filled crevasses appear in these regions. 



\subsubsection{Evolution of Supraglacial Water in the Lower Crevasse Region}

Supraglacial water is observed only in 2021 near Negribreen-Akademikarbreen medial moraine (NAMM)The absence of detected water in the lower crevasse region may reflect the inability of ponding to occur at the surface due to the proximity of the lower crevasse region to the grounding line, as NGS is marine terminating, and more specifically that much of the region contains floating ice. 


%

%
%

\subsection{Ice Surface Velocity Differences}

Small seasonal accelerations of NGS are revealed by differencing the summer and winter velocities prior to the surge (Fig. 16 (a)), which depicts a small region of acceleration in the upper-crevasse province having velocity differences on the scale of 0.5 m/d. Seasonal velocity differences (Fig. 16 (a)) highlight a large region of gradual acceleration in the middle-crevasse province ($\Delta \mathbf{v} \approx 0.35 m/d$) and several clusters of acceleration near the glacier terminus with the highest magnitudes in velocity change ($\Delta \mathbf{v} \approx 0.6 m/d$).

Maps of differences in summer velocity (magnitudes) between subsequent years (Fig. 16 (b)-(f)) show clear regions of inter-annual acceleration or deceleration. 
In the first year of the surge, velocity differences on the scale of 2.5 m/d (Fig. 16 (b)) reflect the initiation of the kinematic surge wave near the glacier terminus. In the subsequent year (2017), surge activity peaks and velocities reach 22 m/d \cite{trantow-2025}. Velocity differences from 2017 depict that the surge-type acceleration has extended to the upper crevasse province. 

In the following years (2018- 2020+) Negribreen experiences several smaller and local accelerations and decelerations, specifically in the lower shear margin (Fig. 16 (d)) and upper shear margin (Fig. 16 (e)). 
Smaller dispersed local accelerations, following a large deceleration of the glacier in 2020, are evident in 2021 in the upper crevasse region, 2022 in the moraine between Negribreen and Ordonnansbreen, and 2023.

\subsection{Spatial Ice Surface Roughness }

Spatial ice surface roughness (SISR), represented as the directional average of the geostatistical parameter, $\mathit{pond}_{res}$, depicts the expansion of crevassing around several roughness centers, see Fig. 16. Ice surface roughness is most directly indicative of regions of active deformation of the ice, which is dependent on both ice material properties and velocity of the glacial body. 
At the beginning of the surge, in 2016, two roughness centers are visible, at the calving front and about 4 km up-glacier from the calving front, with the latter corresponding to the initiating location of the surge, as reported in \cite{trantow-2025}. In 2019, crevassing has extended into the upper crevasse region, and several new roughness centers are present, including the water-filled crevasse province in the upper crevasse region (see section 3.1.2). In later years, 2022 and 2023, the spatial extent of roughness expands around the same centers detected in 2019, with higher overall roughness.

\subsection{Evolution of Surge Dynamics: Combined Analysis of Water, Roughness, and Velocity Differences}

Taken together, the velocity‐difference maps and SISR fields resolve how the stress and strain-rate regimes evolve throughout the surge and how this controls ice-deformation. Positive velocity differences mark zones where ice is accelerating in the flow direction and thus dominated by extensional forcing, whereas negative differences and strong spatial gradients in annual velocity changes indicate deceleration and therefore compressive forcing. Early in the surge (2016-2017), large increases in velocity near the terminus and in the mid-glacier (Figure 16) reflect a kinematic wave with strong extensional stresses at its front. SISR in 2016 shows roughness centers at the calving front and $\sim$4 km up-glacier, with smooth ice upstream (Figure 16), indicating that deformation is initially concentrated in a narrow extensional band that has not yet overprinted the upper crevasse region. By 2019, SISR depicts expanded and intensified roughness around multiple deformation centers, including the upper crevasse province, while velocity differences show a patchwork of smaller accelerations superposed on widespread deceleration. This pattern is interpreted as a transition from a simple surge wave characterized by extensional crevassing to a mixed regime in which earlier extension has fractured the ice and subsequent compressive thickening and shear reactivate and rotate existing crevasse fields. This interpretation is consistent with crevasse classes reported for NGS during the initial years of the surge (2016-2018) \cite{twickleret2025_geoclass2_frontiers, herzfeldet_2024_nnclass_mlp_cnn_mdpi}, in which parallel crevasses begin propagating up-glacier in 2018 and chaos and multidirectional crevasses form at the calving front.

The temporal grids of water occurrence, area, and volume clarify how supraglacial storage in the upper crevasse region lags the mechanical forcing. During rapid acceleration (2016-2017), when extensional stresses and velocity differences are largest, mapped water in the upper crevasse region is short-lived (late June-late July) and has surface areas well below quiescent values ($\sim$0.2-0.3 km$^2$ versus $\sim$0.5 km$^2$ in 2015), and volumes remain modest. The coincidence of strong extension with reduced surface storage suggests that newly opened fractures initially enhance vertical or englacial drainage, so that meltwater is efficiently routed through the surface even as crevasses begin to form. In the early mature phase (2018-2019), the surge front and associated deformation reach the upper crevasse region: SISR shows new roughness centers there, and the binary-occurrence and area grids indicate that water appears earlier (mid-June) and persists into mid-August, albeit with still modest areas and volumes. We interpret this as the stage at which extensional crevassing has created a dense fracture network and the englacial drainage system remains partially efficient, so that water can pond briefly in newly deformed topographic lows before reconnecting to depth.

The late mature phase (2020-2022) is characterized by the strongest coupling between deformation and supraglacial storage in the upper crevasse region. Roughness maps show further expansion and intensification of roughness around the 2019 deformation centers, while the velocity-difference fields indicate that the main surge wave has decelerated and that only localized accelerations persist along margins and tributaries. In the temporal grids (fig. 9), this interval coincides with the sudden increase in surface area and volume of water: areas in the upper crevasse region reach or exceed quiescent values ($\sim$0.55 km$^2$ in late June 2020 and $\sim$0.64 km$^2$ in early September 2021), and volumes peak in late June 2020, late August-early September 2021, and early- to mid-June 2022 (fig. 9). Water is also present over a much longer portion of the melt season (from early May in 2022 through late September in 2021). The timing of these maxima-several years after the strongest extensional accelerations reached the upper crevasse region and contemporaneous with widespread deceleration and high SISR-implies that persistent ponding develops once cumulative deformation has both created complex crevasse topography and, through compressive thickening and shear, decreased hydraulic conductivity between the glacier surface and base. In this state, the upper crevasse province acts as a mechanically deformed, hydraulically inefficient storage zone: extensional phases construct the crevasse architecture, while subsequent compressive and shear phases, together with heterogeneous basal drainage, determine when and where large surface areas and volumes of water can accumulate and persist.

\section{Conclusion}

We show that three remotely observable spatial signatures, (i) spatial ice-surface roughness (SISR) from Sentinel-1, (ii) interannual summer velocity differences, and (iii) supraglacial water occurrence, extent, and persistence from WorldView (with ICESat-2-informed depths), jointly resolve where and when hydrological connectivity, the ability of the glacier to transport supraglacial water to the basal drainage system, changes during a surge. Using these three spatial signatures, we clearly identify three distinct surge phases: (i) rapid acceleration (2016-2017), (ii) mature surge (2018-2021), and (iii) return to quiescence (2022-current(2025)). Maps of annual changes in summer surface velocity depict regions of acceleration and, implicitly, regions of the glacier experiencing compressional or extensional stresses. Spatial ice surface roughness maps (SISR) depict the deformation response as detected from Sentinel-1 SAR data. 
Appearance of water-filled crevasses is common for regions with implied compressional stress (having just experienced acceleration) and  in some cases a secondary acceleration is evident (margins) after ponded water. 
Sudden decreases in volume of water largely reflects ponded water in the upper crevasse region, which we think is connected to the up-glacier extension, followed by compression.
On the Negribreen Glacier System (NGS), the surge propagated up-glacier from the terminus; multiple deformation centers emerged and expanded; and late in the surge, secondary accelerations appeared along margins and in tributaries while surface water became increasingly clustered, voluminous, and persistent into late September. Rapid drainage of a large melt pond during several summers indicates that locally efficient (vertical or englacial) connections can transiently coexist with otherwise inefficient drainage. Together, these patterns imply strongly heterogeneous and clustered connectivity changes that modulate basal effective pressure and sliding through the surge cycle.

\subsection{Key findings for Negribreen Glacier System (NGS)}

We observe glacier-specific dynamics by analysis of geospatial signatures, including the following:

{\it Deformation of the ice surface expands from multiple roughness centers.} SISR maps track an up-glacier expansion from the terminus with multiple roughness centers, indicating that the surge does not advance as a simple single front but rather spatially discrete deformation provinces (crevasse provinces) that later coalesce late in the mature surge phase.

{\it Two kinematic regimes occur during the NGS surge.} Early surge years (2016–2017) show a dominant acceleration that propagates longitudinally from near the glacier terminus to the upper crevasse province; later years (2018–2023) are characterized by smaller, spatially localized accelerations, specifically along the margins of Negribreen, in the upper crevasse province of Negribreen, and in confluent glaciers of NGS.

{\it Supraglacial water is both a response to and driver of acceleration.} Early in the surge, surface water occurrence appears transiently and has covers small areas;
during the late mature surge (2020+), water is highly clustered (recurrent crevasse provinces and lateral moraines), has larger areal extent/volume, and persists later into the season. Episodic rapid drainage events reveal transient reconnection to depth despite overall inefficient drainage.

\subsection{Key findings for polythermal surge theory}

{\it Changes in supraglacial-subglacial hydraulic connectivity occur in clusters.}
Deformation concentrates connectivity changes into spatial clusters (crevasse provinces) that act as hydro-mechanical nodes: when a node transiently connects vertically or englacially, it can inject pulses of surface melt to the bed, lowering $N$ locally and driving secondary accelerations. When it disconnects, surface storage and late-season persistence increase. This clustered switching helps explain the observed patchwork of late-surge accelerations of NGS and could be a mechanism for longer surge phase duration in polythermal glaciers.

{\it Upstream storage of supraglacial water coincides with downstream acceleration earlier in the surge.} In the acceleration phase of NGS, ponding of water in the upper crevasse region coincides with acceleration downstream. In this phase, the area of water present is lower than the quiescent area of water present at the surface. In the mature surge phase, storage of supraglacial water occurs at more localized scales, but spatial coincidence with accelerations is not exact. At least two underlying geophysical processes could be used to explain this phenomena: (i) hydraulic connectivity in the upper crevasse region is low, while extensional crevasses in the middle and lower crevasse regions mechanically increase the hydraulic connectivity, so water that is melted locally or deposited onto the glacier surface is difficult to detect without continuous observations, and (ii) water stored further up-glacier is eventually connected to the base, resulting in increased basal water pressures which enables local forces to accelerate ice downglacier. 
This reconciles gradual surge termination in polythermal glaciers with sustained low $N$ patches rather than an abrupt, systemwide channelization collapse.

{\it Coexistence of drainage states as a stabilizer of long surge durations.}
{\it Efficient and inefficient hydrological drainage systems can coexist during a surge.}
The simultaneous presence of locally efficient conduits (episodic pond drainages) within a dominantly inefficient subglacial and englacial network allows systems to intermittently relieve pressure without fully reorganizing into efficient channelized drainage. We observe rapid drainage of a melt pond in the upper crevasse region (less than one day), despite continued storage of water in other locations of the glacier.

{\it Supraglacial water evolution is consistent with surge phases.}

In the early surge phase, minimal ponding despite high surface velocities implies transient existence of efficient vertical connections and/or rapid routing that prevents surface storage. SISR depicts that, in this phase, roughness centers first appear near the glacier terminus and migrate up-glacier.
In the early mature surge phase, expanding roughness centers coincide with reoccurring water clusters, and it is expected that episodic pulses of water to the bed could sustain localized accelerations outside the main surge signature. 
In the late mature surge phase, we observe the most prolific surface storage in time and space, with recurrent clusters. Extensive deformation of the glacier, driven by earlier accelerations , increases the glacier surface's capacity for snow and firn melt storage and we anticipate that accelerations in the glacier margins and tributary glaciers could be partially driven by transient increasing in hydraulic connectivity to the bed.

\section{Discussion}

Contextualizing the findings of this work in broader theories of glacial hydrology is critical for its applicability to modeling and future observational studies. The volume of supraglacial water is meaningful only in relation to the geometry and expected subglacial water volume needed for frictionless basal sliding. The spatial clustering of supraglacial water is meaningful only in relation to the spatial patterns of acceleration and deformation.

A sudden influx of water could potentially contribute to the initiation of a surge for temperate surges \cite{trantow-2024}, overwhelming the drainage system before sufficiently large efficient channels can develop to remove the large influx of water, resulting in a reduced basal effective pressure. While a hydraulic surge trigger is unlikely for polythermal surge glaciers, precipitation observations at the Svalbard Lufthavn  and reports of mudslides and extreme rainfall \cite{lapointe-2024, das-2025} indicate unusually large amounts of precipitation in January, July, October and November of 2016, which could have accelerated the surge further, after it began in May 2016.

\subsection{Limitations}

\subsubsection{Data Availability and Coverage}

{\it Temporal gaps in imagery.} Several windows without cloud-free WorldView scenes (e.g., late June–early August 2018; sparse clear scenes in 2023) limit within-season continuity and bias the seasonal “first appearance” and “last persistence” of supraglacial water toward available acquisitions. Consequently, our timing inferences (early onset vs. late persistence) should be interpreted as lower/upper bounds rather than exact dates.

{\it Spatial gaps in ICESat-2 sampling.} Although nine reference ground tracks cross NGS, true coincidences of ATLAS strong beams with water-filled features are rare. Depth constraints rely on (i) one direct coincidence and (ii) interpolation from DDA-ice crevasse depths along tracks to features off-track. This introduces spatial representativeness error where crevasse geometry varies over short distances.

\subsubsection{Data Analysis Methods}

While image thresholding and volume derivation from ICESat-2 derived crevasse depths provides a reasonable, engineering-type estimate of water area and volume in the summer season of a given year for the Negribreen Glacier System (NGS), these approaches are highly sensitive to user error as demonstrated in the sensitivity analysis of total water area estimates. Hence, this approach could be useful for creating initial estimates for specific case studies of glaciers, but is not scalable and it would be expected that vastly different results could result from varying user interpretations of appropriate image thresholds. While several tools were developed to help select the most appropriate spectral threshold for each image, ultimately, the threshold is left to user discretion. 

In this study, extensive visual validation of water features gives us confidence in our final water area estimates, but volumes are difficult to verify as no in situ depth measurements have been made on NGS, or any surge glacier for that matter. Future work should include use of the Densisty Dimension Algorithm for ice surfaces 2 (DDA-ice-2) \cite{DDA-ice-2}, which penetrates both water and ice, to resolve two surfaces and validate the assumptions made in this work (water depth is equal to half crevasse depth).

\subsubsection{Water Volumes as Conservative Estimates}

Volumes are conservative engineering estimates: (a) water depth is taken as ½ crevasse depth where direct water bottoms are unavailable; (b) geometry is idealized with a triangular-prism factor; (c) depth is interpolated from along-track DDA-ice to polygons. These choices likely underestimate volumes in convex or pond-like features and overestimate volumes in very shallow/brim-filled crevasses. A full uncertainty budget (threshold variance, interpolation length-scale, shape factor) and Monte Carlo propagation would formalize confidence intervals in future work. Future work should also include comparison of the spectral classification approach demonstrated here to other image classification approaches such as \cite{twickleret2025_geoclass2_frontiers} and a statistical approach to correlation of the geophysical processes reported by spatial ice surface roughness, velocity changes, and supraglacial water occurrence.

\section{Data Availability}

ICESat-2 ATLAS/ICESat-2 L2A Global Geolocated Photon Data (Version 7) was downloaded from NASA's Earthdata portal using a python download script provided by the National Snow and Ice Data Center \cite{icesat2-atl03-access}.

Data access support and geospatial services for this work was provided provided by the Polar Geospatial Center under NSF-OPP awards 1043681, 1559691, and 2129685. WorldView Maxar (c) imagery [2015-2023] was accessed using the Polar Geospatial Center's FRIDGE portal. 
Arctic-DEM's for shadow correction were also obtained from the Polar Geospatial Center \cite{Arctic-DEM}.


Sentinel-1 SAR images were downloaded from Copernicus Browser (\url{browser.dataspace.copernicus.eu}), which contains a freely available complete archive of Sentinel-1 data, in addition to many other data collections. Velocity was derived using ESA's SNAP toolbox \cite{SNAP}.

NCEP/DOE Reanalysis II data provided by the NOAA PSL, Boulder, Colorado, USA, from their website at \url{https://psl.noaa.gov}.

Precipitation observed at the Svalbard Airport were provided by Norsk KlimaServiceSenter, which are available for download at \url{https://seklima.met.no}.


%
%
%
%
%
%
%
%

%
%
%
%
%

\bibliographystyle{igs}
\bibliography{watermap}

\end{document}


\begin{figure}[H]
\begin{minipage}[b]{.9\textwidth}
\centering
\fbox{\includegraphics[width=\textwidth,trim=1 0 0 0, clip]{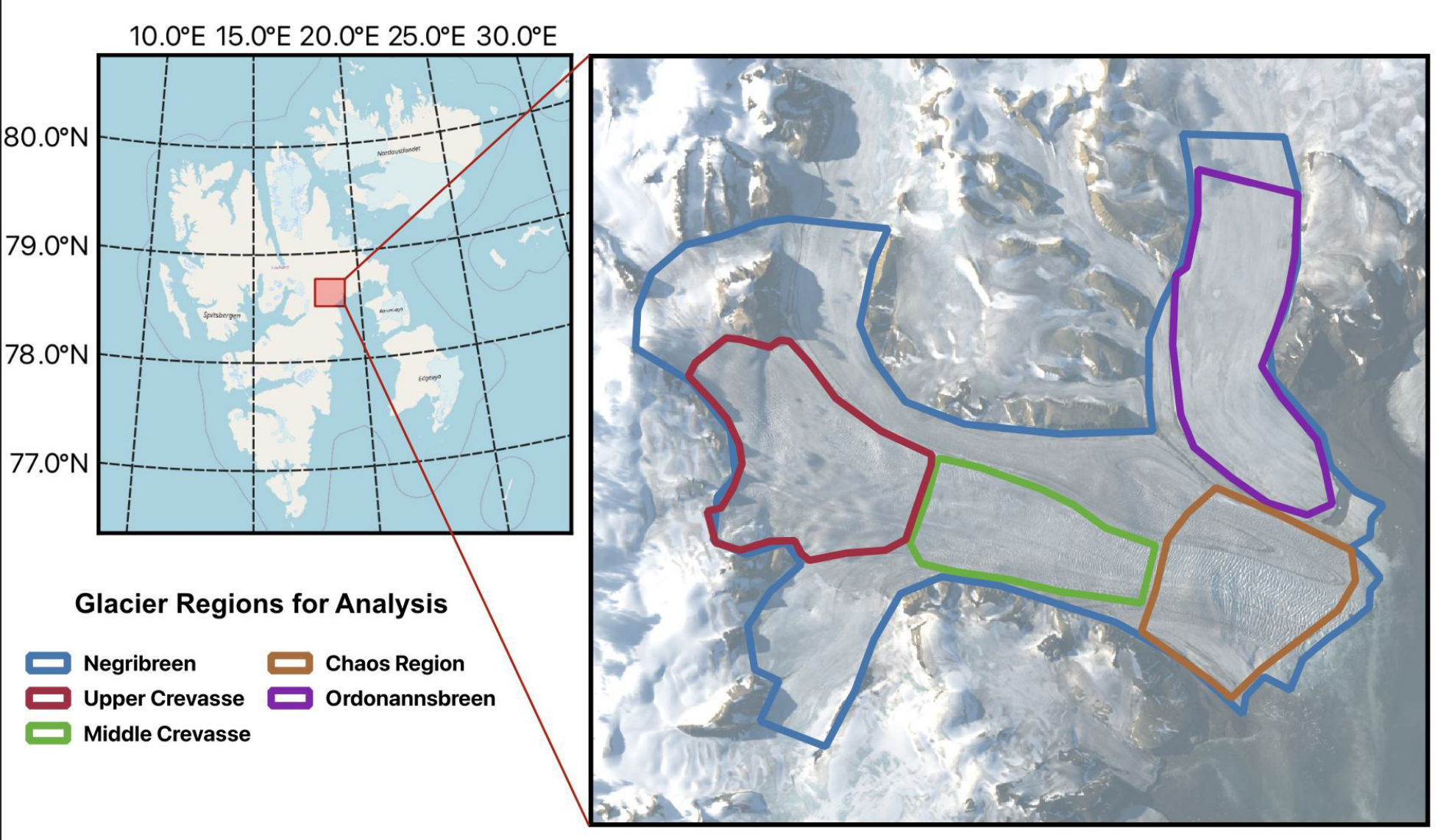}}
\caption{Relative location of Negribreen Glacier System (NGS) in Svalbard (left) and the glacier regions of NGS used in this analysis (right). Glacier regions consist of a portion of the tributary glacier, Ordonannsbreen (purple), an upper crevasse region (red), and middle crevasse region (green), and lower crevasse region(orange).
}
\end{minipage}
\end{figure}

\begin{figure}[H]
\begin{center}
\includegraphics[width=0.8\textwidth]{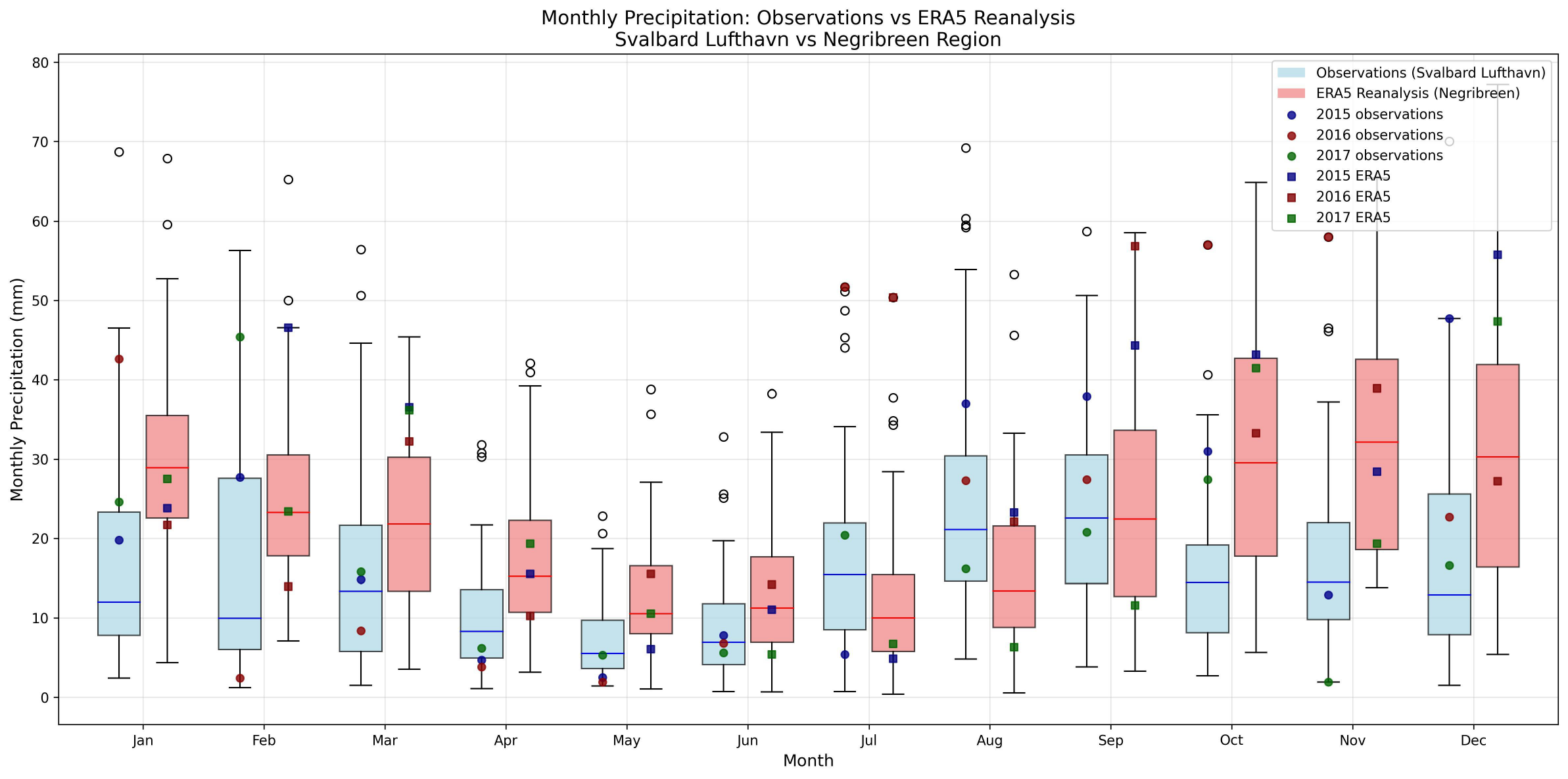}
\caption{Monthly precipitation observations (blue) at the Svalbard Airport 1912-2024 (\url{https://seklima.met.no}) and modeled monthly precipitation from NCEP Reanalysis II 1979-2025 (red) \cite{NCEP-II}.}
\end{center}
\end{figure}

\begin{figure}[H]
\includegraphics[width=0.8\textwidth]{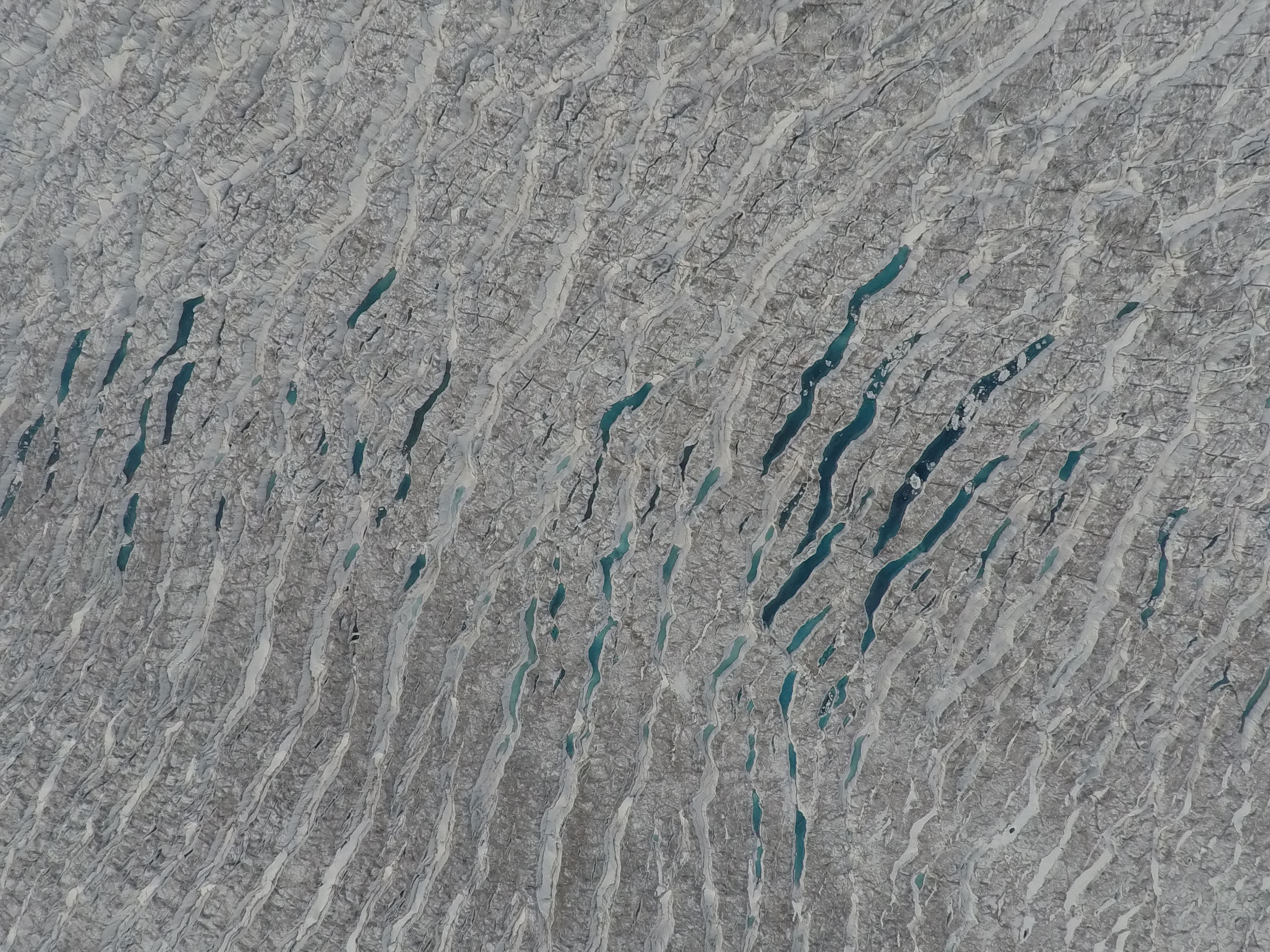}
\caption{Water filled crevasses on Negribreen Glacier System from a helicopter, August 2019. Photo credit: Ute Herzfeld.}

\end{figure}

\begin{figure}[H]
\includegraphics[width=0.32\textwidth]{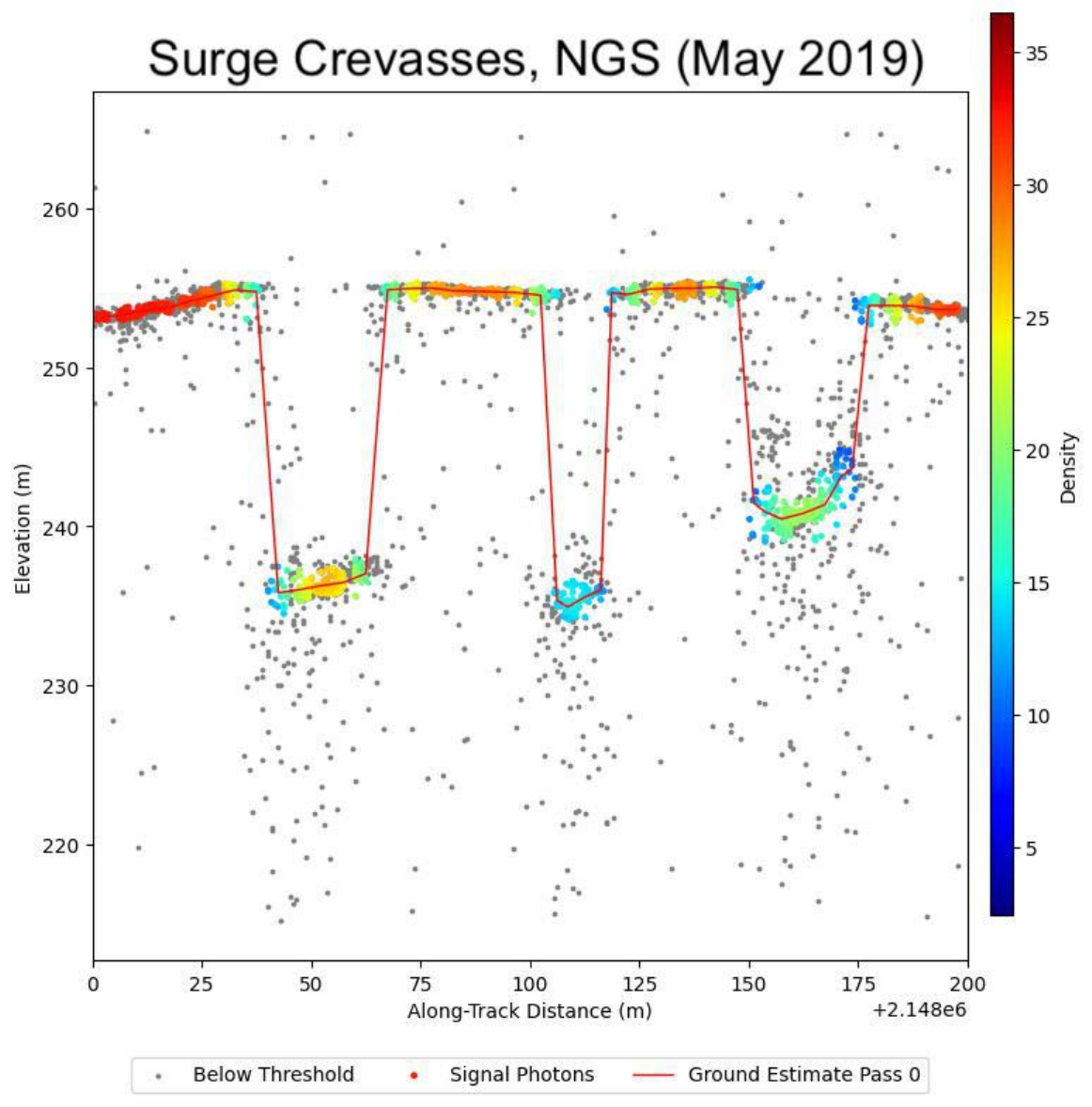}
\includegraphics[width=0.33\textwidth]{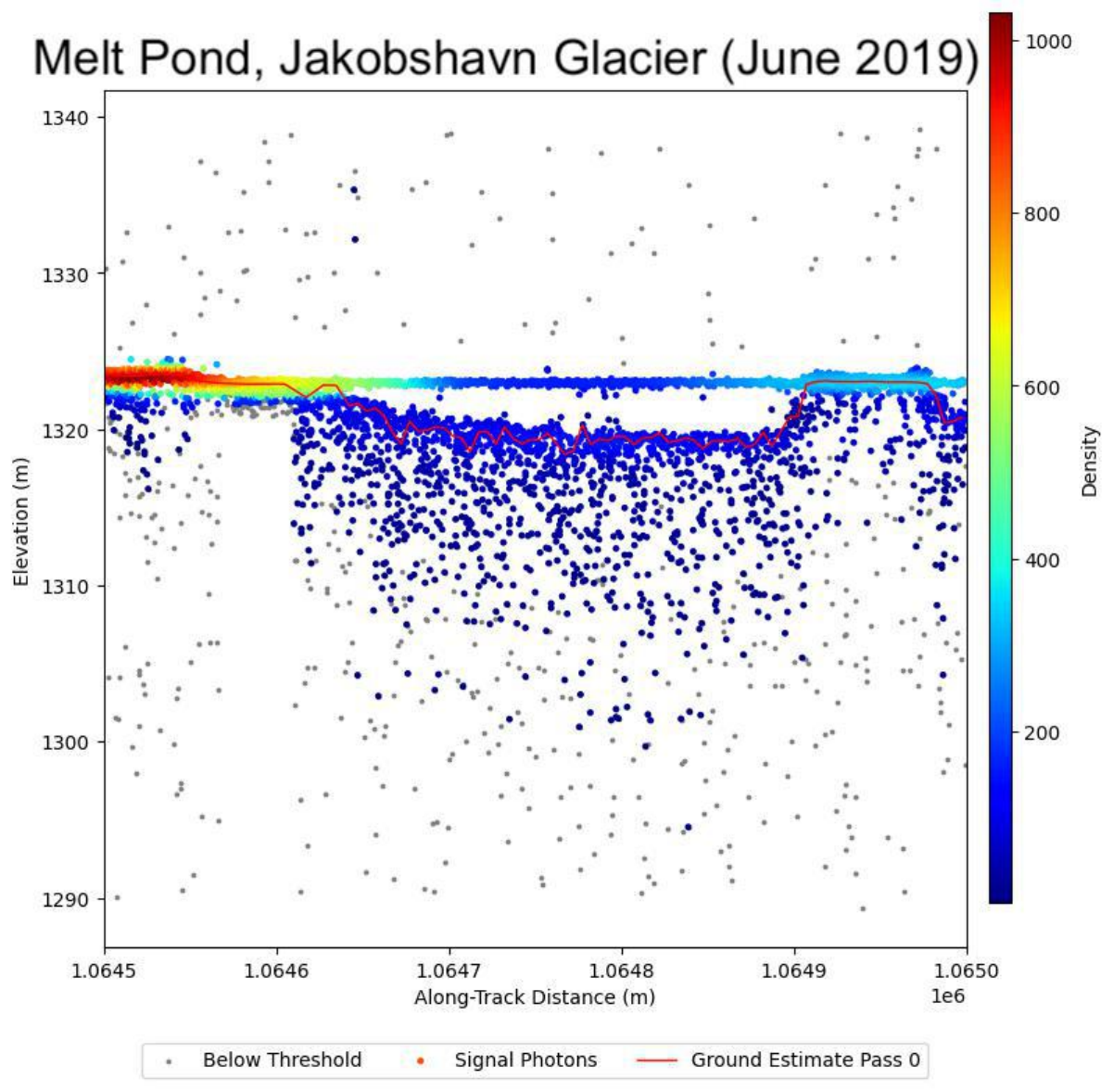}
\includegraphics[width=0.32\textwidth]{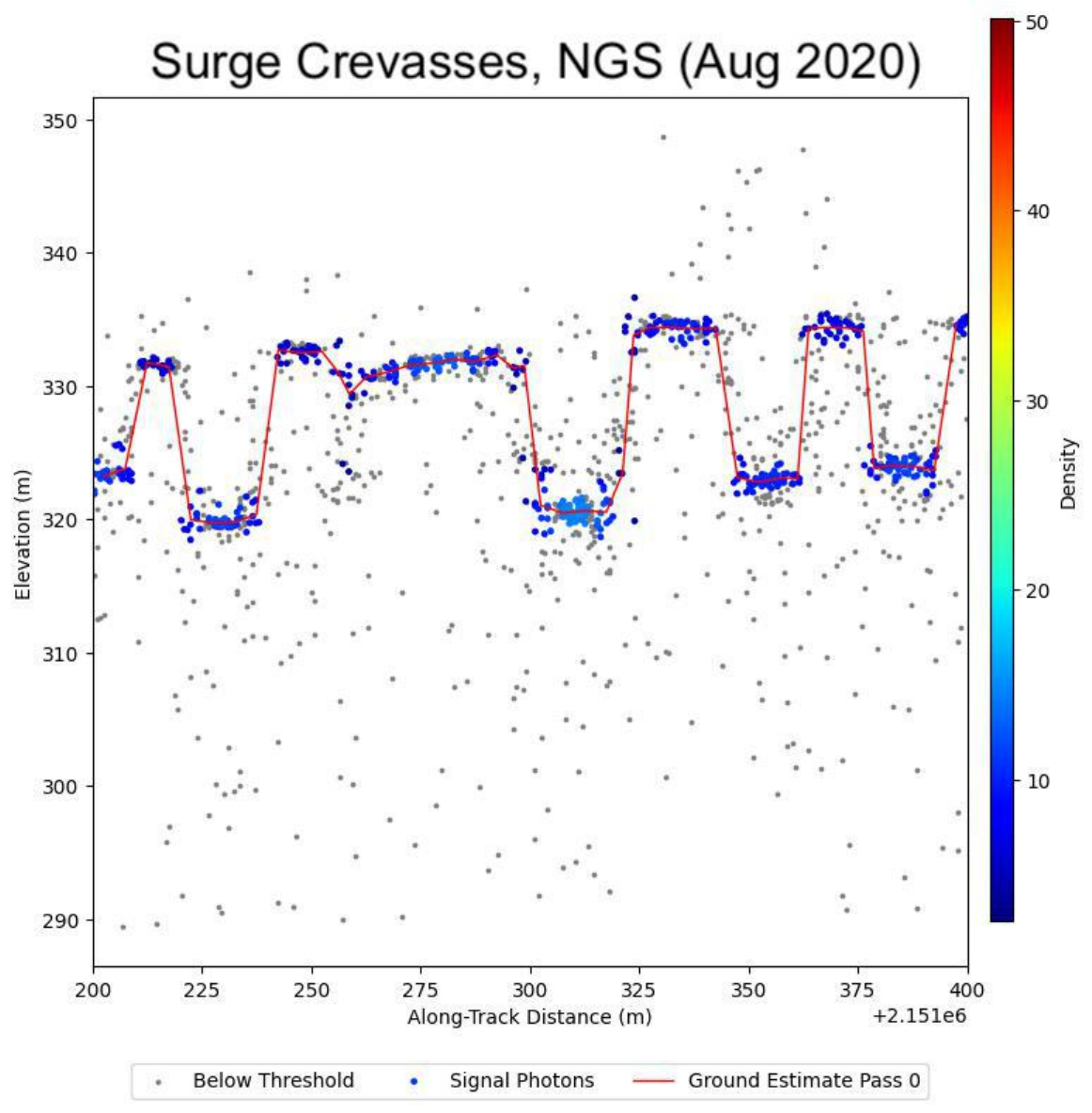}
\caption{Typical scale of supraglacial water features, as observed in ICESat-2 Altimetry, processed with the Density Dimension Algorithm for Ice Surfaces \cite{herzfeld_icesat2.crev}. (a) Newly opened surge crevasses on Negribreen, May 7, 2019 (ATL03 Granule 594, Track GT3L), (b) Melt pond on the surface of the Jakobshavn Isbr\ae, Greenland Ice Sheet, June 6, 2019 (ATL03 Granule 1161, Track GT3L) exemplifies a cross section of a melt pond similar in scale to melt ponds observed in upper Negribreen, and (c) crevasses on Negribreen which could contain water or firn, August 3, 2020, (ATL03 Granule 594, track GT1L). For (a) and (c), each x axis tick is 25m along track. For (b), each x axis tick is 100 m along track.}
\end{figure}

\begin{figure}[H]
\centering
\begin{minipage}[b]{\textwidth}
\includegraphics[width=\textwidth]{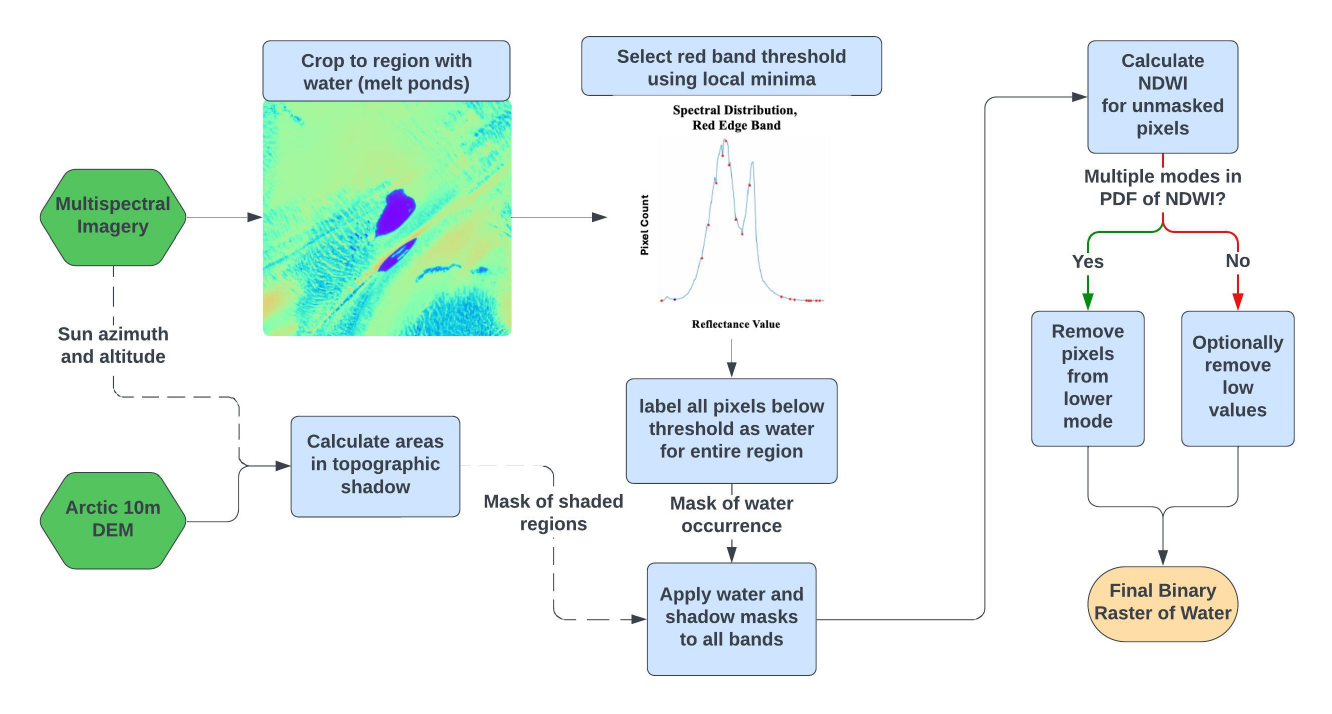}
\caption{Image labeling process for extraction of pixels containing water.
Flow chart depicting the process for binary classification of water in multispectral Maxar WorldView imagery. For a given multispectral image, the classification process proceeds as follows: 
the sun azimuth and altitude are obtained from the image's metadata; 
areas in topographic shadow, according to a line of sight analysis with Arctic 10m DEM, are masked from image; 
a subset of the image, cropped to a region with known water features, is used to plot the overall spectral distribution for the red edge, blue and NDWI bands; 
peaks from the red edge distribution are used as initial guesses for the classification, and the user selects the one best matching the image subset; 
pixels with intensity values higher than the selected threshold in the red edge band are masked out (set to 0);
the NDWI spectral distribution across all pixels in the classified image is plotted allowing the user to optionally set another threshold in the NDWI spectrum; 
all values lower than this NDWI threshold are masked; the final binary raster (1 = water, 0 = no water) is obtained. }
\end{minipage}
\end{figure}

\begin{figure}[H]
\includegraphics[trim=7.5cm 0 7cm 0, width=0.99\textwidth, clip]{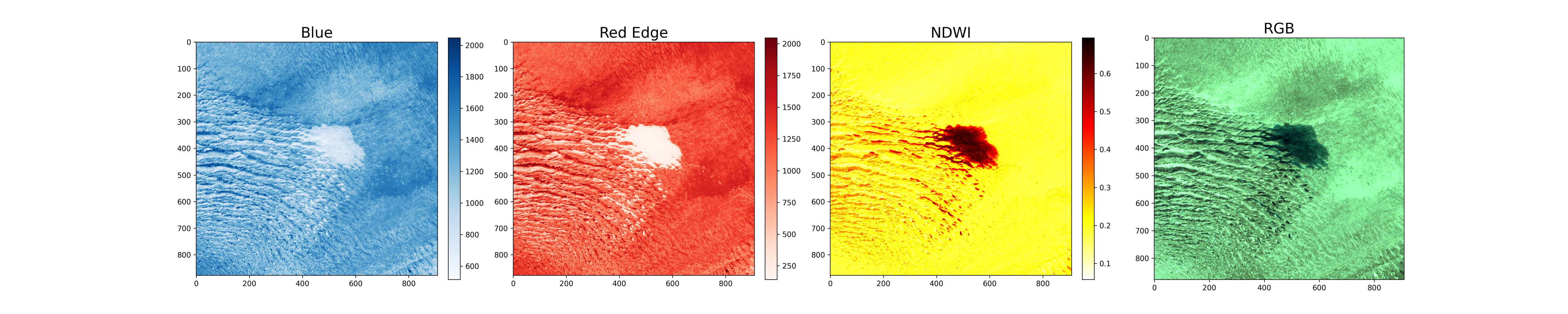}
\caption{Images of intensity in red, blue, and ndwi (a,b, and c respectively) WV spectral bands over a field of crevasses, filled with water. (a) Pixels containing water are generally darker, or lower in reflectance in the red band. (c) NDWI is calculated as expressed in Equation 1. With the red (a) and blue (b) values. Note how pixels containing water have both high NDWI and low red, thus both low NDWI and high red can be used to remove surface types such as ice and shadows.}
\end{figure}

\begin{figure}[H]
\includegraphics[width = 0.66\textwidth]{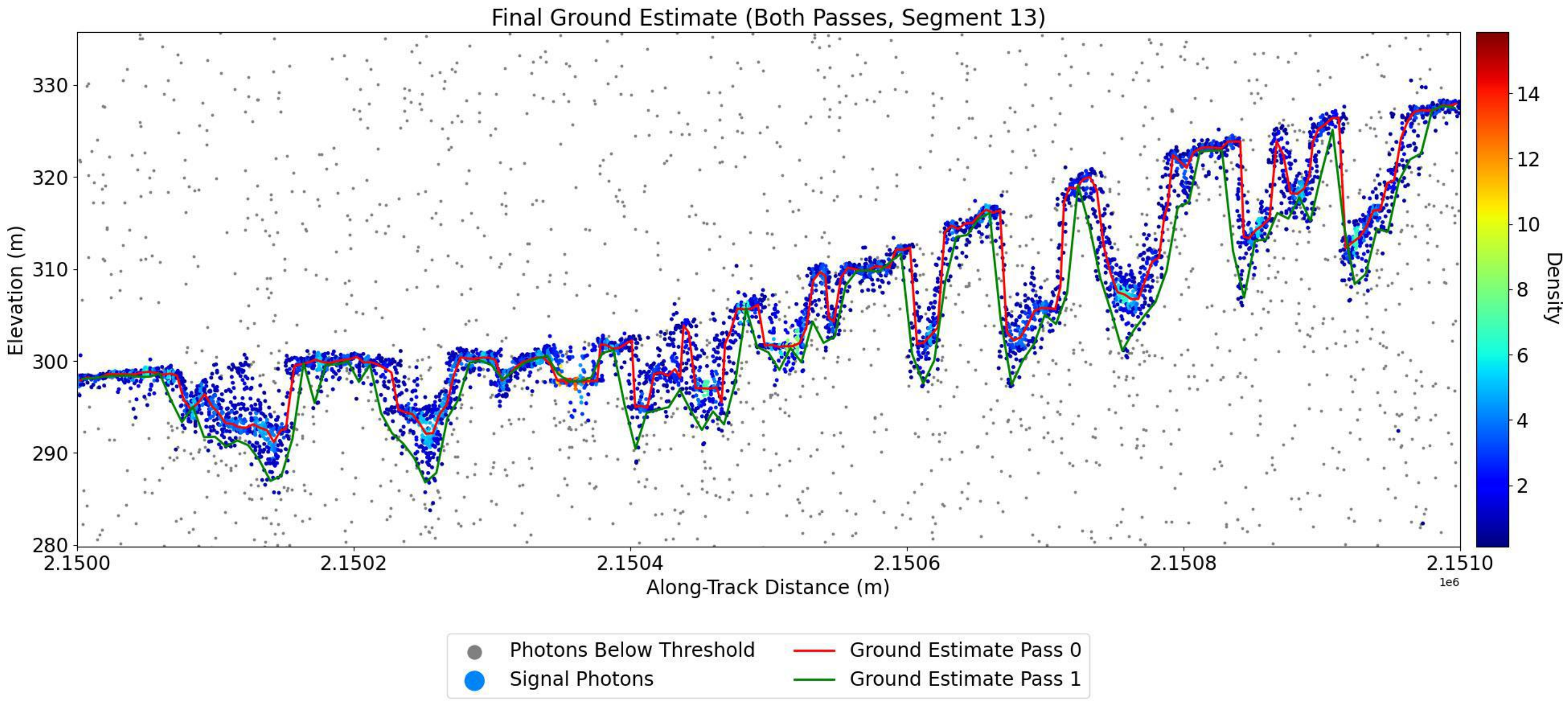}
\includegraphics[width = 0.31\textwidth, trim=17cm 2.25cm 16cm 3cm, clip]{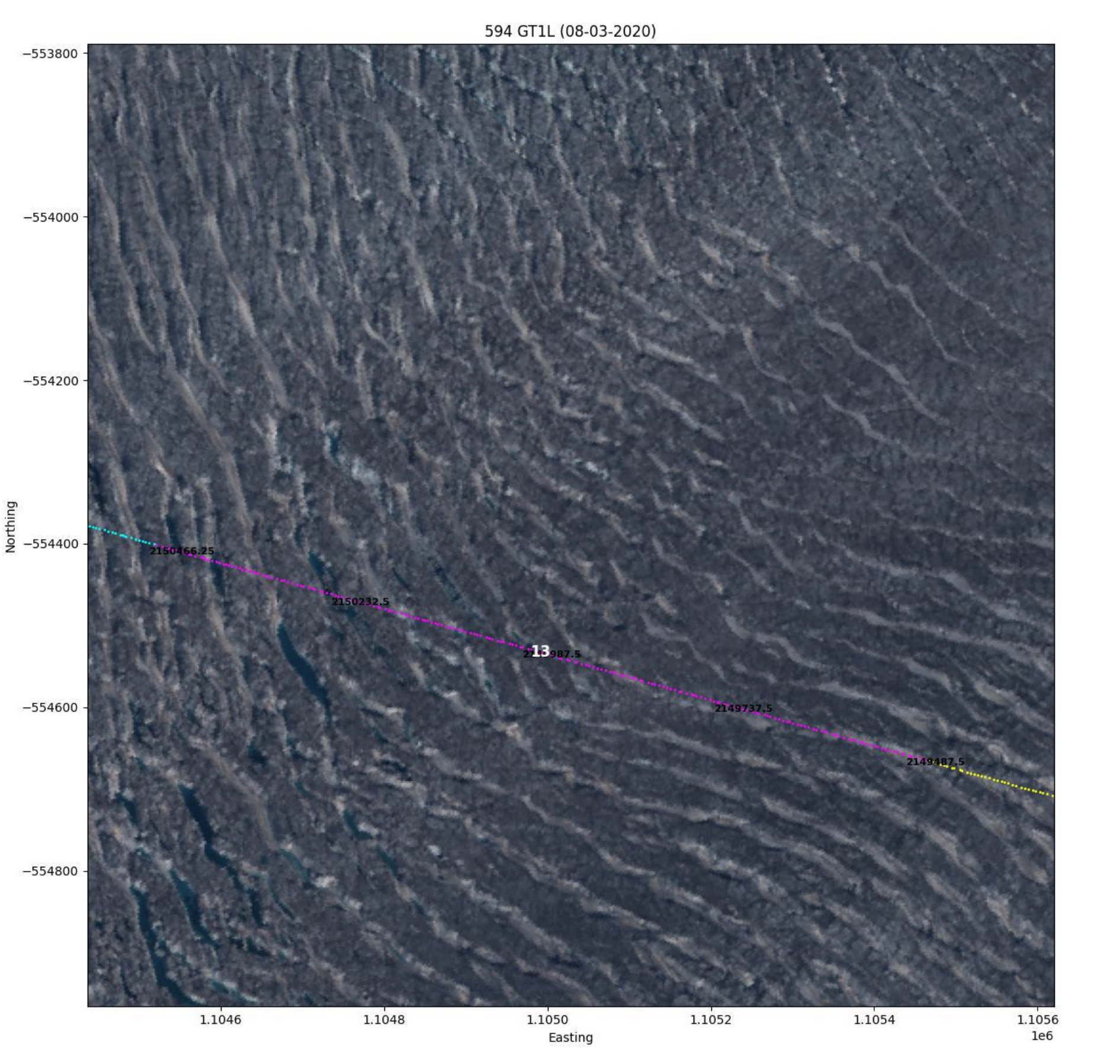}

\caption{Ice surface heights determined using the DDA-Ice-2 Algorithm for ice surfaces \cite{DDA-Ice-2} which coincide with water filled crevasses, giving more insight into the depth of water in a crevasse feature. (a) depicts the final ground estimate from the algorithm for the granule specified below, for track GT1L. (b) depicts the coincidence of the track with the water features, as they are viewable from a WorldView Image. The WV image and ICESat-2 data were collected two days apart, Aug 1, 2020 and Aug 3, 2020 respectively. \\
WV Image: WV02\_20200801124218\_10300100AA721300\_20AUG01124218-M1BS-504570334070\_01\_P004\
ATL03 granule name: ATL03\_20200803060754\_05940803\_006\_01.h5 \\
}
\end{figure}

\begin{figure}[H]
\includegraphics[width=0.9\textwidth]{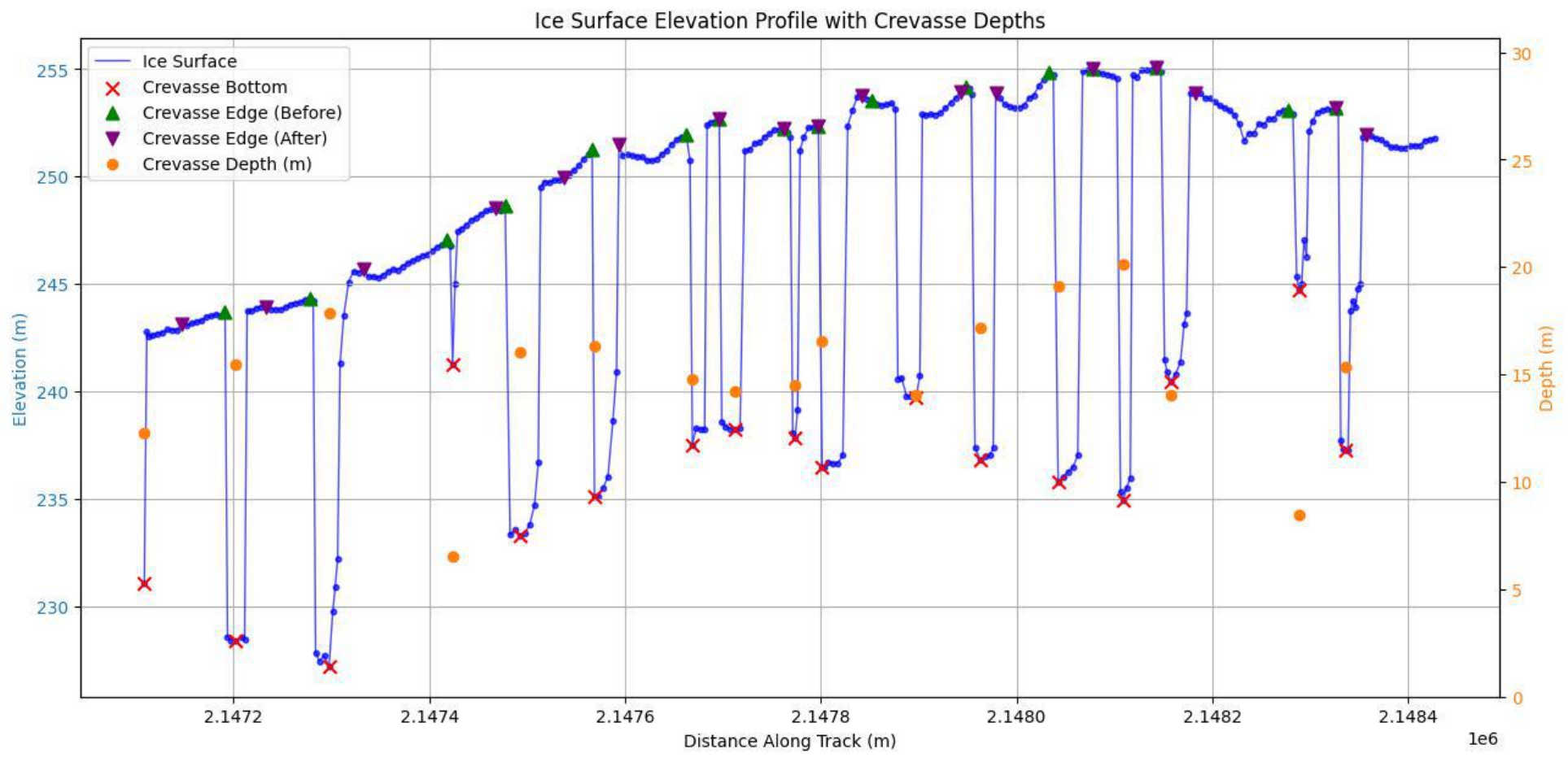}
\caption{Local minima and maxima are detected in the elevation profile from the DDA-ice algorithm ice surface elevation product. Minima are labeled as red x's, maxima are labeled as triangles, and the derived depth is also plotted, with the depth axis on the right side of the plot. }
\end{figure}

\begin{figure}[H]
\includegraphics[width=0.98\textwidth]{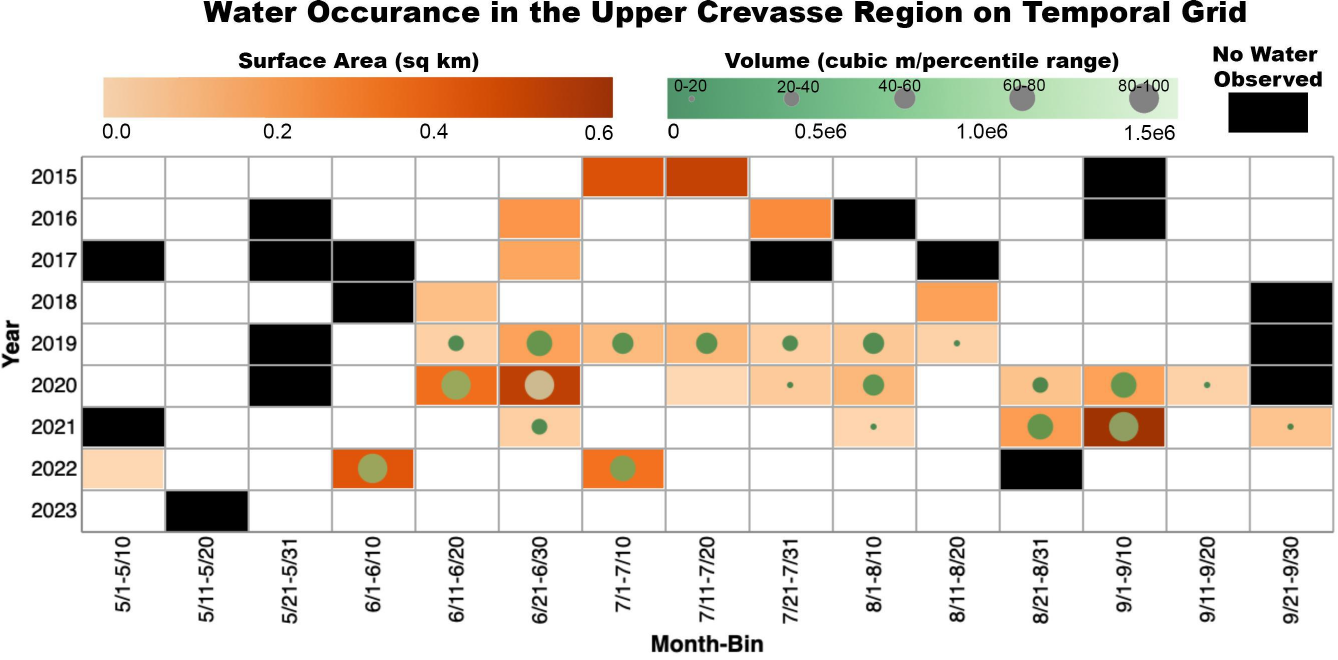}
\caption{Water occurrence, area mapped, and volume calculated organized on a temporal grid, with years 2015-2023 on the y axis and 10-day date ranges on the x axis. Black cells represent no water observed, meaning an image was available, but no water was visible. Surface area of mapped water scales with orange. Volume calculated scales by percentiles (circle sizes) and with green (circle colors).}
\end{figure}

\begin{figure}[H]
\centering

\begin{minipage}{15cm}
\includegraphics[width = 0.32\textwidth, trim=2.9cm 5.5cm 4.5cm 5cm, clip]{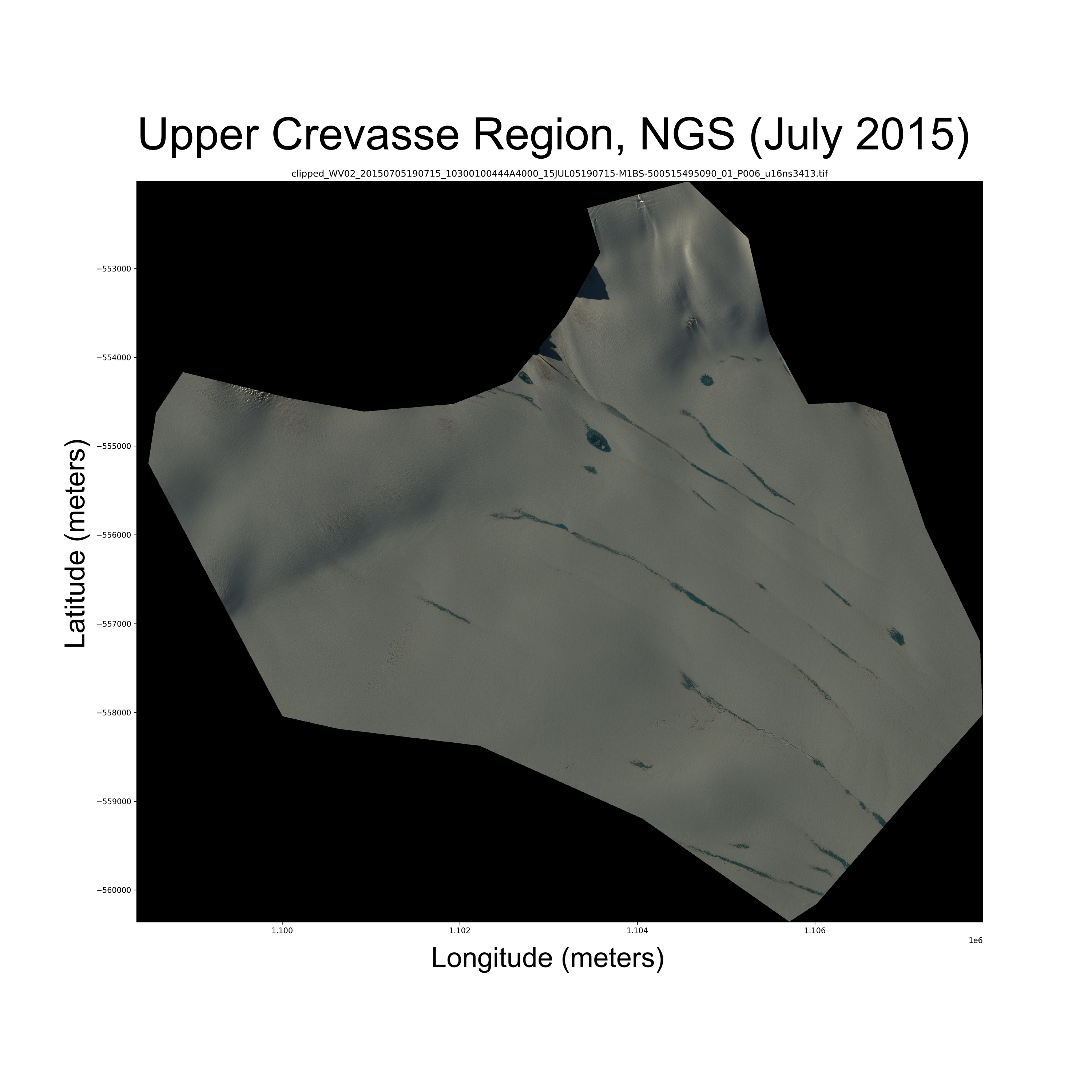}
\includegraphics[width = 0.3\textwidth, trim=3cm 4cm 4.5cm 4cm, clip]{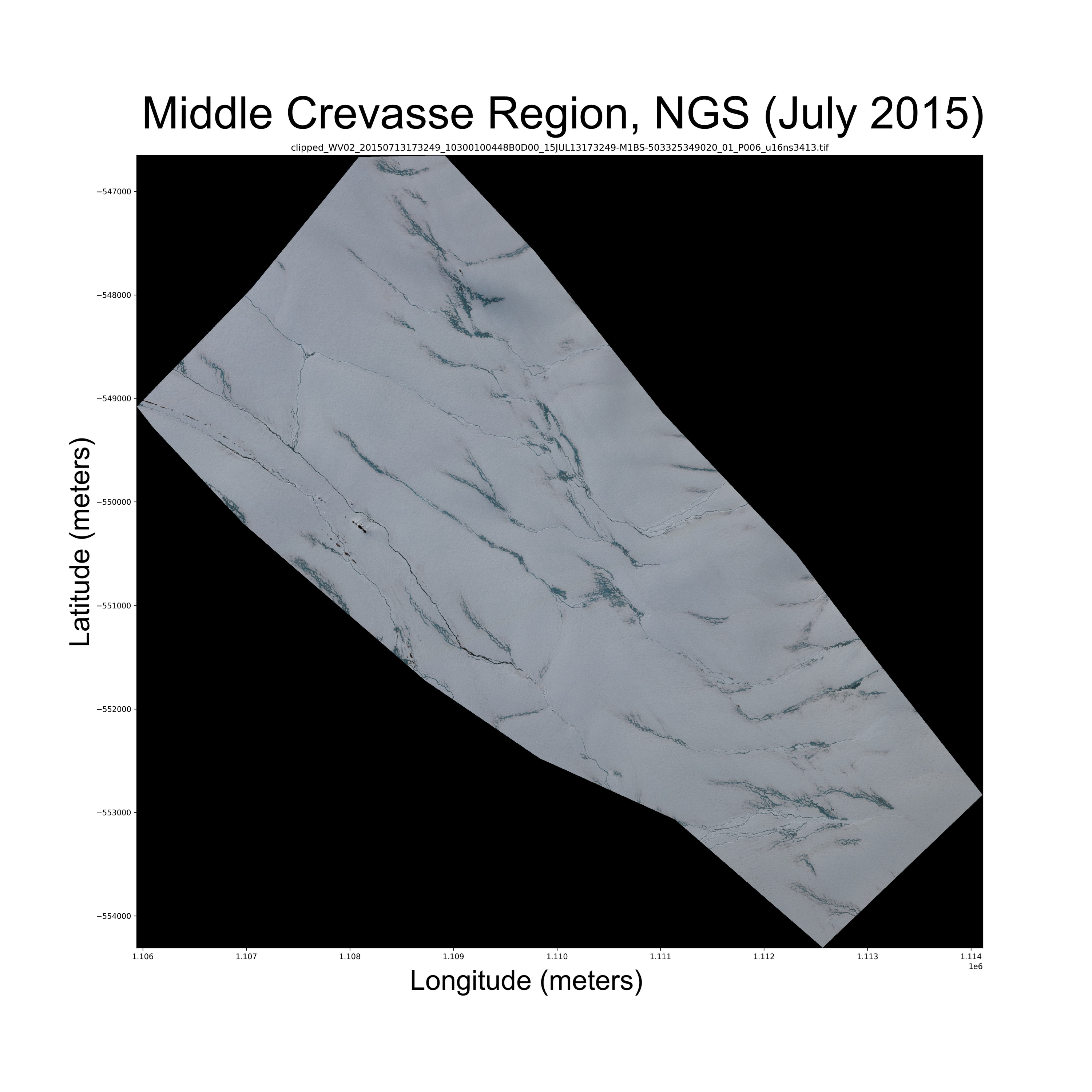}
\includegraphics[width = 0.26\textwidth, trim=5cm 3cm 6cm 3cm, clip]{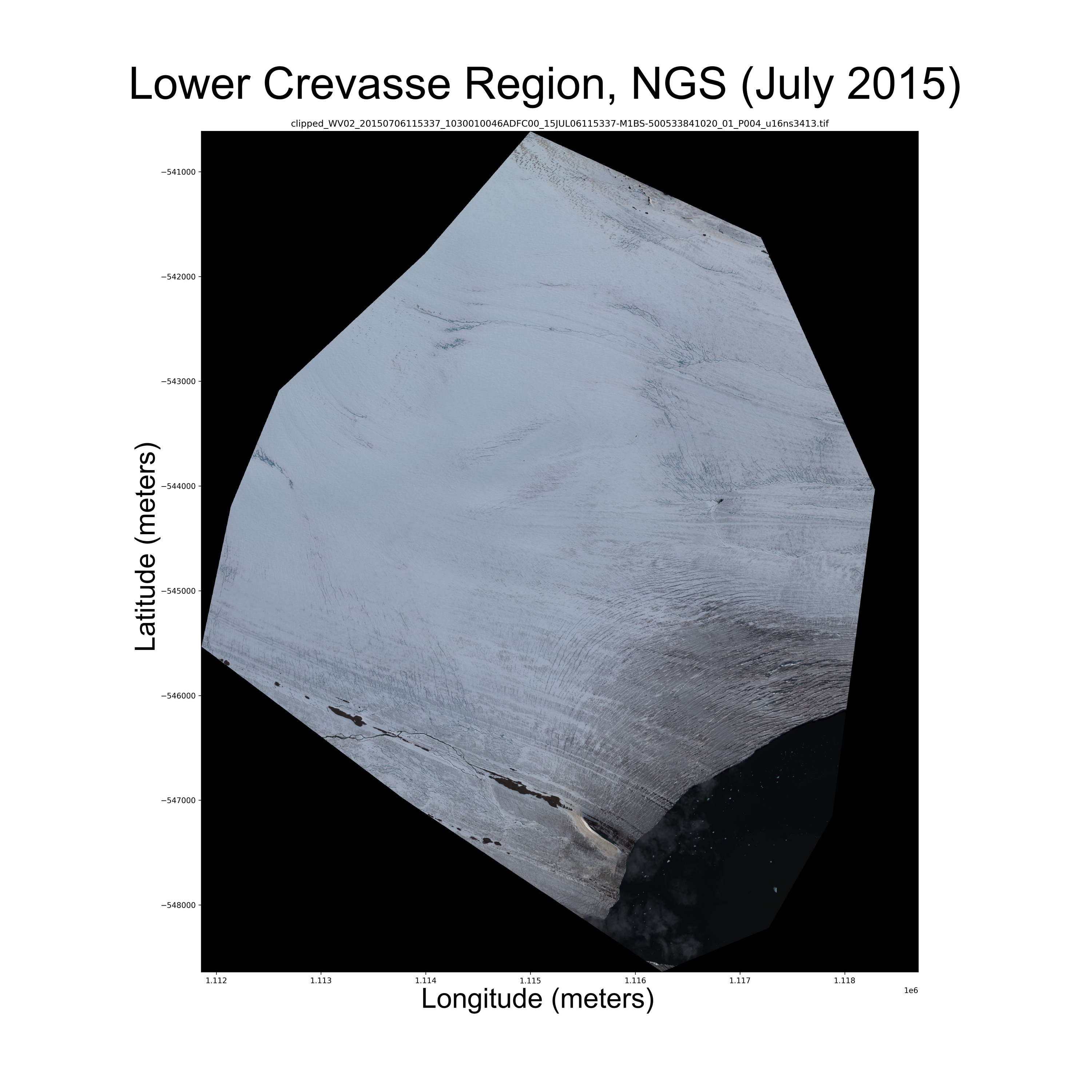}
\end{minipage}
\caption{WorldView (WV) imagery over Negribreen prior to its surge for the upper crevasse region, July 5, 2015 (a), middle region, July 13, 2015 (b) and lower region, July 6, 2015 (c). Supraglacial water takes the form of melt streams. }
\end{figure}

\begin{figure}[H]
\includegraphics[width=0.45\textwidth,trim=50 125 50 125, clip]{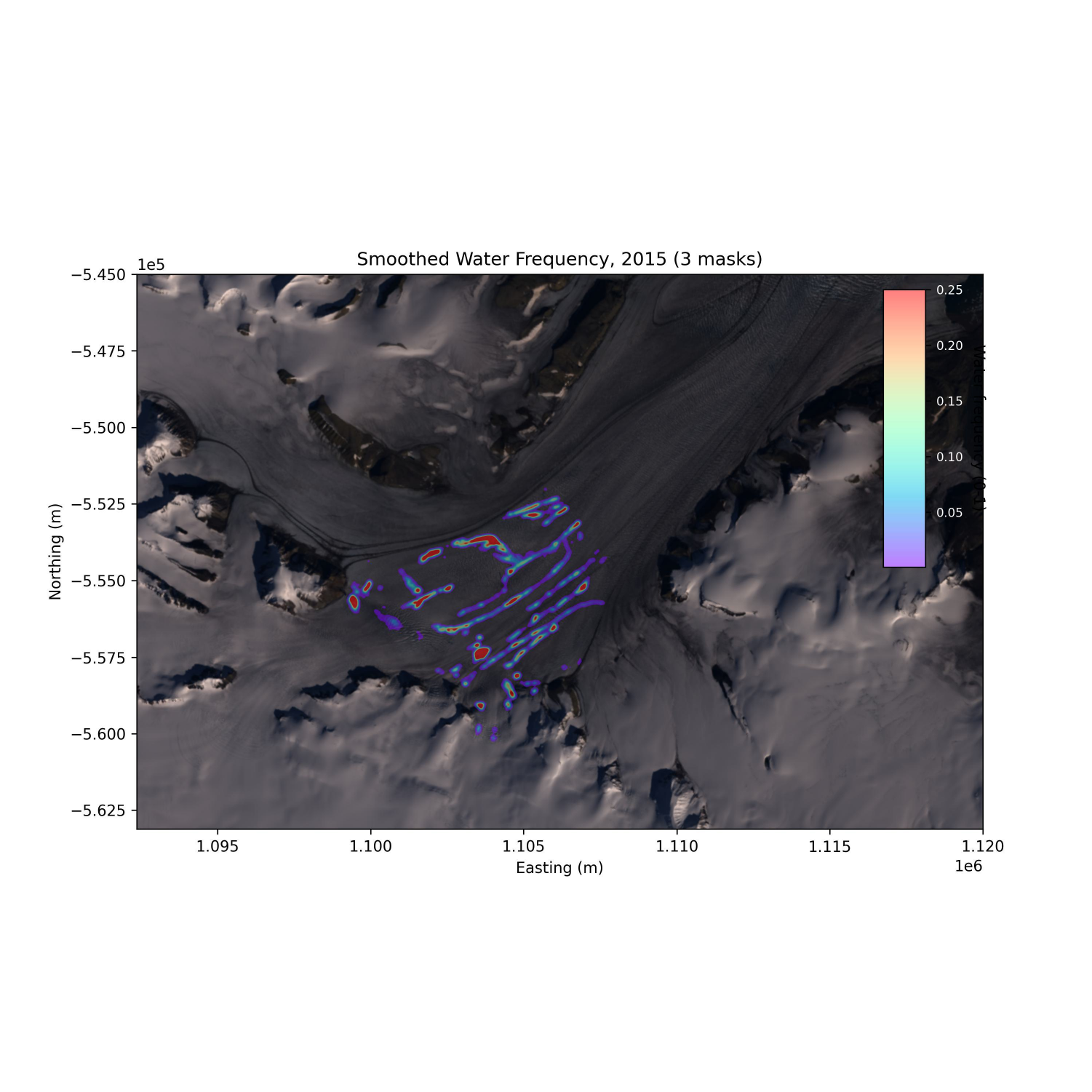}
\includegraphics[width=0.45\textwidth,trim=50 125 50 125, clip]{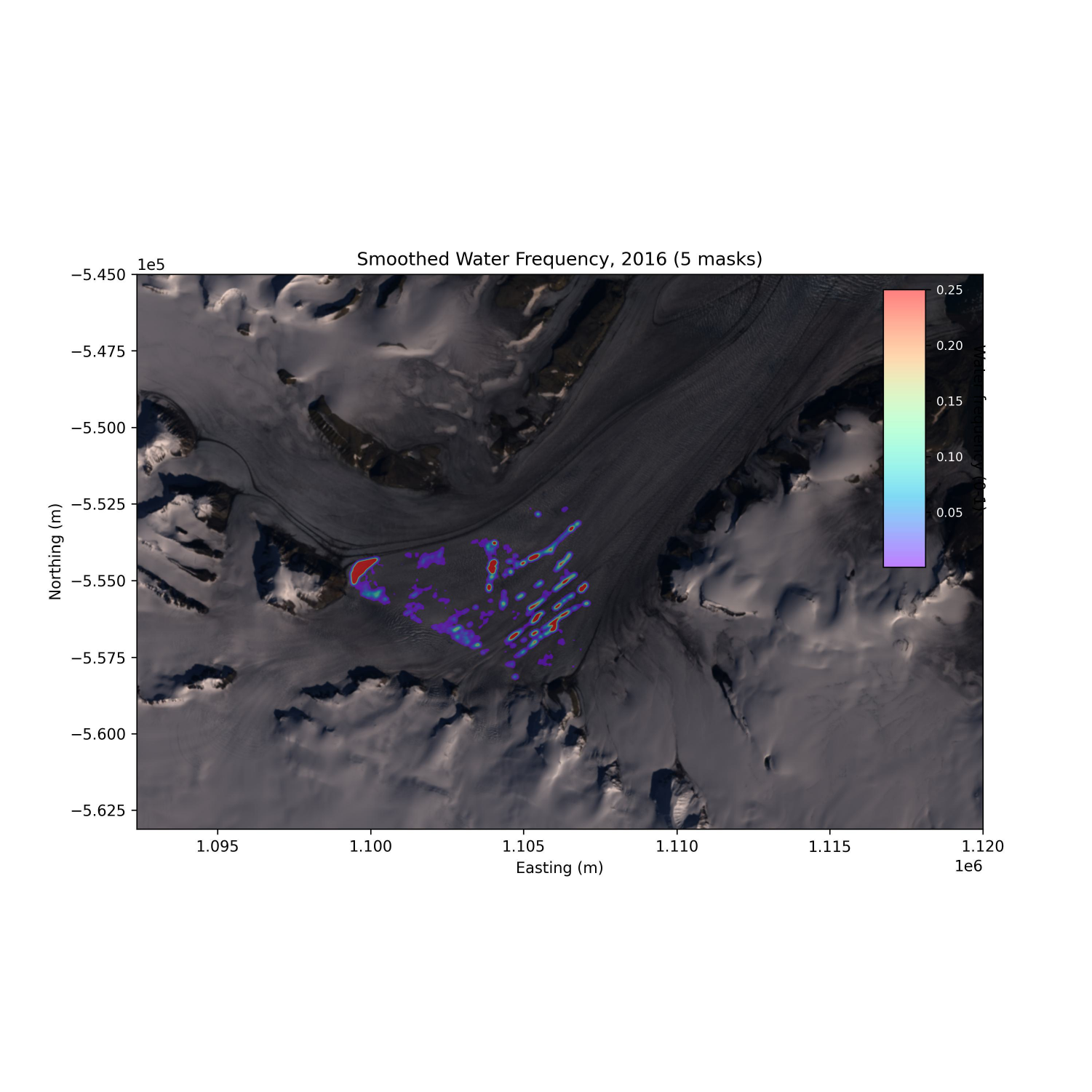}

\includegraphics[width=0.45\textwidth,trim=50 125 50 125, clip]{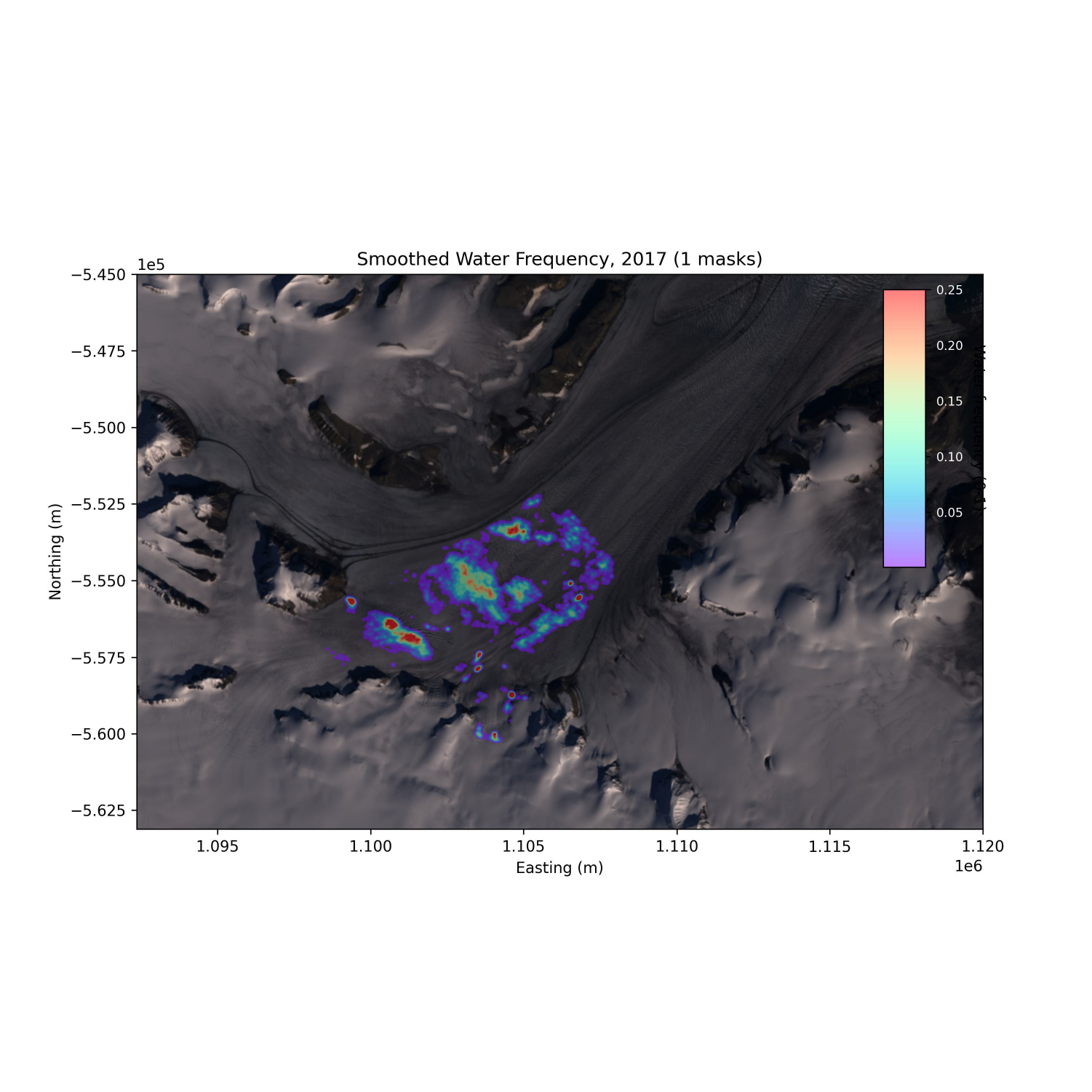}
\includegraphics[width=0.45\textwidth,trim=50 125 50 125, clip]{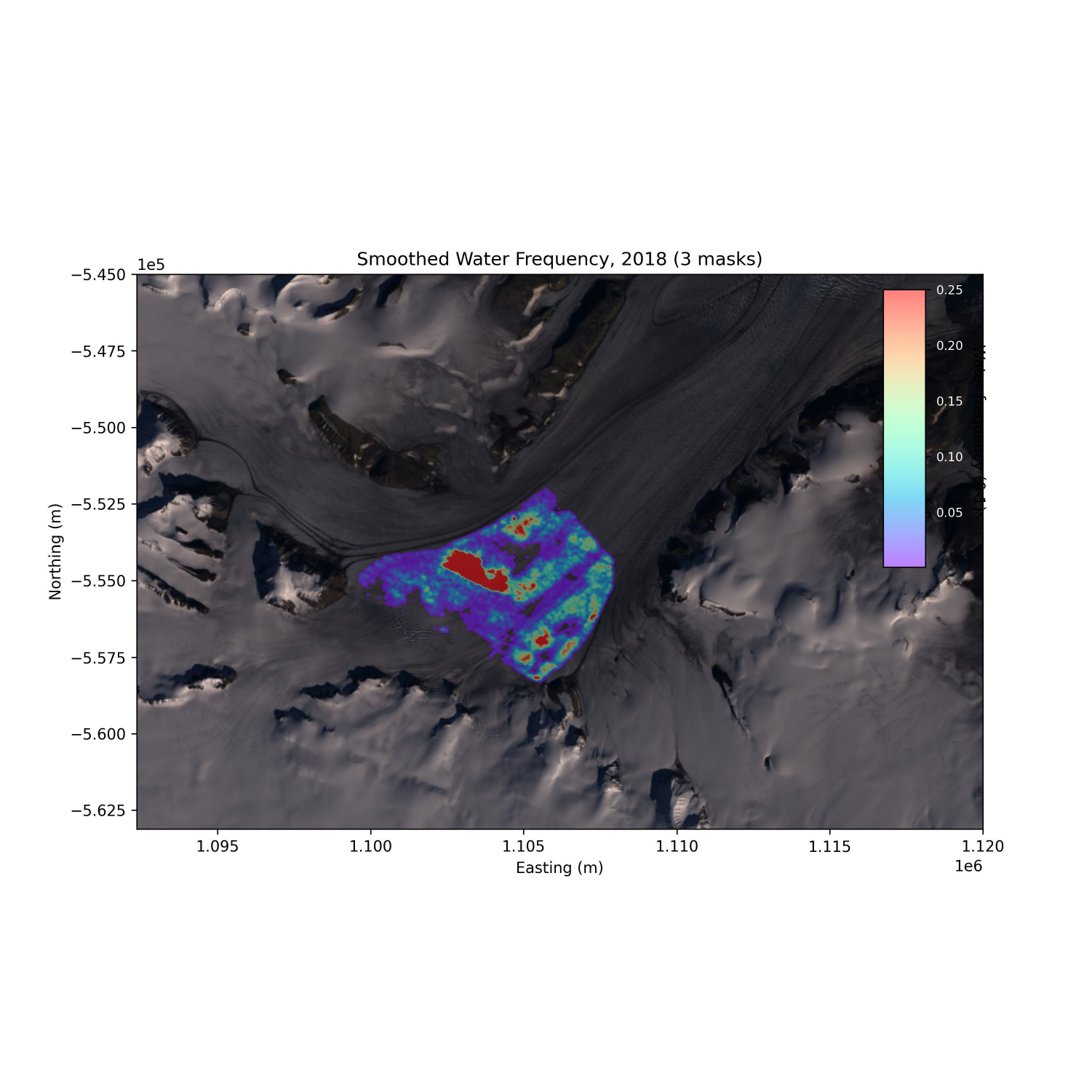}

\includegraphics[width=0.45\textwidth,trim=50 125 50 125, clip]{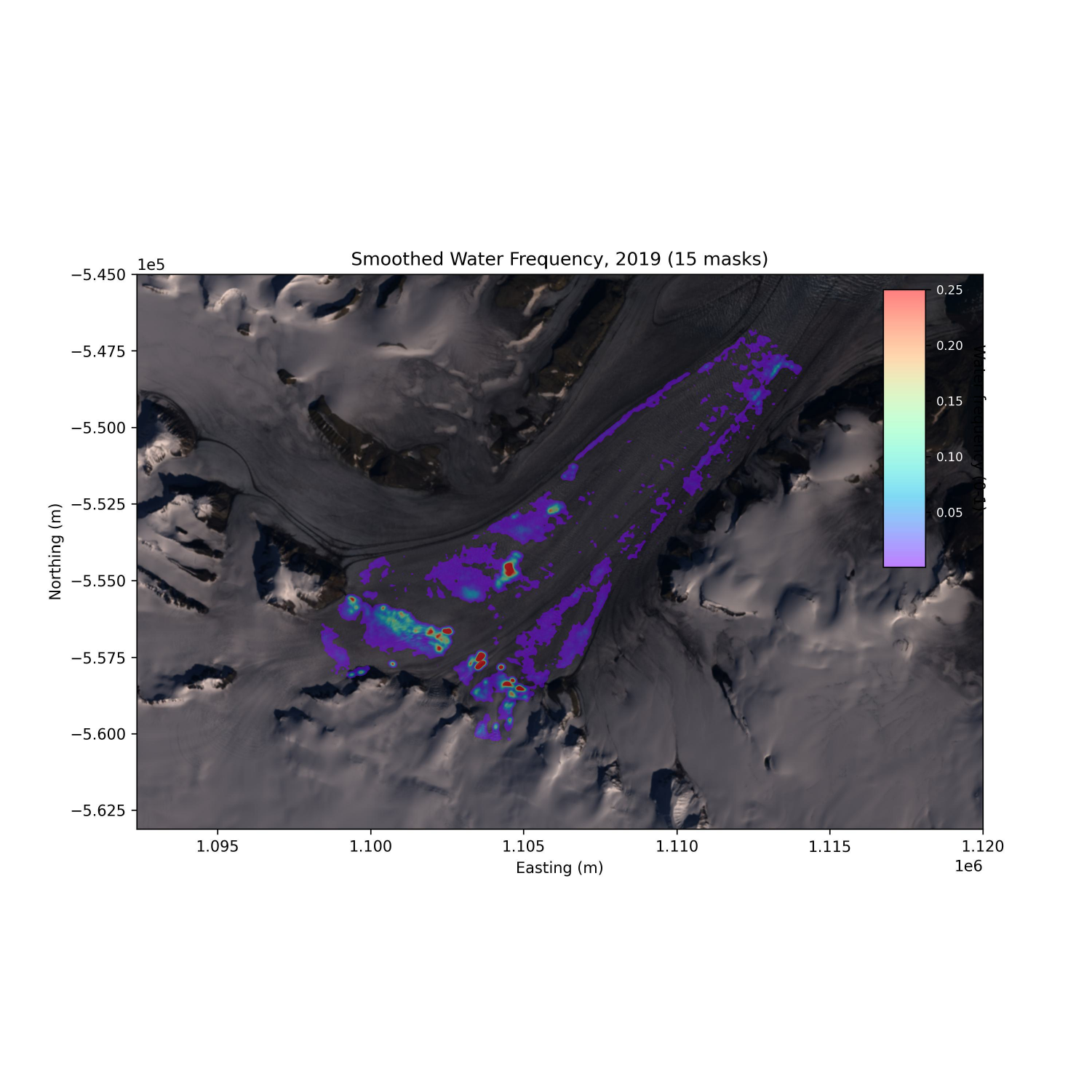}
\includegraphics[width=0.45\textwidth,trim=50 125 50 125, clip]{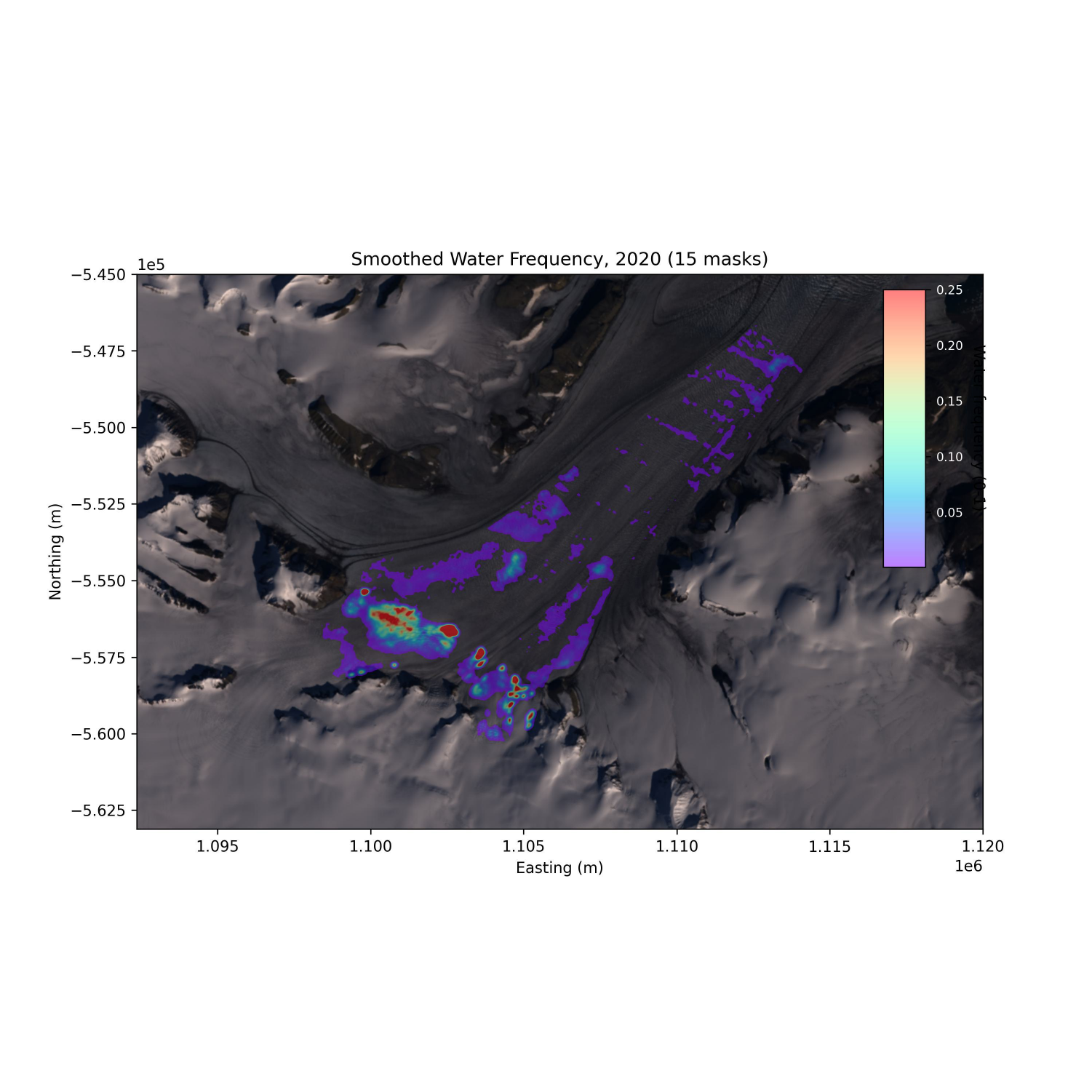}

\includegraphics[width=0.45\textwidth,trim=50 125 50 125, clip]{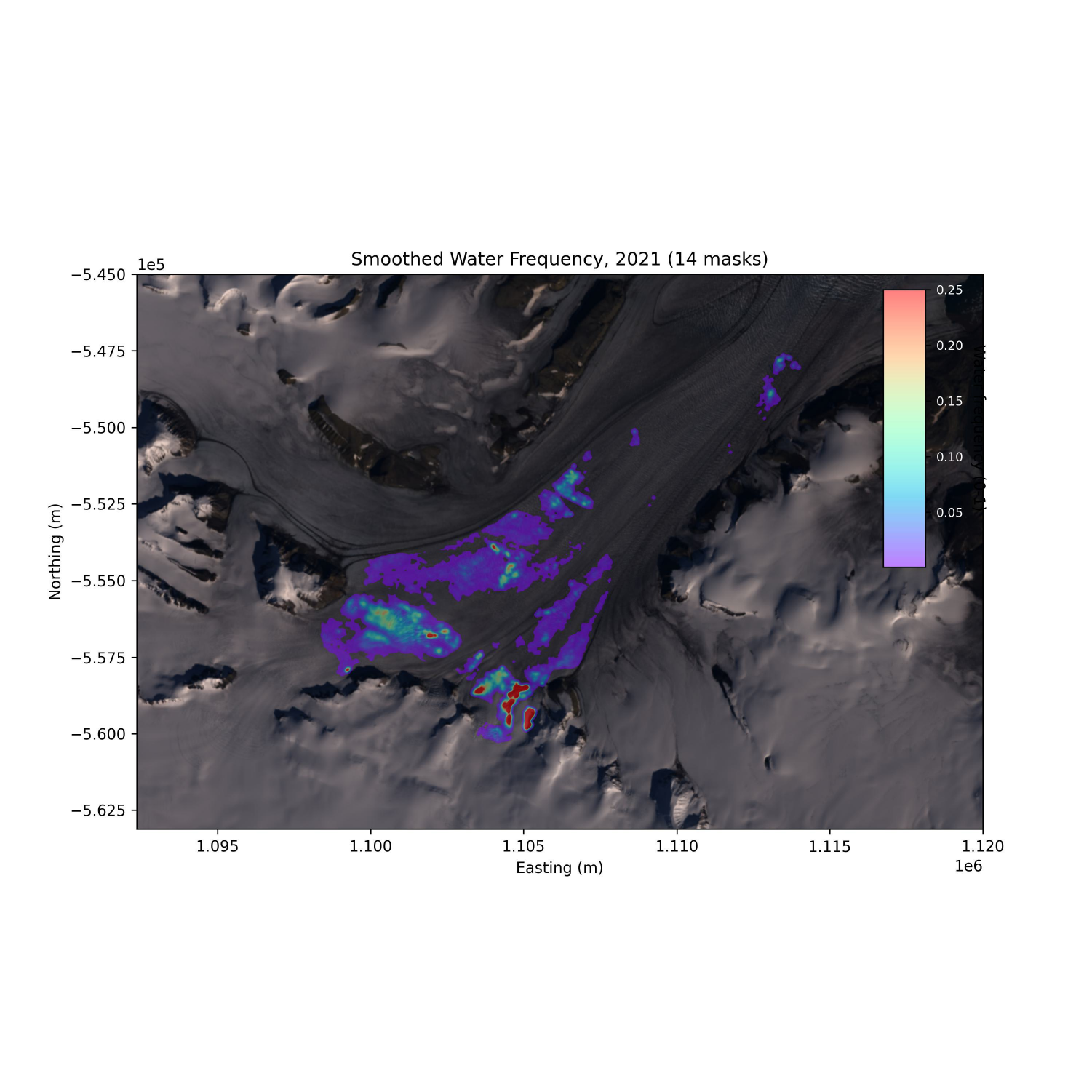}
\includegraphics[width=0.45\textwidth,trim=50 125 50 125, clip]{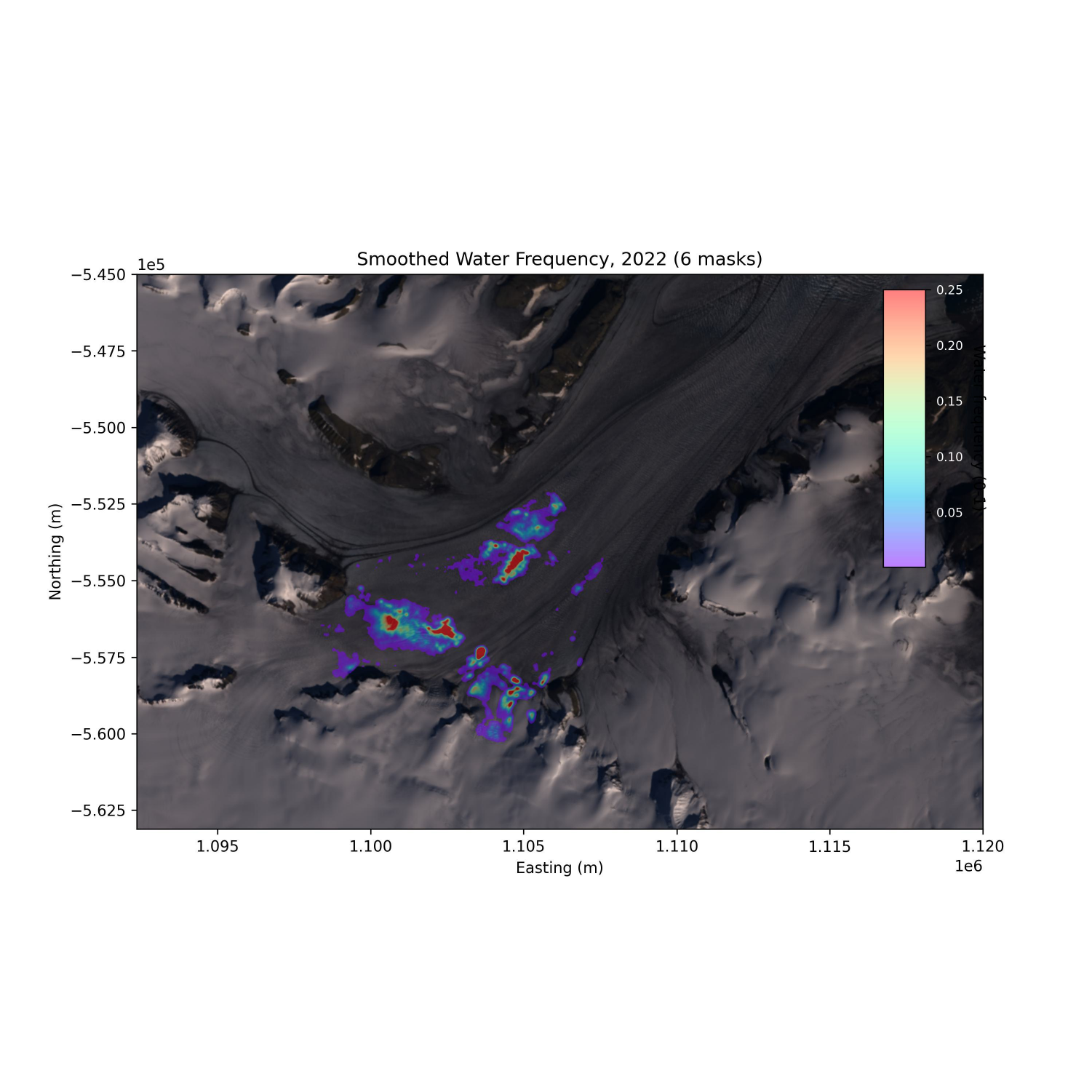}

\caption{Supraglacial watermaps of the upper and middle crevasse regions of NGS, averaged and smoothed with a guassian filter for visibility for each observed year during the surge (2016-2022). Higher values indicate higher probability of water occurrence.}
\end{figure}

\begin{figure}[H]
\includegraphics[width=0.245\textwidth,trim=5cm 5cm 3cm 5cm, clip]{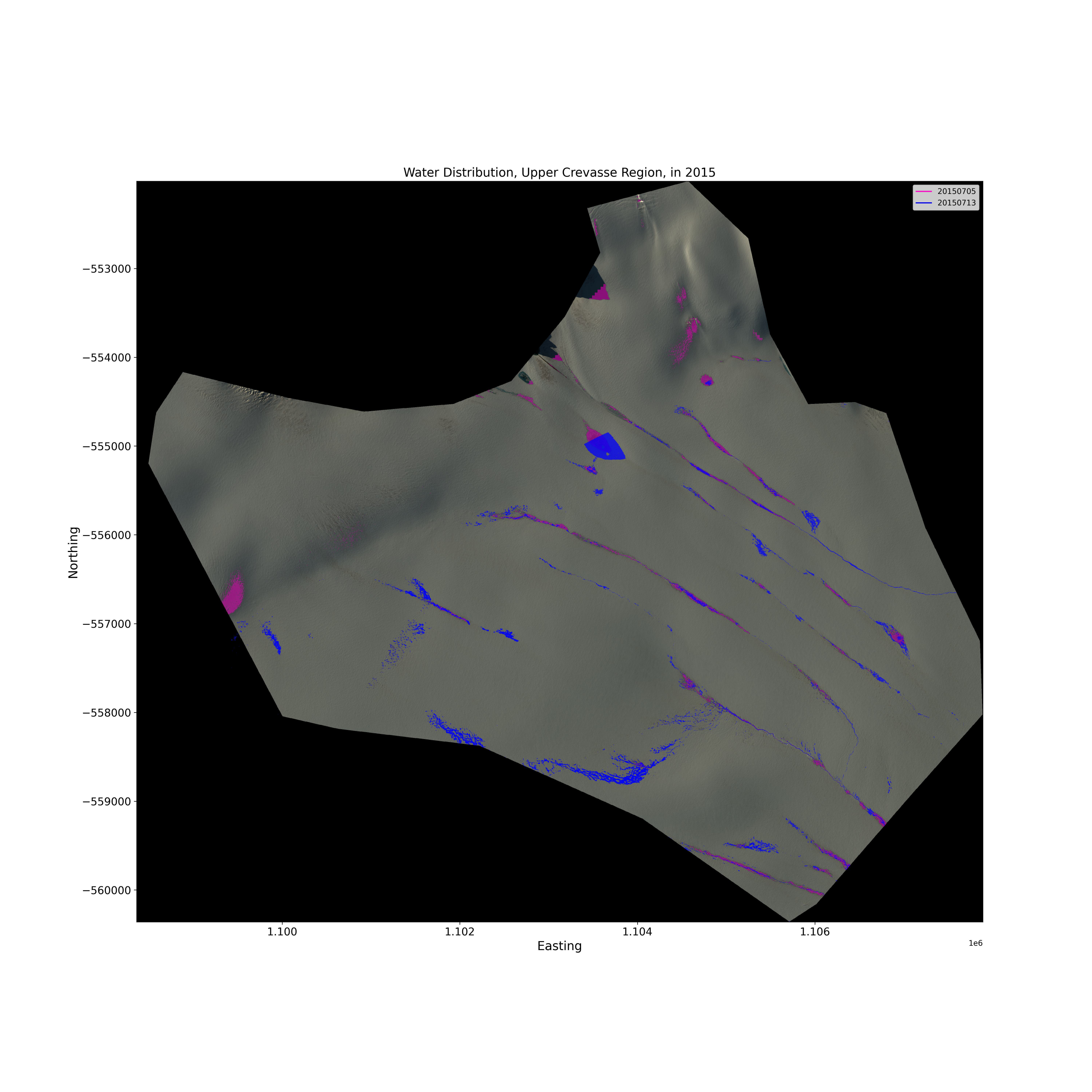}
\includegraphics[width=0.245\textwidth,trim=5cm 5cm 3cm 5cm, clip]{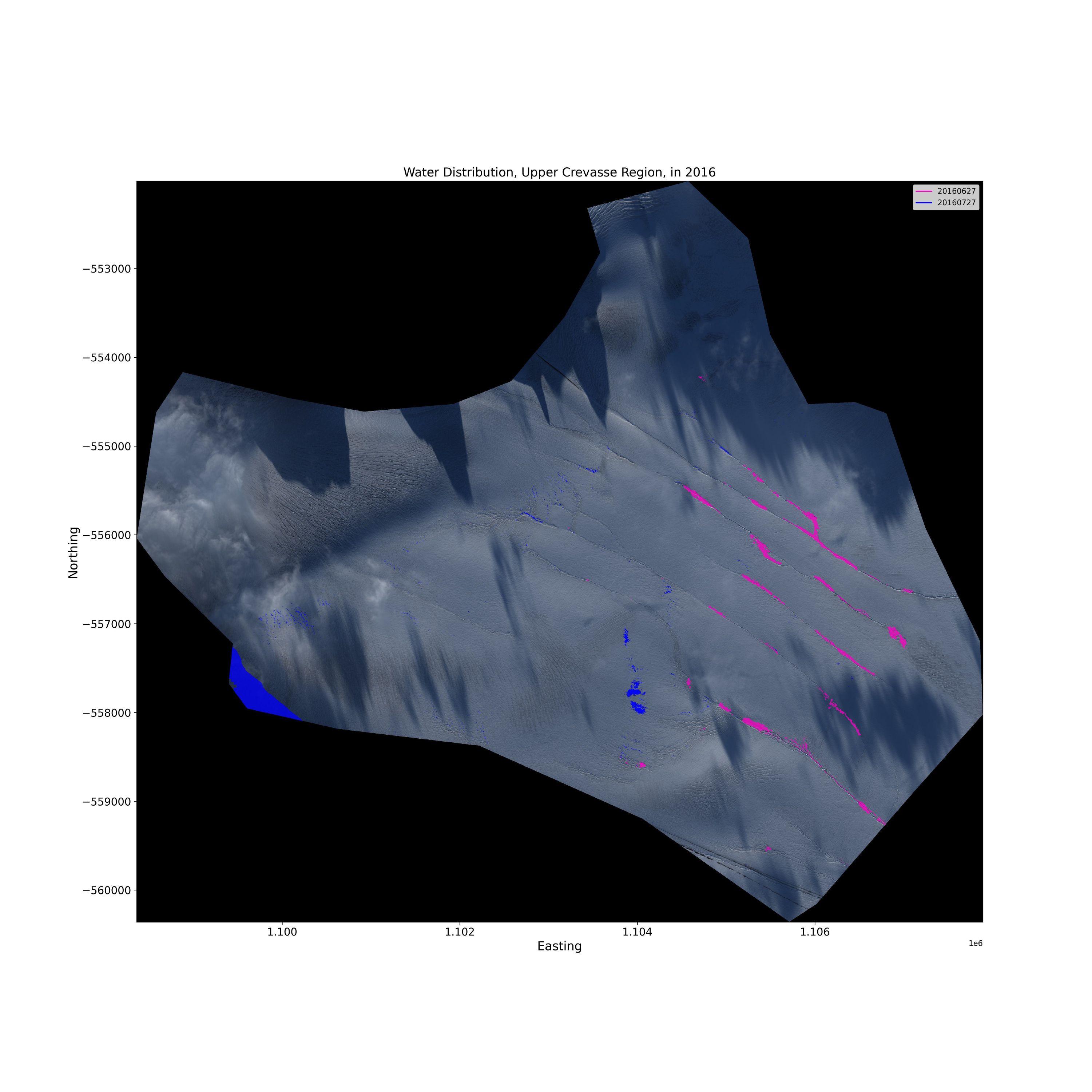}
\includegraphics[width=0.245\textwidth,trim=5cm 5cm 3cm 5cm, clip]{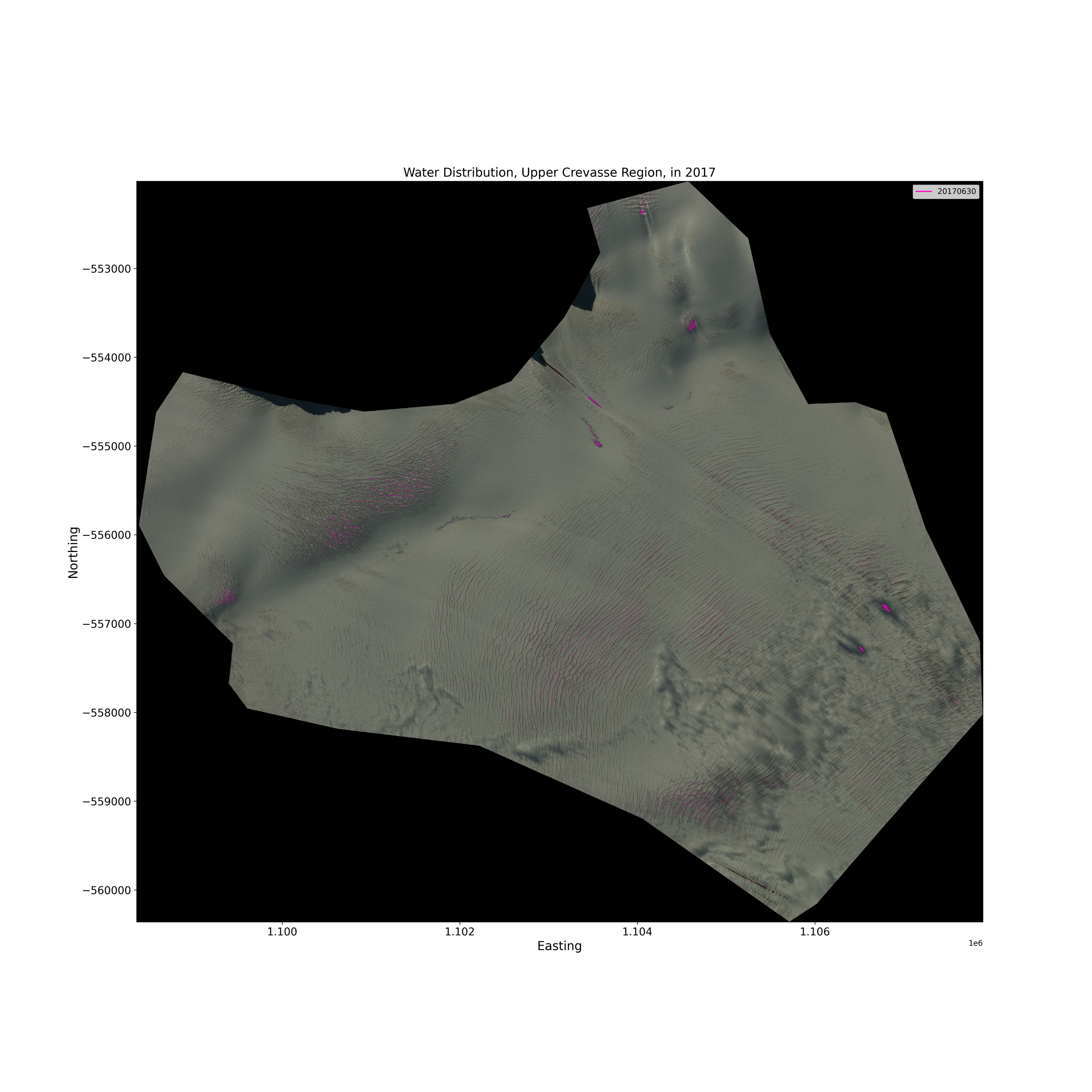}
\includegraphics[width=0.245\textwidth,trim=5cm 5cm 3cm 5cm, clip]{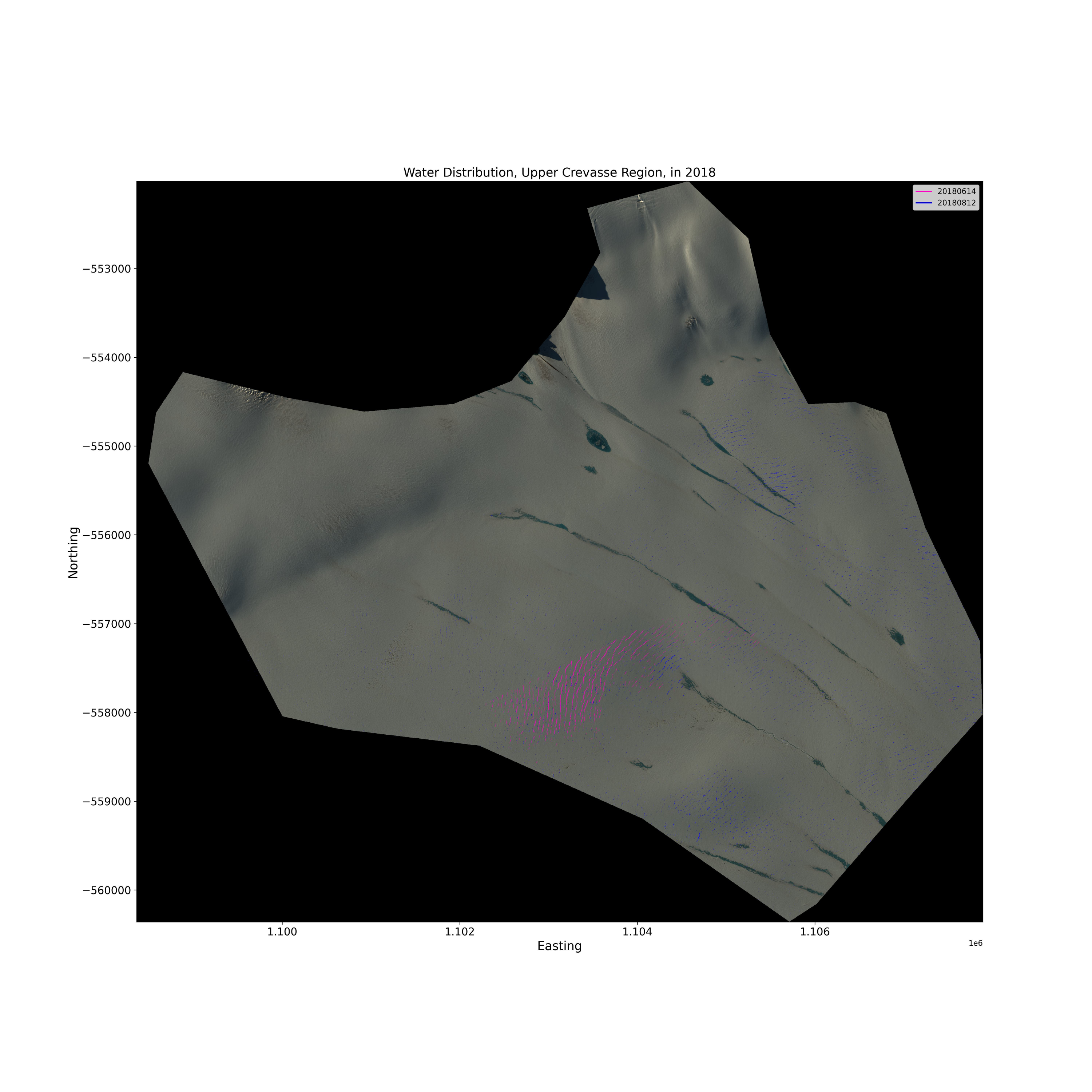}

\includegraphics[width=0.245\textwidth,trim=5cm 5cm 3cm 5cm, clip]{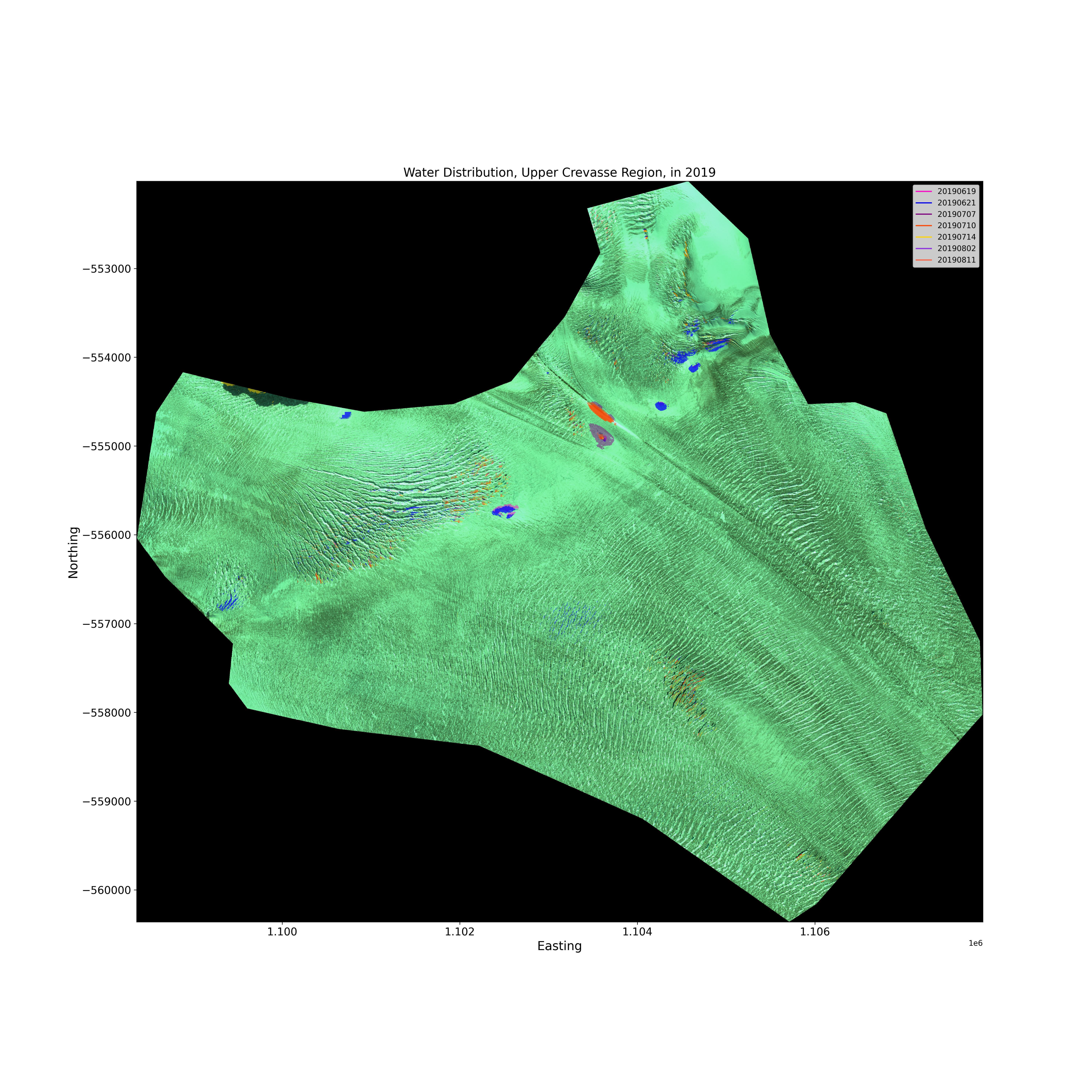}
\includegraphics[width=0.245\textwidth,trim=5cm 5cm 3cm 5cm, clip]{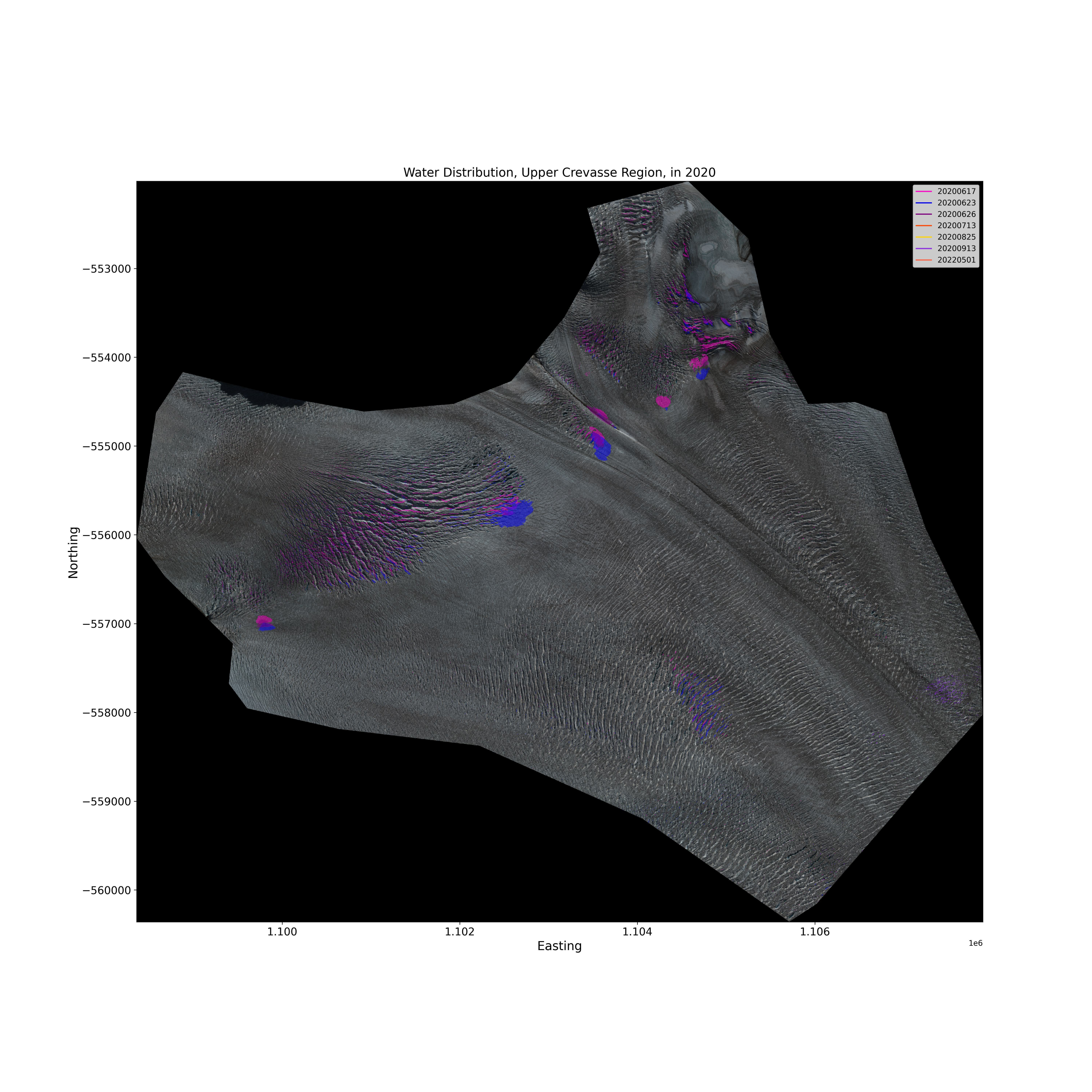}
\includegraphics[width=0.245\textwidth,trim=5cm 5cm 3cm 5cm, clip]{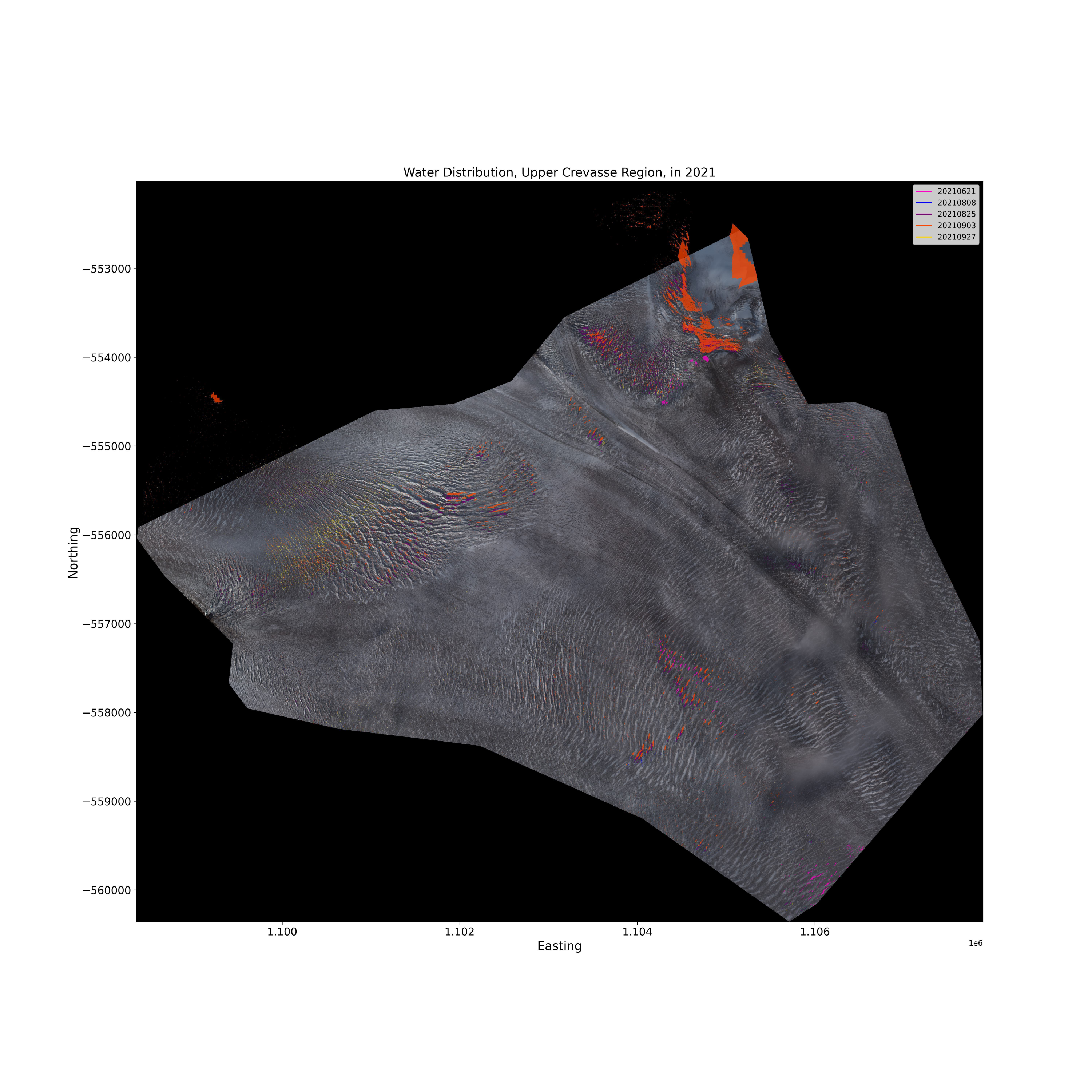}
\includegraphics[width=0.245\textwidth,trim=5cm 5cm 3cm 5cm, clip]{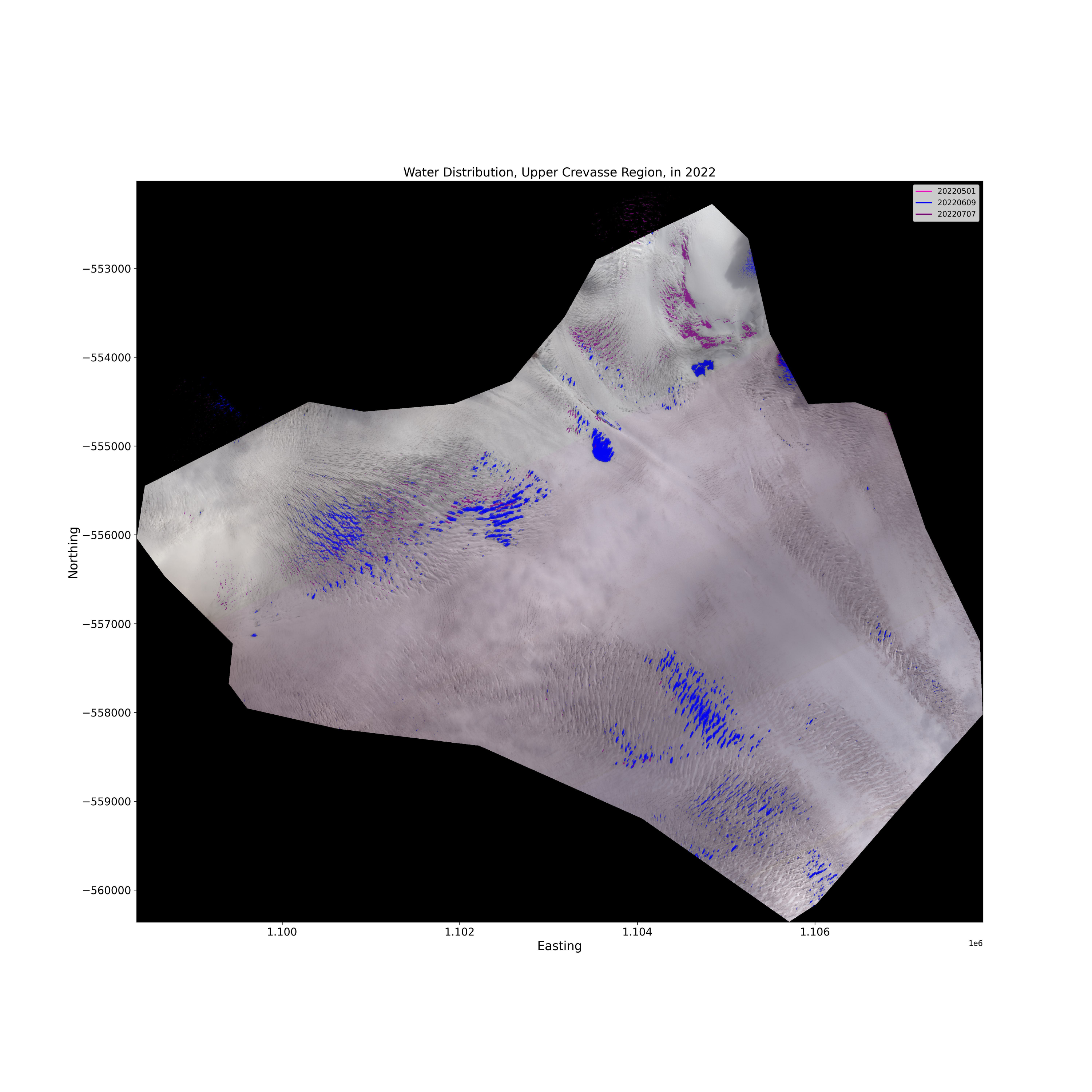}

\caption{Maps of supraglacial water in the upper crevasse region for (a) 2015, (b) 2016, (c) 2017, (d) 2018, (e) 2019 (f), 2020, (g) 2021, and (h) 2022. Pixels classified as containing water are colored. }
\end{figure}

\begin{figure}[H]
\includegraphics[width = 0.98\textwidth]{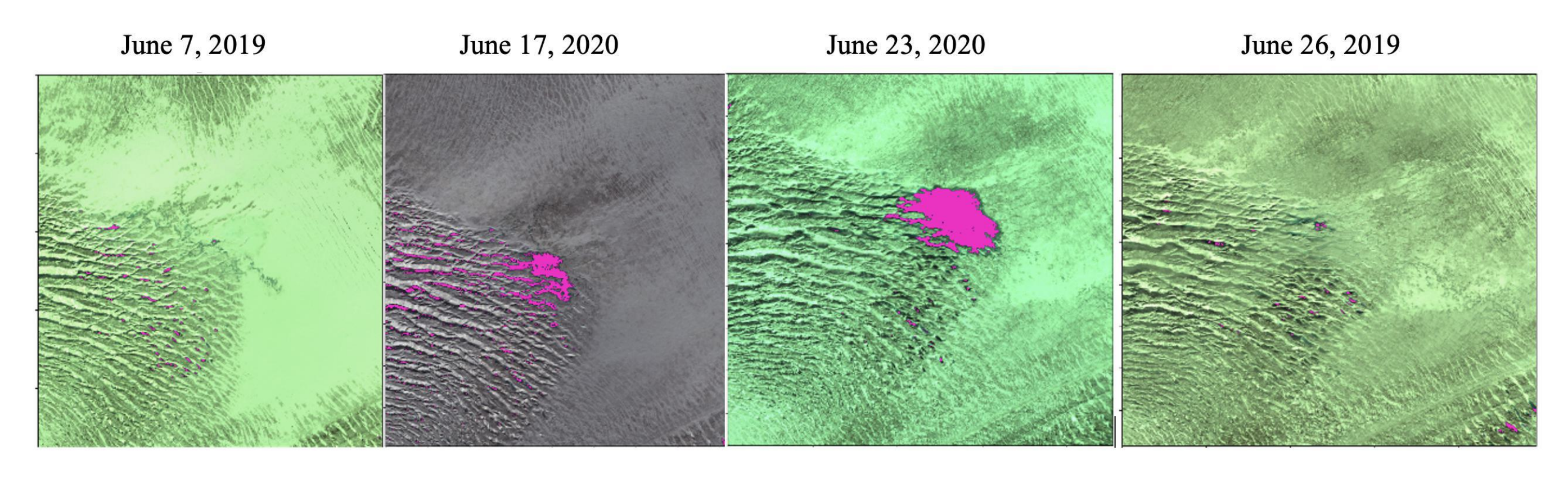} 
\caption{Formation and drainage of large melt pond in the upper crevasse region. It's quick drainage between June 23 and June 26, 2020 implies that the glacier has a partially efficient drainage mechanism. Water filled crevasses are present around the melt pond, even after its drainage.}
\end{figure}

\begin{figure}[H]
\includegraphics[width = 0.5\textwidth, trim=4cm 5cm 4.5cm 6cm, clip]{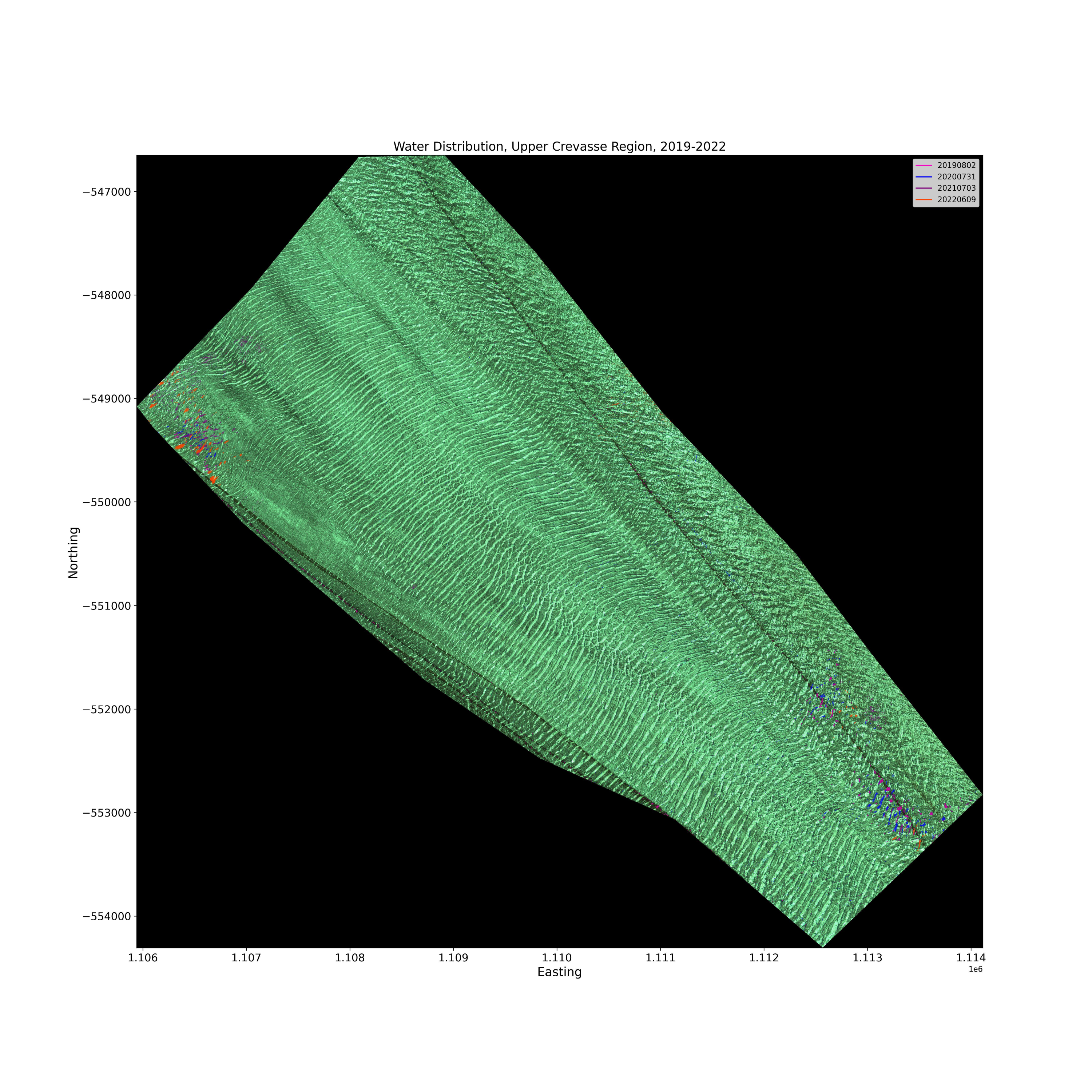}
\includegraphics[width = 0.5\textwidth, trim=5cm 4cm 6cm 4cm, clip]{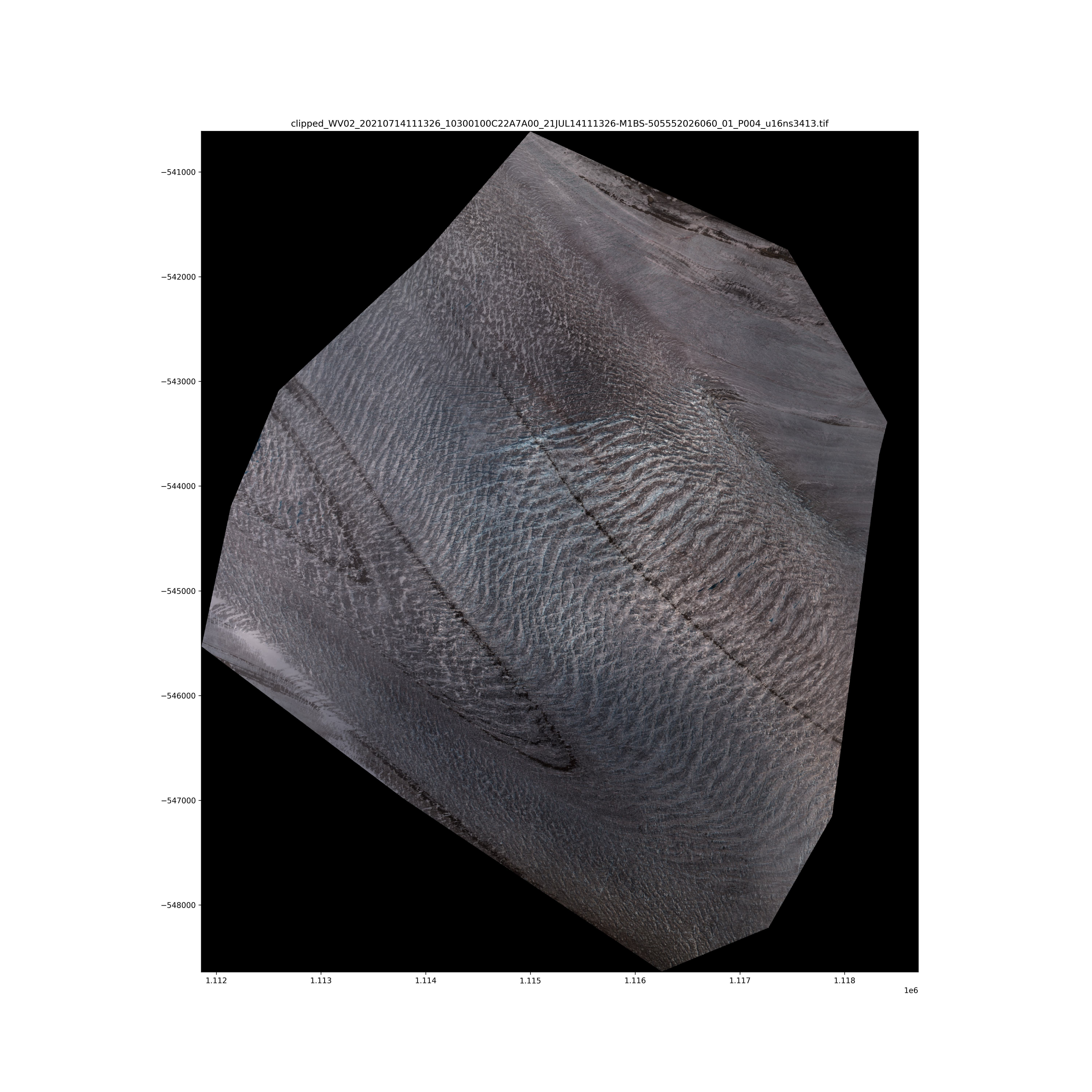}
\caption{All WV images for which supraglacial water in observed in the middle crevasse region. August 2, 2019 (pink) ; July 31, 2020 (blue); July 3, 2021 (magenta); June 9, 2022 (orange).
Occurrences of water-filled crevasses in the middle crevasse region (left) and lower crevasse region (right). Occurrence of supraglaical water in the upper crevasse region is observed for 2019-2022 (August 2, 2019 (pink) ; July 31, 2020 (blue); July 3, 2021 (magenta); June 9, 2022 (orange)). Occurrence of water in the lower crevasse region on July 14, 2021 but not in other images.}
\end{figure}

\begin{figure}[H]
\includegraphics[width=0.33\textwidth,trim=5cm 0 5cm 1cm, clip]{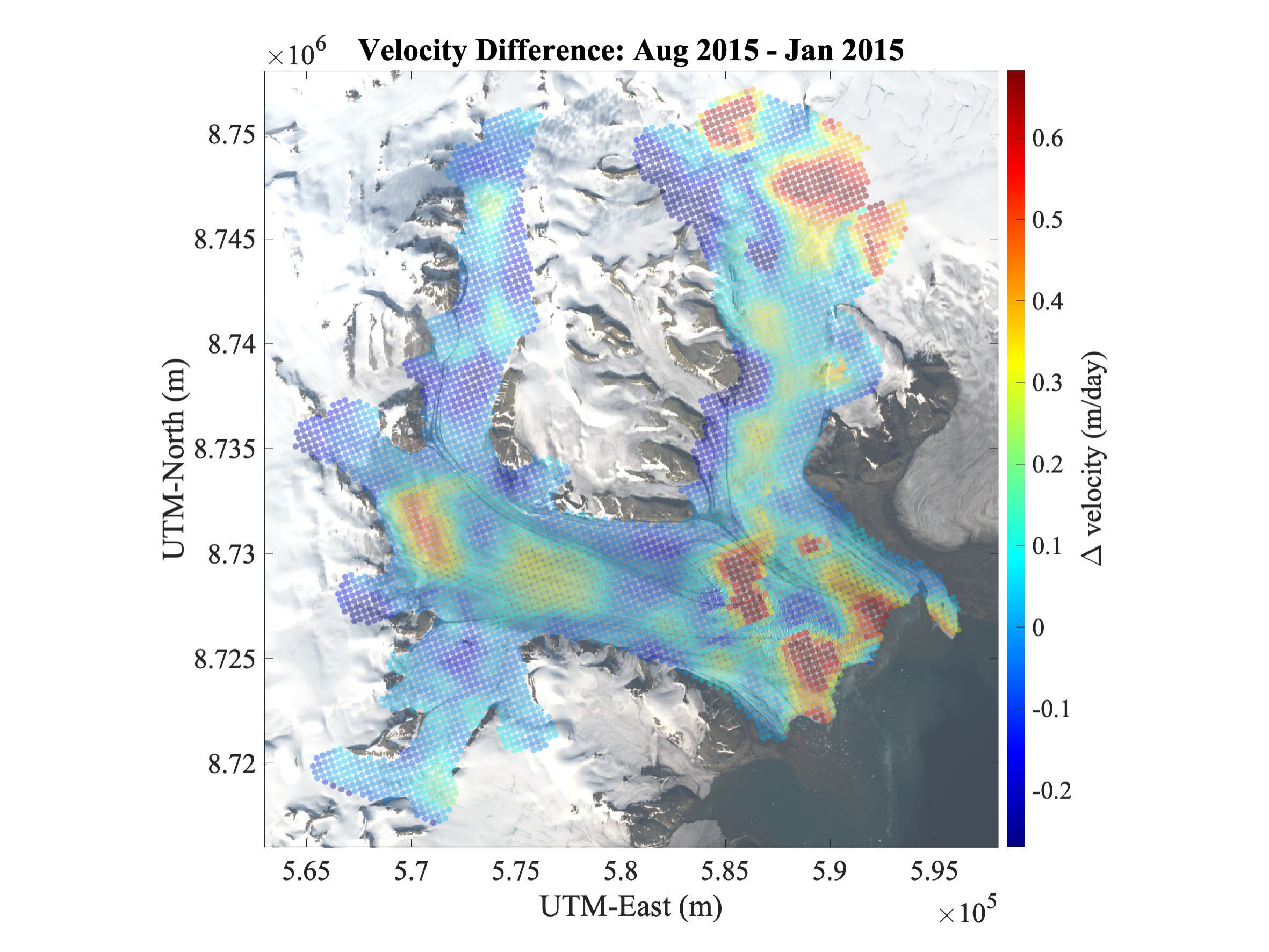}
\includegraphics[width=0.33\textwidth,trim=5cm 0 5cm 1cm, clip]{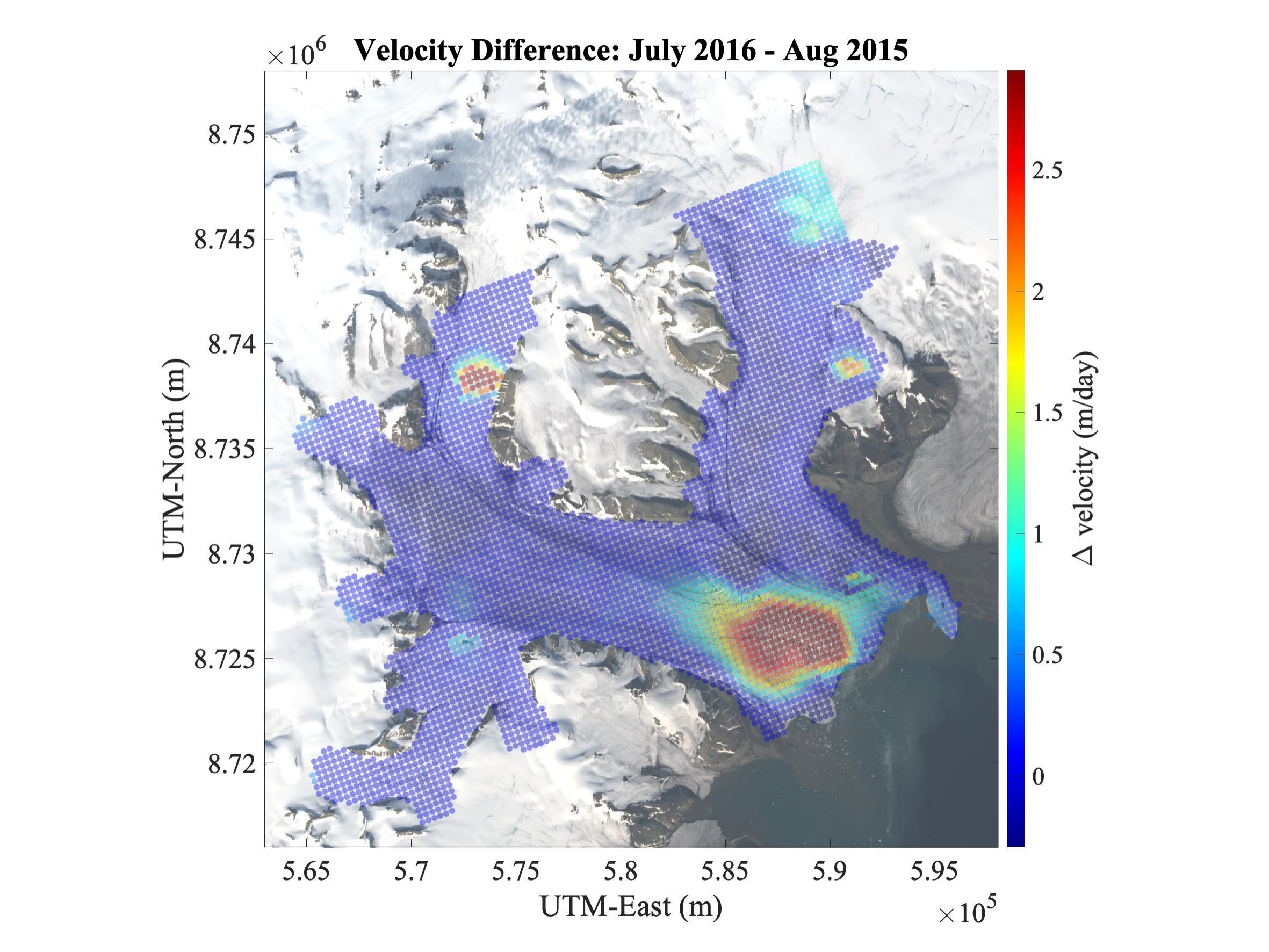}
\includegraphics[width=0.33\textwidth,trim=5cm 0 5cm 1cm, clip]{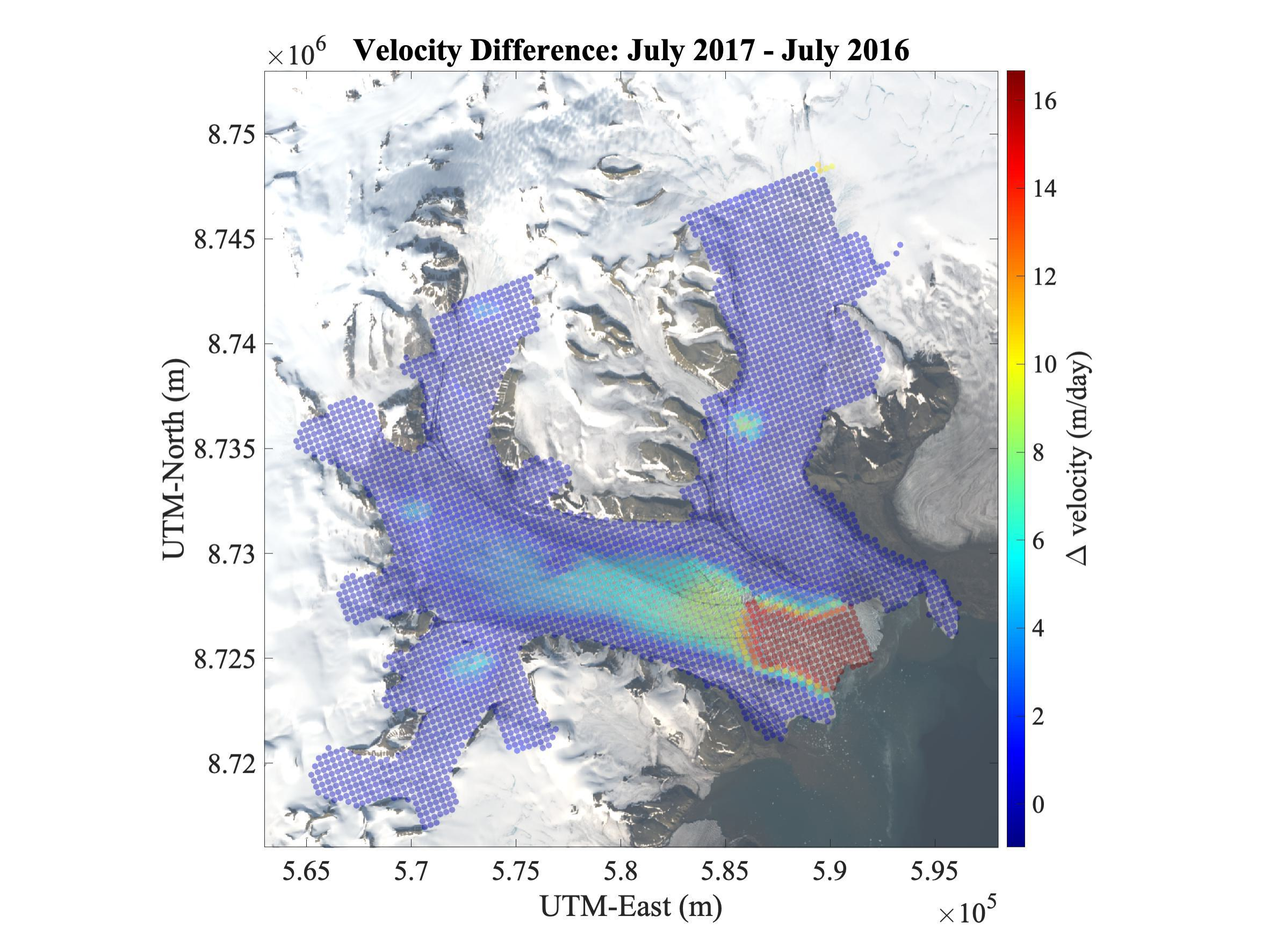}	

\includegraphics[width=0.33\textwidth,trim=5cm 0 5cm 1cm, clip]{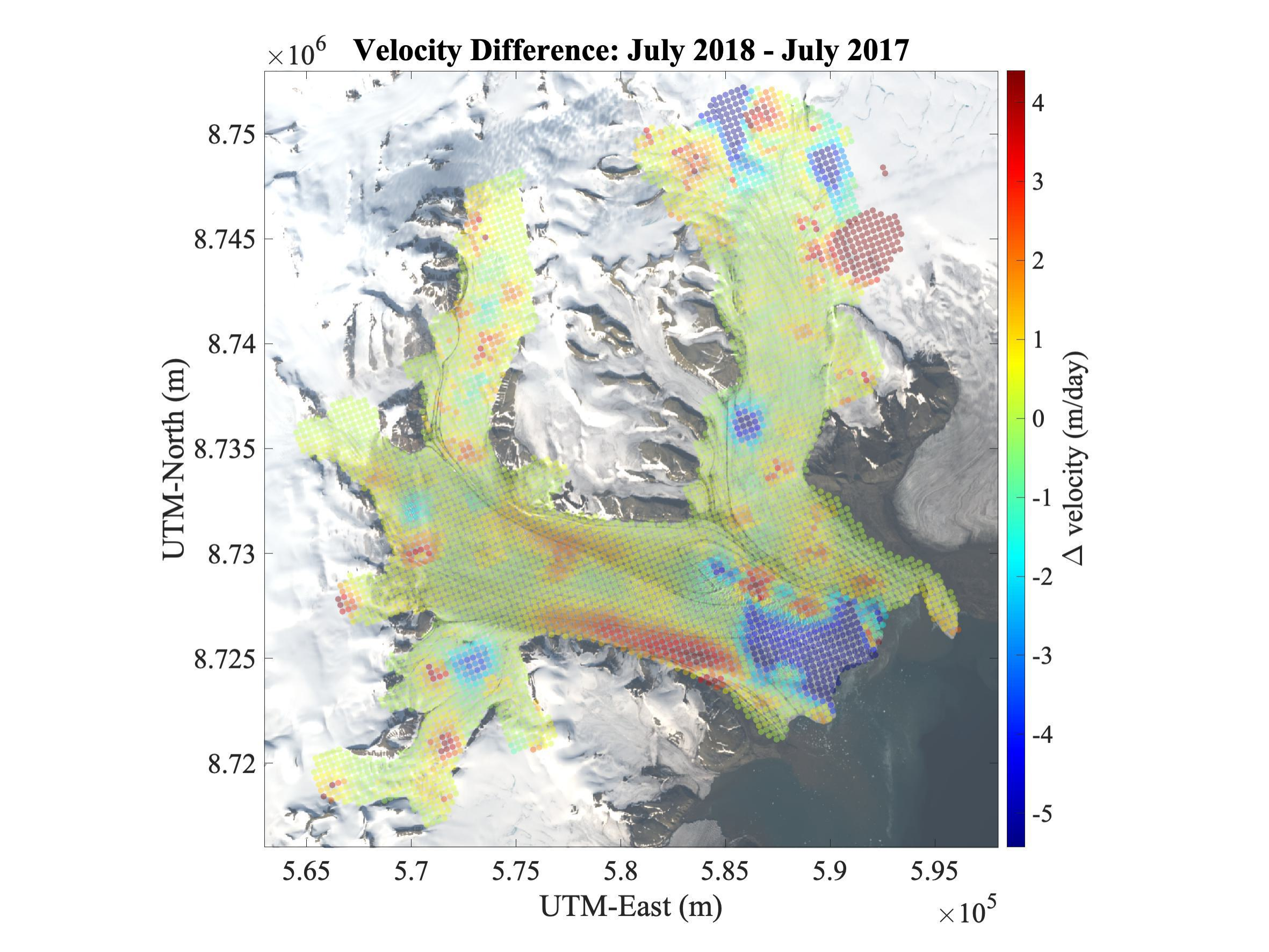}
\includegraphics[width=0.33\textwidth,trim=5cm 0 5cm 1cm, clip]{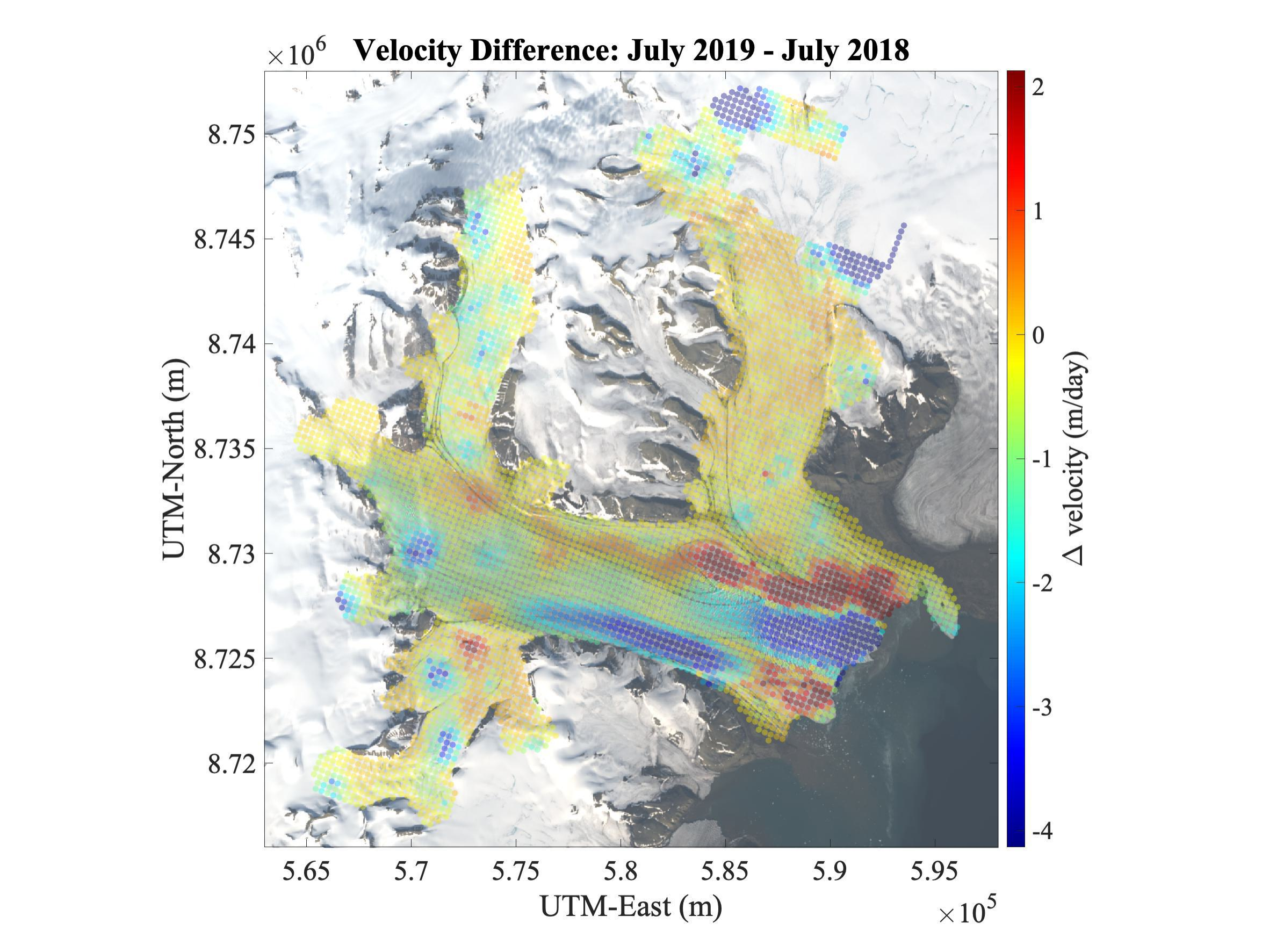}
\includegraphics[width=0.33\textwidth,trim=5cm 0cm 5cm 1cm, clip]{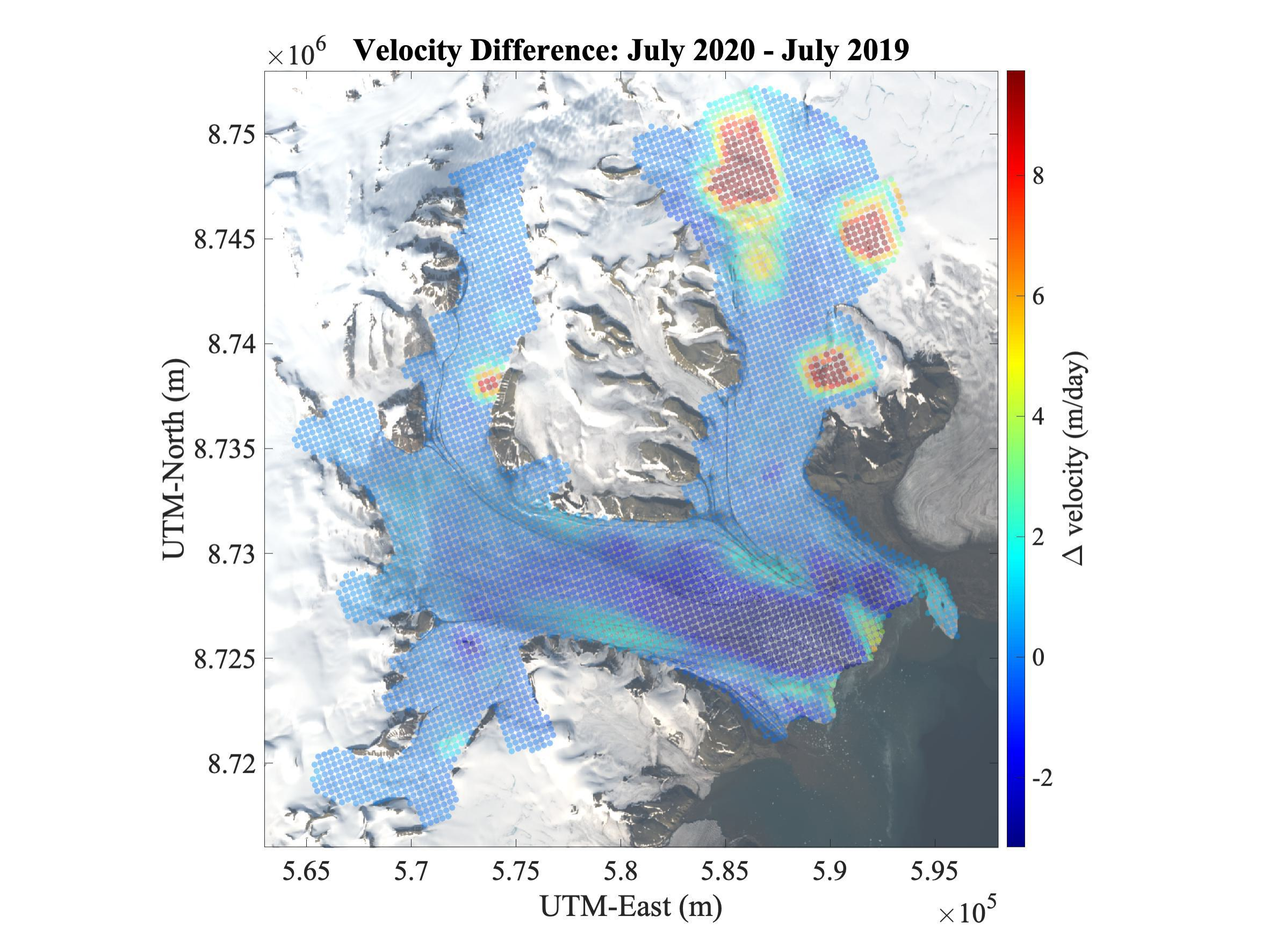}

\includegraphics[width=0.33\textwidth,trim=13.5cm 0 13.5cm 0cm, clip]{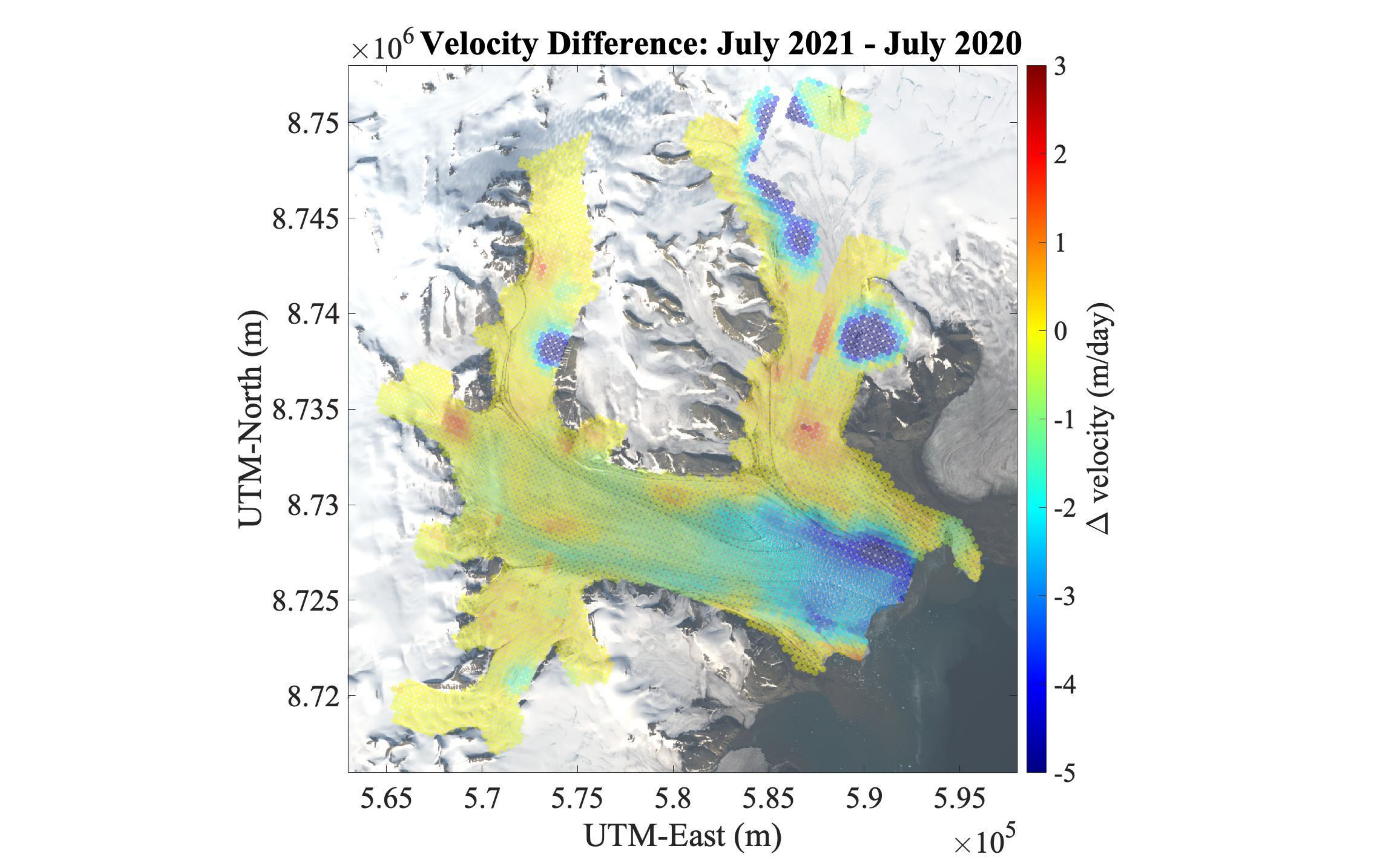}
\includegraphics[width=0.33\textwidth,trim=13.5cm 0 13.5cm 0cm, clip]{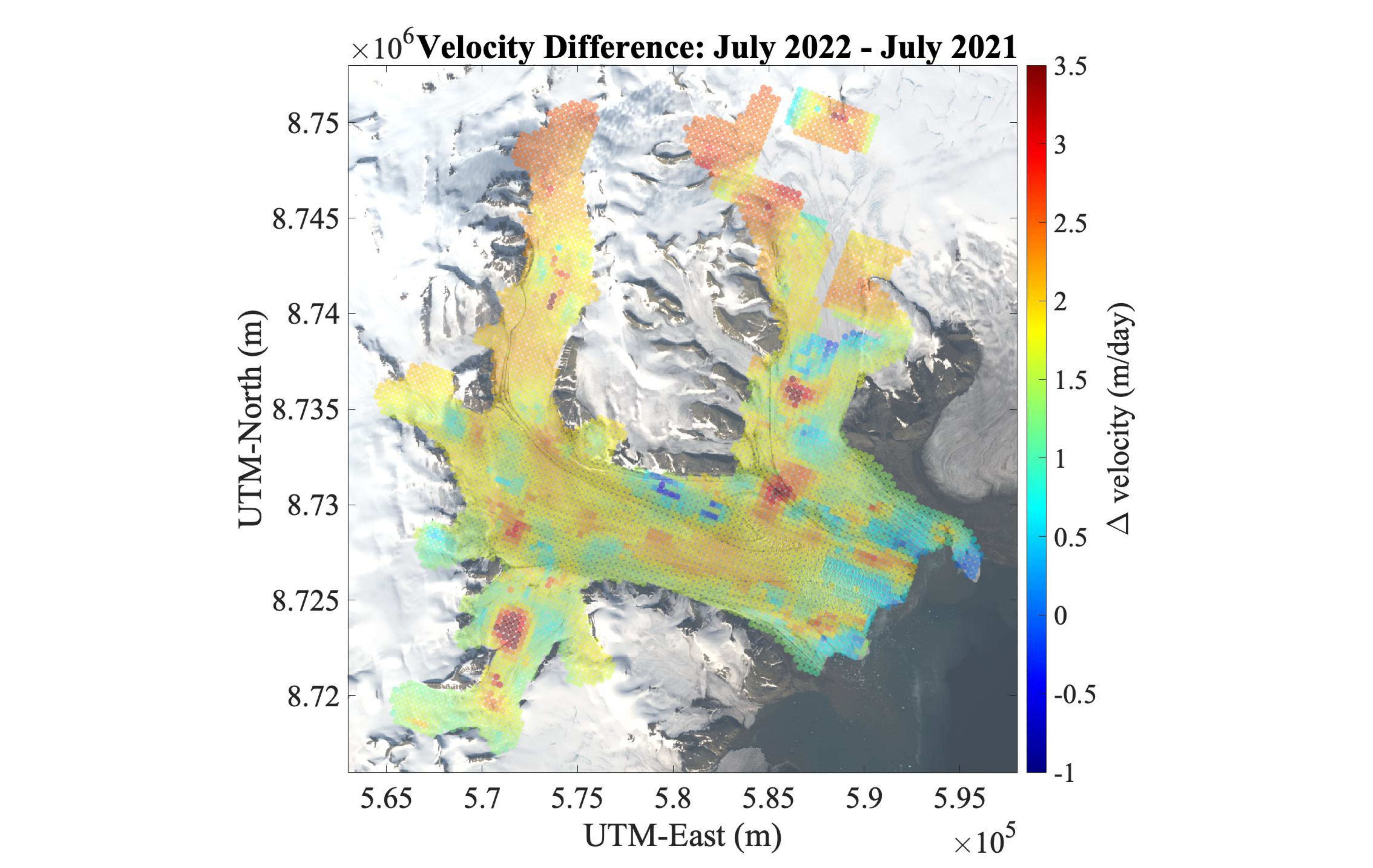}
\includegraphics[width=0.33\textwidth,trim=5cm 0.5cm 5cm 1cm, clip]{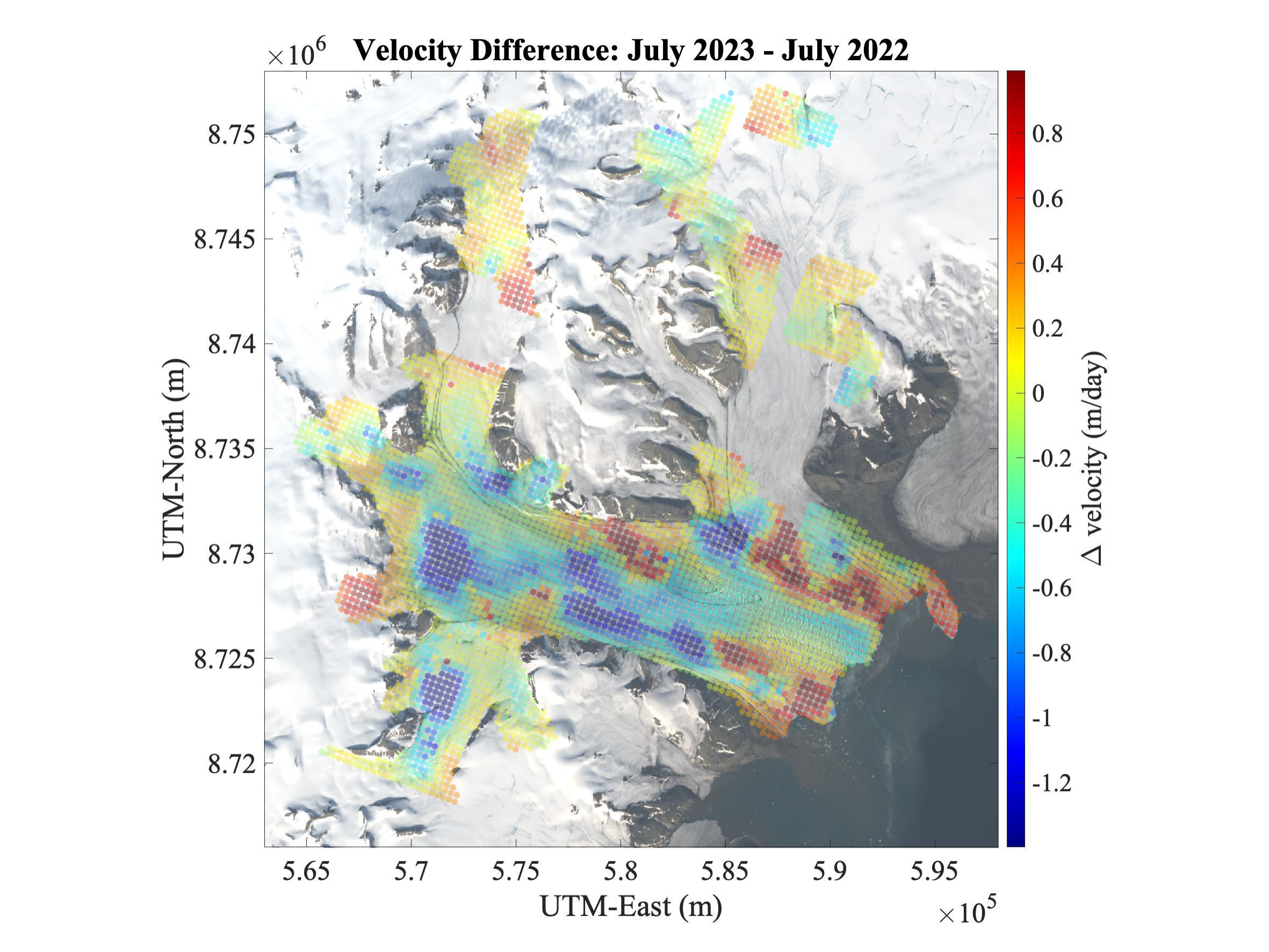}
\caption{Velocity Field Differences reflect the location of the kinematic surge wave. (a) Seasonal accelerations prior to the surge. (b) The surge wave begins near the terminus of the glacier. (c) The surge wave has propagated towards the source area and velocities are higher overall. (d) Wave has completed and secondary accelerations are present, especially on the southern side of Negribreen. (e) Secondary localized accelerations continue, especially on the northern side of Negribreen. (f) The fastest-moving regions begin to slow, local accelerations in tributary glaciers resemble the seasonal accelerations in (a).}
\end{figure}

 \begin{figure}[H]

  \includegraphics[width = .45\textwidth,trim=4.5cm 1.8cm 6.5cm 2.2cm, clip]{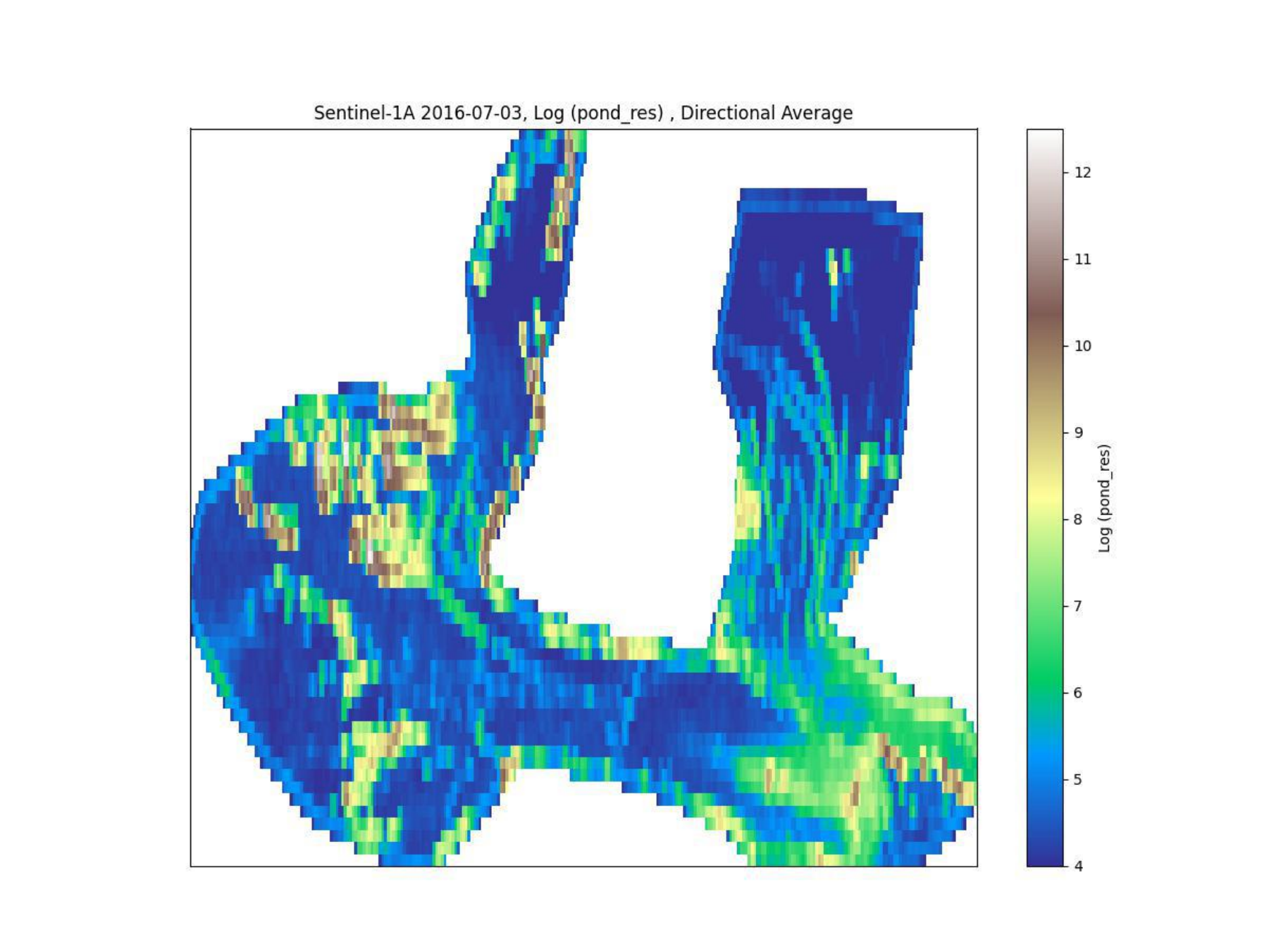}
  \includegraphics[width = .51\textwidth,trim=4.5cm 1.8cm 3.8cm 2.2cm, clip]{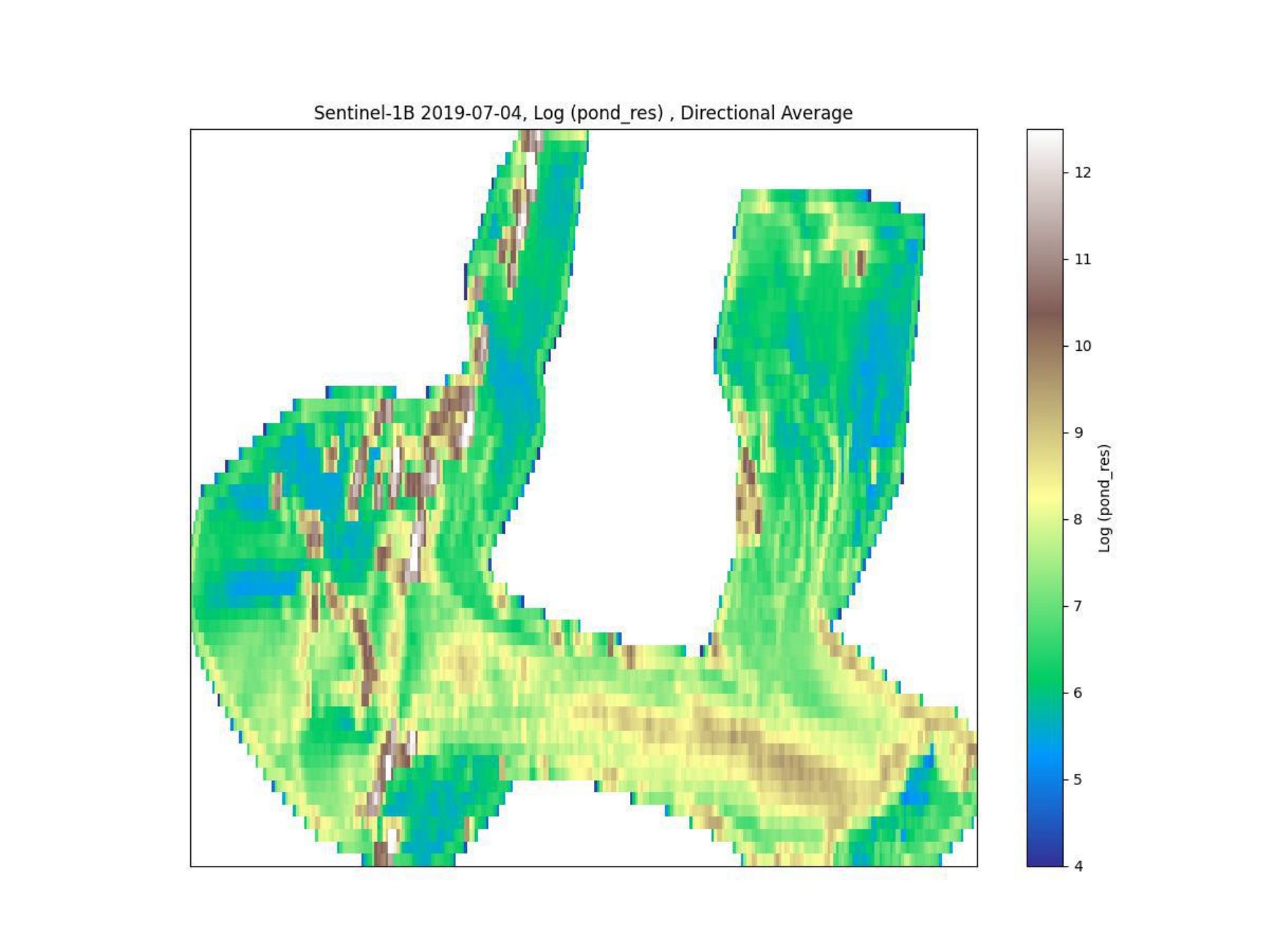}
  
  \includegraphics[width = .45\textwidth,trim=4.5cm 1.8cm 6.5cm 2.2cm, clip]{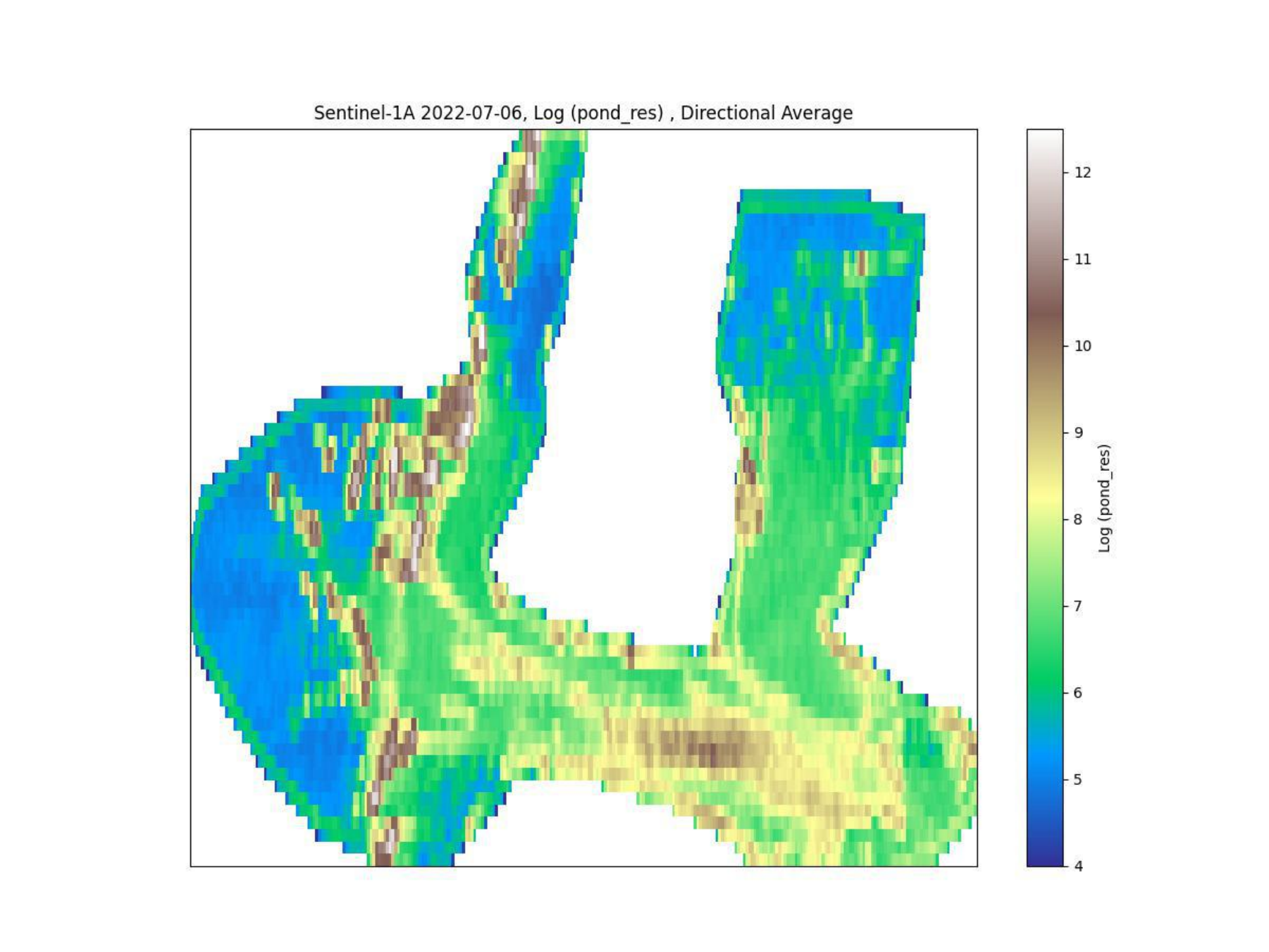}
  \includegraphics[width = .51\textwidth,trim=4.5cm 1.8cm 3.8cm 2.2cm, clip]{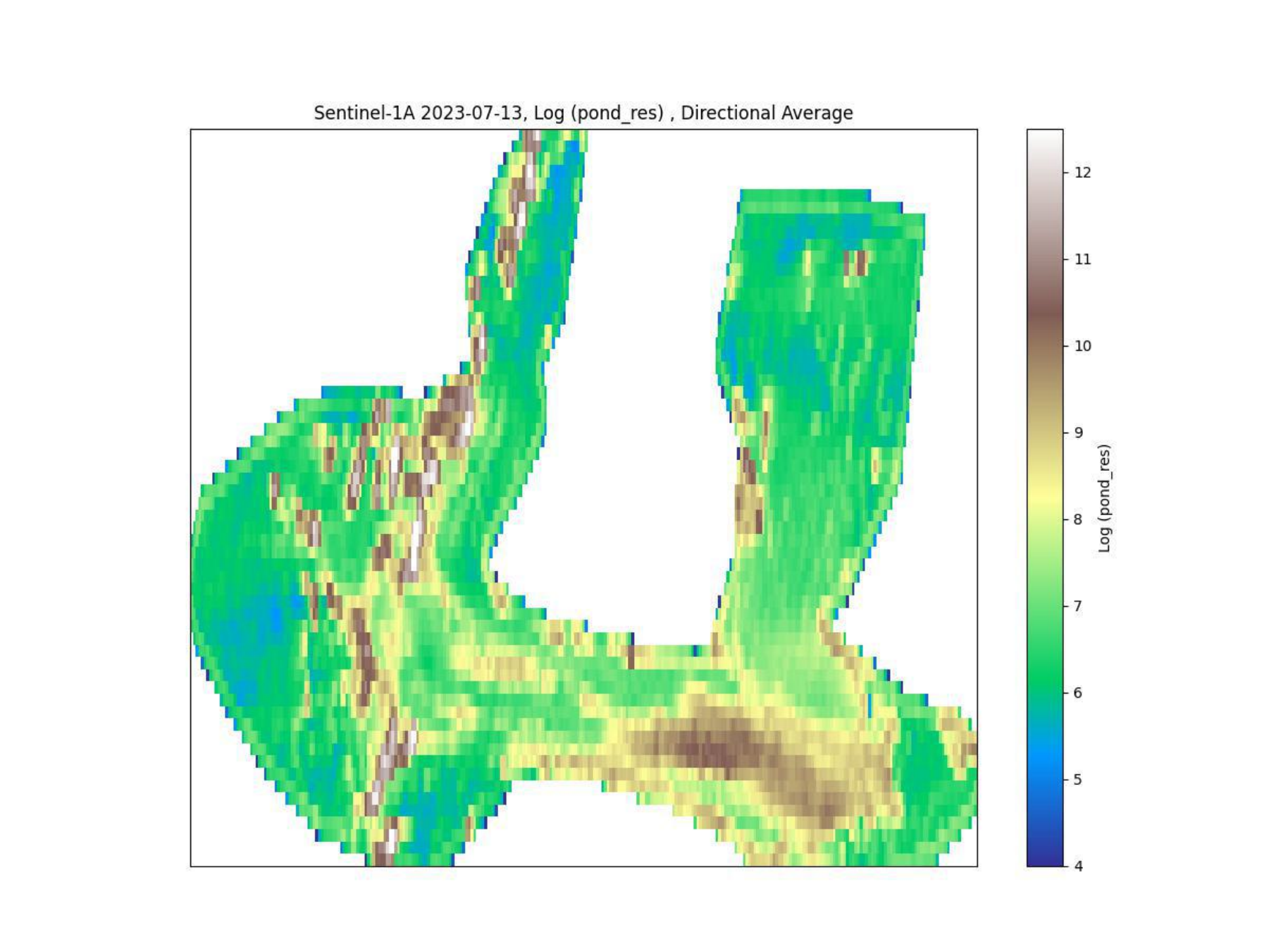}

 \caption{Ice surface roughness of Negribreen, at different stages of the surge, is represented by the geostatistical parameter, residual pond, indicated by $pond_{res}$ in equation 3. Residual pond is calculated for Sentinel-1 SAR imagery which has been clipped to the extent of the Negribreen Glacier System. Here, we plot the directional average of residual pond from 8 angles (0, 22.5, 45, 67.5, 90, 112.5, 135, 157 degrees) on the log scale.  \\
 (a) Roughness of NGS, July 3, 2016. Derived from Sentinel-1 SAR, HH polarization, \\
 image name: S1A\_IW\_GRDH\_1SSH\_20160703T154521\_20160703T154546\_011986\_0127DD\_64DD \\
  (b) Roughness of NGS, July 4, 2019. Derived from Sentinel-1 SAR, VV polarization, \\
  image name: S1B\_IW\_GRDH\_1SDV\_20190704T061445\_20190704T061510\_016980\_01FF2C\_3139 \\
 (c) Roughness of NGS, July 6, 2022. Derived from Sentinel-1 SAR, VV polarization, \\
 image name: S1A\_IW\_GRDH\_1SDV\_20220706T061545\_20220706T061610\_043976\_053FD2\_3FF9 \\
 (d) Roughness of NGS, July 13, 2023. Derived from Sentinel-1 SAR, VV polarization, \\
 image name: S1A\_IW\_GRDH\_1SDV\_20230713T061551\_20230713T061616\_049401\_05F0B7\_52A5}
 \end{figure}

\bibliographystyle{apalike}
\bibliography{watermap}